\documentclass{aa}
\makeatletter
\renewcommand*\aa@pageof{, page \thepage{} of \pageref*{LastPage}}
\makeatother
\usepackage{graphicx}
\usepackage{placeins}
\usepackage{txfonts}
\usepackage{hyperref}
\usepackage{xcolor}
\usepackage[font=small]{caption}

\begin{document}

\title{Gas thermodynamics meets galaxy kinematics:\\Joint mass measurements for eROSITA galaxy clusters}
\titlerunning{Gas thermodynamics meets galaxy kinematics}
\authorrunning{P. Li et al.}

   \author{Pengfei Li
          \inst{1,2}
          \and
          Ang Liu\inst{3}
          \and
          Matthias Kluge\inst{3}
          \and
          Johan Comparat\inst{3}
          \and
          Yong Tian\inst{4}
          \and
          Mariana P. Júlio\inst{2,5}
          \and
          Marcel S. Pawlowski\inst{2}
          \and
          Jeremy Sanders\inst{3}
          \and
          Esra Bulbul\inst{3}
          \and
          Axel Schwope\inst{2}
          \and
          Vittorio Ghirardini\inst{6,3}
          \and
          Xiaoyuan Zhang\inst{3}
          \and
          Y. Emre Bahar\inst{3}
          \and
          Miriam E. Ramos-Ceja\inst{3}
          \and
          Fabian Balzer\inst{3}
          \and
          Christian Garrel\inst{3}
          }

   \institute{School of Astronomy and Space Science, Nanjing University, Nanjing, Jiangsu 210023, China\\
              \email{pli@nju.edu.cn}
              \and
              Leibniz-Institute for Astrophysics,
              An der Sternwarte 16, 14482 Potsdam, Germany
              \and
              Max Planck Institute for Extraterrestrial Physics, Giessenbachstrasse 1, 85748 Garching, Germany 
              \and
              Institute of Astronomy, National Central University, Taoyuan 32001, Taiwan
              \and
              Institut f\"{u}r Physik und Astronomie, Universit\"{a}t Potsdam, Karl-Liebknecht-Straße 24/25, D-14476 Potsdam, Germany
              \and
              INAF, Osservatorio di Astrofisica e Scienza dello Spazio, via Piero Gobetti 93/3, 40129 Bologna, Italy
                     }

   \date{Received xxx; accepted xxx}
 
  \abstract
{The mass of galaxy clusters is a critical quantity for probing cluster cosmology and testing theories of gravity, but its measurement could be biased given assumptions are inevitable in order to make use of any approach. In this paper, we employ and compare two mass proxies for galaxy clusters: thermodynamics of the intracluster medium and kinematics of member galaxies. We select 22 galaxy clusters from the cluster catalog in the first SRG/eROSITA All-Sky Survey (eRASS1) that have sufficient optical and near-infrared observations. We generate multi-band images in the energy range of (0.3, 7) keV for each cluster, and derive their temperature profiles, gas mass profiles and hydrostatic mass profiles using a parametric approach that does not assume dark matter halo models. With spectroscopically confirmed member galaxies collected from multiple surveys, we numerically solve the spherical Jeans equation for their dynamical mass profiles. Our results quantify the correlation between dynamical mass and line-of-sight velocity dispersion, $\log M_{\rm dyn} = (1.296 \pm 0.001)\log(\sigma_{\rm los}^2r_{\rm proj}/G) - (3.87\pm0.23)$, with an rms scatter of 0.14 dex. We find the two mass proxies lead to roughly the same total mass, with no observed systematic bias. As such, the $\sigma_8$ tension is not specific to hydrostatic mass or weak lensing shears, but also appears with galaxy kinematics. Interestingly, the hydrostatic-to-dynamical mass ratios decrease slightly toward large radii, which could possibly be evidence for accreting galaxies in the outskirts. We also compare our hydrostatic masses with the latest weak lensing masses inferred with scaling relations. The comparison shows the weak lensing mass is significantly higher than our hydrostatic mass by $\sim$110\%. This might explain the significantly larger value of $\sigma_8$ from the latest measurement using eRASS1 clusters than almost all previous estimates in the literature. Finally, we test the radial acceleration relation (RAR) established in disk galaxies. We confirm the missing baryon problem in the inner region of galaxy clusters using three independent mass proxies for the first time. As ongoing and planned surveys are providing deeper X-ray observations and more galaxy spectra for cluster members, we expect to extend the study to cluster outskirts in the near future.
}

   \maketitle


\section{Introduction}
Studying the dynamics of galaxy clusters provides a useful way to constrain the nature of dark matter, as the cold dark matter model predicts that dark matter halos are in a universal profile \citep{Navarro1996}. It also helps test the universality of scaling relations that are mostly established in galaxies, such as the Baryonic Tully-Fisher relation \citep{McGaugh2000} and the radial acceleration relation \citep[RAR,][]{McGaugh2016PRL, Lelli2017, Li2018}. Both scientific objectives require robust measurements of cluster mass profiles, which is non-trivial, as it often suffers from both technical and observational difficulties. 

\begin{table*}[t]
        \centering
        \caption{Properties of the 22 selected clusters.}
        \label{tab:Cluster}
        \begin{tabular}{ccccccccccc}
                \hline
                 1eRASS Cluster & z$_{\rm spec}$\ \  & ${\rm RA_{BCG}}$ & ${\rm Dec_{BCG}}$ & ${\rm RA_{X-ray}}$ & ${\rm Dec_{X-ray}}$ & $N_{\rm spec}$ & $N_{\rm eff}$ & $R_{\rm 500, wl}$ & $M_{\rm 500, wl}$ & $M_{\rm 500, hydro}$\\
                     & & (deg.) & (deg.) & (deg.) & (deg.) & & & (kpc) &$(10^{13}M_\odot)$ & $(10^{13}M_\odot)$\\
                \hline
J004049.8-440743 & 0.3493 &  10.2082 & -44.1307 &  10.2077 & -44.1298 & 51 & 43.1 & 1250$^{+40}_{-35}$ & 79.9$^{+8.0}_{-6.6}$ & 26.9$^{+38.9}_{-12.8}$\\
J004207.1-283154 & 0.1075 &  10.5370 & -28.5357 &  10.5328 & -28.5334 & 45 & 39.4 & 1258$^{+37}_{-28}$ & 62.6$^{+5.7}_{-4.1}$ & 28.6$^{+18.9}_{-9.0}$\\
J024339.2-483339 & 0.4983 &  40.9121 & -48.5608 &  40.9115 & -48.5612 & 33 & 26.6 & 1248$^{+38}_{-28}$ & 94.7$^{+8.9}_{-6.4}$ & 31.8$^{+37.1}_{-14.3}$\\
J034656.2-543854 & 0.5286 &  56.7311 & -54.6486 &  56.7329 & -54.6475 & 45 & 37.3 & 1048$^{+37}_{-35}$ & 58.1$^{+6.4}_{-5.7}$ & 24.5$^{+30.5}_{-12.3}$\\
J043817.8-541917 & 0.4218 &  69.5737 & -54.3223 &  69.5736 & -54.3218 & 55 & 48.8 & 1292$^{+38}_{-28}$ & 95.9$^{+8.9}_{-6.0}$ & 85.6$^{+51.5}_{-32.2}$\\
J052806.2-525951 & 0.7681 &  82.0222 & -52.9981 &  82.0220 & -52.9973 & 33 & 25.0 & 752$^{+50}_{-61}$ & 28.6$^{+6.0}_{-6.4}$ & 16.1$^{+35.0}_{-9.7}$\\
J055942.9-524851 & 0.6087 &  89.9301 & -52.8242 &  89.9285 & -52.8224 & 42 & 33.6 & 984$^{+52}_{-68}$ & 52.8$^{+8.9}_{-10.2}$ & 30.1$^{+44.1}_{-17.6}$\\
J073220.0+313748 & 0.1705 & 113.0846 &  31.6335 & 113.0828 &  31.6301 & 78 & 62.4 & 1358$^{+37}_{-32}$ & 84.2$^{+7.1}_{-5.8}$ & 15.3$^{+8.3}_{-3.7}$\\
J073721.1+351739 & 0.2094 & 114.3372 &  35.2949 & 114.3365 &  35.2947 & 56 & 47.5 & 1297$^{+44}_{-37}$ & 76.5$^{+8.0}_{-6.4}$ & 33.4$^{+39.6}_{-16.0}$\\
J080056.9+360324 & 0.2866 & 120.2367 &  36.0565 & 120.2375 &  36.0577 & 97 & 80.4 & 1360$^{+39}_{-33}$ & 96.0$^{+8.3}_{-6.8}$ & 62.9$^{+70.9}_{-33.6}$\\
J082317.9+155700 & 0.1529 & 125.8304 &  15.9627 & 125.8297 &  15.9578 & 66 & 57.9 & 991$^{+43}_{-38}$ & 32.2$^{+4.4}_{-3.6}$ & 15.2$^{+30.3}_{-9.4}$\\
J084257.5+362208 & 0.2818 & 130.7399 &  36.3665 & 130.7412 &  36.3663 & 125 & 107.6 & 1481$^{+36}_{-31}$ & 123.1$^{+9.4}_{-7.6}$ & 107.3$^{+69.7}_{-47.1}$\\
J090131.5+030055 & 0.1936 & 135.3795 &  3.0157 & 135.3786 &   3.0149 & 51 & 41.8 & 1091$^{+39}_{-38}$ & 44.8$^{+4.9}_{-4.5}$ & 30.9$^{+20.5}_{-9.7}$\\
J101703.2+390250 & 0.2048 & 154.2652 &  39.0471 & 154.2645 &  39.0487 & 77 & 63.2 & 1373$^{+30}_{-38}$ & 90.2$^{+6.1}_{-7.3}$ & 15.5$^{+5.8}_{-3.9}$\\
J115518.0+232422 & 0.1422 & 178.8250 &  23.4049 & 178.8257 &  23.4057 & 59 & 54.6 & 1422$^{+33}_{-28}$ & 93.9$^{+6.7}_{-5.4}$ & 48.7$^{+35.1}_{-17.7}$\\
J121741.6+033931 & 0.0773 & 184.4214 &  3.6559 & 184.4221 &   3.6578 & 61 & 56.8 & 1252$^{+33}_{-22}$ & 60.1$^{+4.9}_{-3.0}$ & 37.5$^{+15.9}_{-8.9}$\\
J125922.4-041138 & 0.0843 & 194.8438 & -4.1961 & 194.8441 &  -4.1944 & 61 & 55.5 & 1323$^{+31}_{-32}$ & 71.4$^{+5.2}_{-5.1}$ & 49.5$^{+27.7}_{-15.7}$\\
J130252.8-023059 & 0.0838 & 195.7191 & -2.5164 & 195.7180 &  -2.5165 & 58 & 51.2 & 1014$^{+23}_{-26}$ & 32.1$^{+2.3}_{-2.3}$ & 14.1$^{+24.9}_{-5.1}$\\
J213056.8-645842 & 0.3163 & 322.7342 & -64.9779 & 322.7398 & -64.9789 & 37 & 31.3 & 1079$^{+56}_{-39}$ & 49.5$^{+8.2}_{-5.1}$ & 32.2$^{+42.3}_{-18.4}$\\
J213536.8-572622 & 0.4268 & 323.9060 & -57.4419 & 323.9046 & -57.4404 & 29 & 27.5 & 1021$^{+66}_{-65}$ & 47.7$^{+9.8}_{-8.6}$ & 15.9$^{+33.9}_{-8.9}$\\
J213800.9-600758 & 0.3188 & 324.5035 & -60.1317 & 324.5031 & -60.1333 & 28 & 26.5 & 1320$^{+34}_{-48}$ & 91.0$^{+7.2}_{-9.6}$ & 34.3$^{+55.7}_{-17.8}$\\
J235137.0-545253 & 0.3838 & 357.9086 & -54.8817 & 357.9078 & -54.8830 & 29 & 26.5 & 1032$^{+66}_{-57}$ & 46.8$^{+9.5}_{-7.4}$ & 20.5$^{+40.9}_{-12.1}$\\
                \hline
        \end{tabular}
        \tablefoot{The sample is selected from \citet{Bulbul2024}. The weighted spectroscopic redshift $z_{\rm spec}$, the coordinates of BCGs (${\rm RA_{BCG}}$, ${\rm Dec_{BCG}}$), and best-fit X-ray centers (${\rm RA_{X-ray}}$, ${\rm Dec_{X-ray}}$) are from \citet{Kluge2024} and \citet{Bulbul2024}, respectively. N$_{\rm spec}$ is the number of available spectroscopic member galaxies, while $N_{\rm eff}$ is the corresponding effective number of galaxy spectra, taking into account membership probabilities. The cluster radius $R_{\rm 500, wl}$ and total mass $M_{\rm 500, wl}$ are derived from scaling relations, as presented in \citet{Ghirardini2024}. $M_{\rm 500,hydro}$ is the hydrostatic mass derived in this paper.}
\end{table*}

The commonly employed dynamical tracer to measure cluster mass profiles is the hot cluster gas via the assumption of hydrostatic equilibrium. Hot cluster gas emits X-ray photons, so they can be measured with X-ray observations. The first all-sky X-ray telescope ROSAT led to the construction of the Northern (NORAS) catalog of 495 clusters \citep{Bohringer2000} and the Southern (REFLEX) catalog of 447 clusters \citep{Bohringer2004}. There are smaller sky-coverage but deeper observations with the XMM-Newton telescope, such as the XMM-Newton Cluster Survey \citep[XCS,][]{Mehrtens2012} for 503 galaxy clusters and the XMM-Newton Cluster Archive Super Survey \citep[X-CLASS][]{Koulouridis2021} for 1646 galaxy clusters. Deep follow-up programme are limited to tens to hundreds of galaxy clusters, such as the X-COP project \citep{Eckert2017} and the CHEX-MATE project \citep{CHEX-MATE2021}. A big milestone is made by the latest all-sky survey telescope, eROSITA \citep{Predehl2021}, which has observed more than 1.2 million X-ray sources \citep{Merloni2024}, and identified more than 26 thousand extended sources \citep{Bulbul2024} in the first SRG/eROSITA All-Sky Survey (eRASS1). More than 12 thousand extended sources have been identified by optical observations \citep{Kluge2024}, leading to the largest catalog of galaxy clusters with X-ray data available. The significant increase in the number of X-ray selected clusters suggests that studying cluster dynamics in a statistical way is becoming more and more promising. 

Cluster galaxies have also been used as dynamical tracers assuming they are in dynamical equilibrium. The total cluster mass can be estimated by building a scaling relation between the total velocity dispersion of cluster galaxies and the virial mass via simulations \citep{Biviano2006, Munari2013,Saro2013}. To derive dynamical mass profiles, one needs to solves the Jeans equation \citep[e.g.][]{Binney2008}. However, this approach suffers two technical difficulties. First, dynamical mass is related to radial velocity dispersion and 3-D galaxy distribution through spherical Jeans equation \citep{Binney2008}, but observationally only projected spatial distributions and line-of-sight velocity dispersion of member galaxies are measurable. Second, there is a serious degeneracy between dynamical mass and velocity anisotropy. Degeneracy is a general problem in astronomy, which can lead to unrealistic estimations of parameters and even result in absurd conclusion \citep[e.g. see][]{Li2021}. \citet{Mamon2013, Biviano2013} proposed to tackle these problems by parameterizing mass profiles, tracer distribution functions, and velocity anisotropy profiles using 3D functions. These functions are determined by matching their projected forms to observed phase-space information. \citet{Li2023} also used parameterized profiles but with more flexible functions, and addressed the mass-velocity anisotropy degeneracy by introducing additional constraints that involve two virial shape parameters, namely the fourth order of velocity dispersion \citep{Merrifield1990}. As a result, the dynamical mass profiles can be well derived \citep[see Appendix B in][]{Li2023}. 

Both approaches assume galaxy clusters are relaxed, which cannot be independently, robustly verified. An interesting fact is that the two approaches assume different relaxations, one on hot intracluster gas via thermodynamics, and the other on member galaxies through kinematic motions. It is therefore interesting to compare these two mass estimators in a statistical sense. This is particularly interesting because of the $\sigma_8$ tension \citep{Planck2014XXIX}, since the cluster mass measurements using hot gas are thought to be biased lower due to non-thermal flows \citep[e.g.][]{Lau2009, Vazza2009, Battaglia2012, Nelson2014, Shi2015,Biffi2016} and thereby lead to lower matter fluctuation compared with that from cosmic microwave background fluctuations \citep{Planck2016XIII}.

In this paper, we select a subsample of galaxy clusters from the cluster catalog in the first all-sky survey of eROSITA \citep{Bulbul2024} based on their optical properties \citep{Kluge2024}. We measure their cluster mass profiles using both hot intracluster gas and cluster galaxies as dynamical tracers and make a systematic comparison between these two mass proxies. The paper is organized as follows: Section \ref{sec:sample} describes our sample selection criteria; Section \ref{sec:gasthermo} \& \ref{sec:galkin} present the results of X-ray analysis and kinematics analysis, respectively; Section \ref{sec:masscomparison} compares the mass profiles measured with different estimators; Section \ref{sec:RAR} shows a test for the universality of the RAR; Section \ref{sec:discuss} discusses the results and concludes the paper.

\section{Sample selection}\label{sec:sample}

In this work, we select galaxy clusters from the eRASS1 cluster catalog \citep{Bulbul2024}. The main selection criterion is that there are both X-ray data and a sufficient number of member galaxies with spectroscopic redshift information, so that both gas thermodynamics and galaxy kinematics are applicable to measure cluster mass profiles. Since eROSITA clusters are X-ray selected, the major limitation comes from the effective number of member galaxies, defined as the summation of membership probability of cluster galaxies, 
\begin{equation}
    \lambda=\sum_i p_{mem,i},
\end{equation}
where $p_{\rm mem,i}$ denotes the membership probability of individual galaxies. This definition differs from the optical richness \citep{Kluge2024} by a scaling factor, which compensates for masked area and limited observational depth. Since we study galaxy kinematics, completeness is less important \citep[see appendix A in][]{Li2023}. As such, we do not need the true richness, which requires a complete sample. However, we shall distinguish photometric and spectroscopic richness. Photometric richness is simply the effective number of galaxies that are brighter than a threshold magnitude, while spectroscopic richness refers to spectroscopically confirmed galaxies. Photometric richness is generally larger than spectroscopic richness, given spectra are more expensive to obtain. We determine galaxy number density based on photometric richness, since it is more complete; while we measure velocity dispersion using spectroscopic redshifts, given they are more accurate. To ensure robustness, we use 25 effective galaxies to calculate the line-of-sight velocity dispersion for each radial bin, and require at least two bins for galaxy number density and one bin for velocity dispersion. Therefore, our major limitation is the availability of galaxy spectra, which significantly reduces the size of our sample.

In order to assume spherical symmetry, we select only clusters that can be approximately modeled with a single center, namely requiring the offset between the X-ray center and optical center smaller than 60 kpc, following \citet{Tian2021}. X-ray centers are determined by fitting X-ray images using the public code \texttt{MBProj2D} by \citet{Sanders2018} and the best-fit X-ray centers are given in \citet{Bulbul2024}. Optical centers correspond to the positions with the minimum gravitational potential due to cluster galaxies, or equivalently the positions with the highest luminosity weighted galaxy number density \citep{Rykoff2014, Rykoff2016, Kluge2024}. We identify the brightest cluster galaxies (BCGs) based on their z-band magnitude, and select clusters with their BCGs in the optical centers. Given these criteria, although X-ray and optical data are analyzed using their own centers, we do not distinguish the centers of the derived mass profiles throughout the paper.

\begin{figure}[h!]
    \centering
    \includegraphics[scale=0.35]{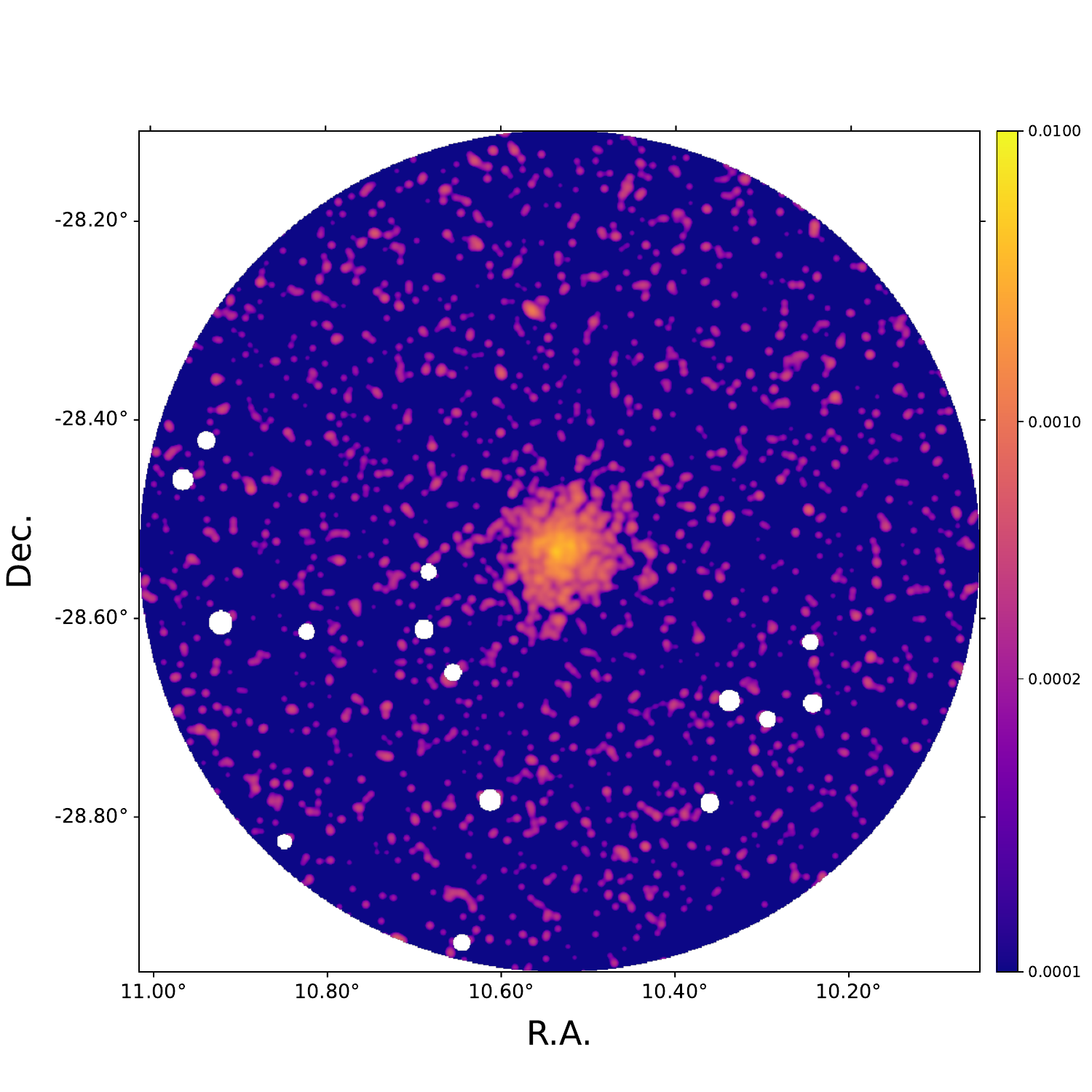}
    \caption{The X-ray image of 1eRASS J004207.0-283154 with exposure map corrected in the energy range of (0.3-7.0) keV, smoothed with Gaussian filters. Small white circles show the masked regions that are selected using the X-ray source catalog \citep{Merloni2024} and the cluster catalog \citep{Bulbul2024}. The color bar labels the exposure-map corrected X-ray photon counts.}
    \label{fig:image}
\end{figure}

We exclude clusters that are undergoing merging, such as cluster J090911.9+105835, as their velocity distribution histograms of member galaxies present clear double peaks. In summary, we select clusters from the eRASS1 cluster catalog by \citet{Bulbul2024} based on three criteria: (1) there are at least 25 effective spectroscopic galaxies and 50 effective photometric galaxies; (2) their BCG position, X-ray and optical centers are roughly consistent; (3) clusters are not merging. Eventually, our sample includes 22 clusters. Their properties and other names in various published catalogs are given in Table \ref{tab:Cluster} and \ref{tab:clusternames}, respectively. The sample covers a large range in redshift, up to $\sim$0.77. The data quality also varies significantly, from around 50 photons to more than 2000 photons in the band of (0.2, 2.3) keV. Though the current sample is small compared to the large cluster sample in the eRASS1 catalog, we expect to significantly expand the sample in a few years with the help of dedicated observations by the black hole mapper of SDSS-V \citep{Almeida2023} and the galaxy cluster survey of 4MOST \citep{Finoguenov2019, deJong2019}. The X-ray data quality will also be 5 times better once the eRASS:5 is made available.

\begin{figure*}
    \centering
    \includegraphics[scale=0.36]{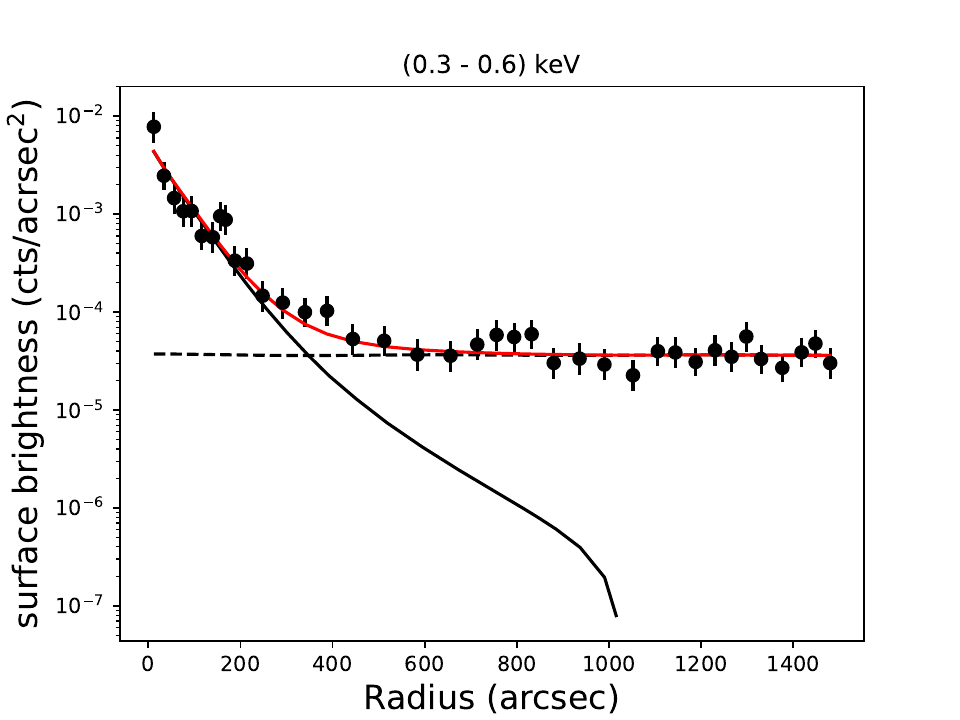}\includegraphics[scale=0.36]{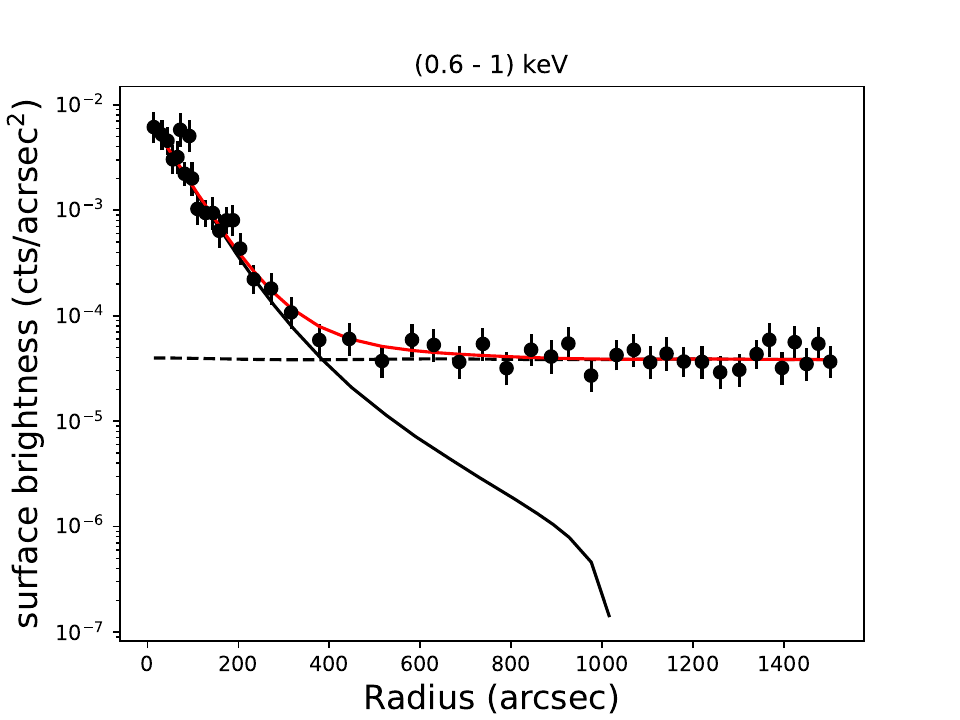}\includegraphics[scale=0.36]{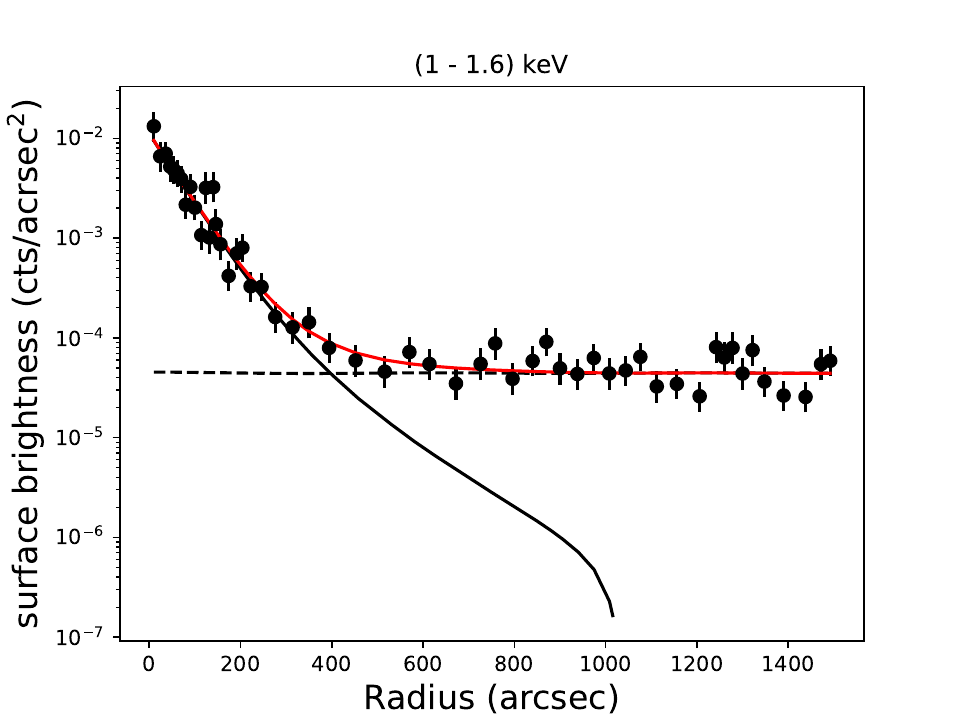}\\
    \includegraphics[scale=0.36]{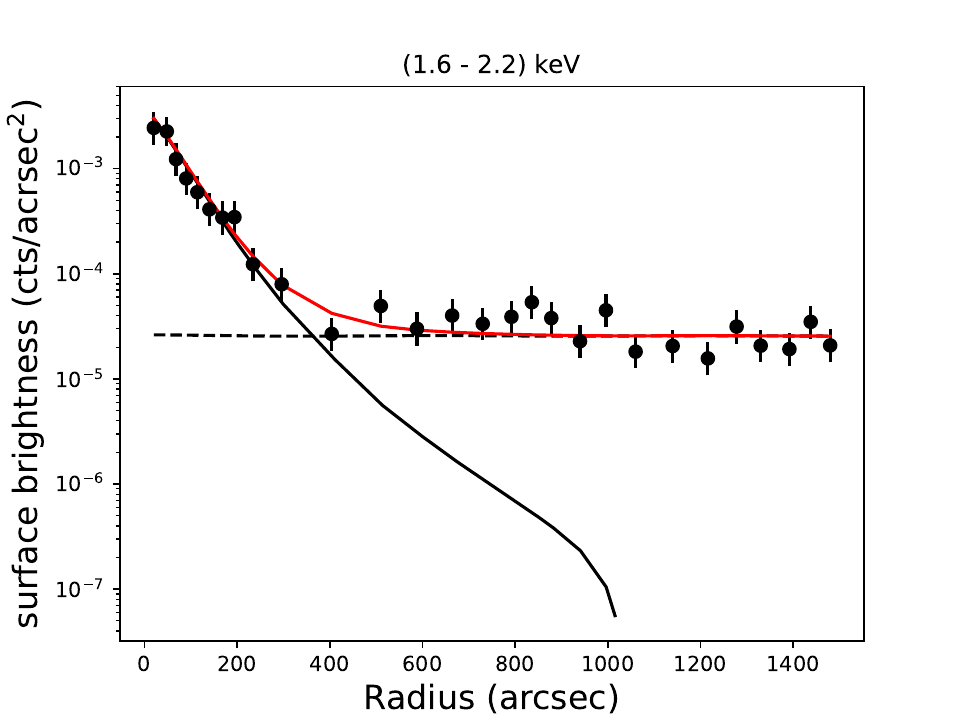}\includegraphics[scale=0.36]{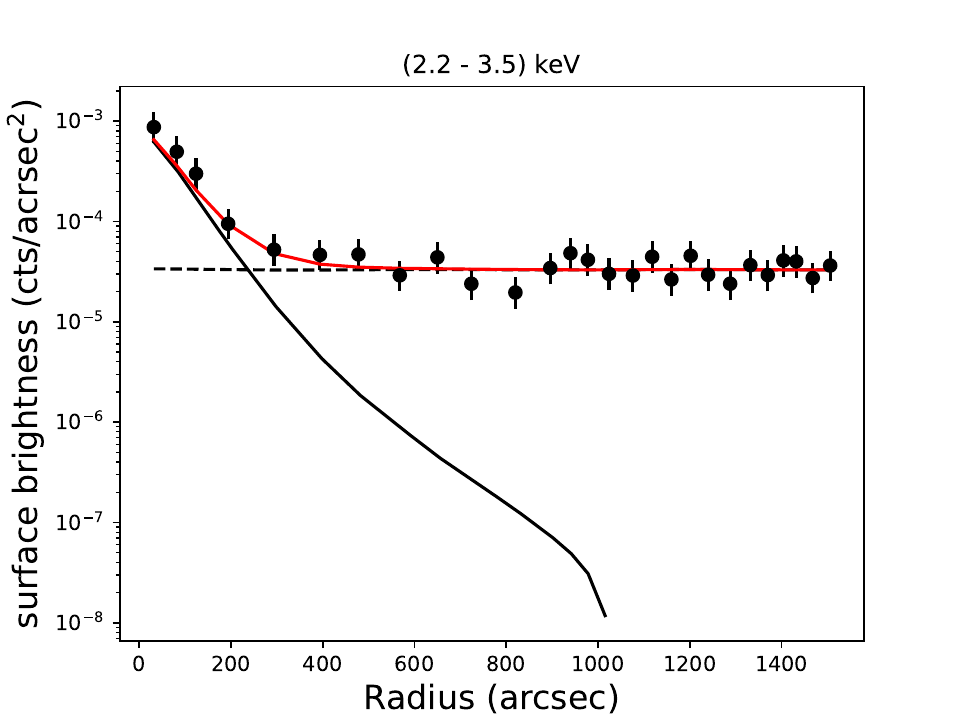}\includegraphics[scale=0.36]{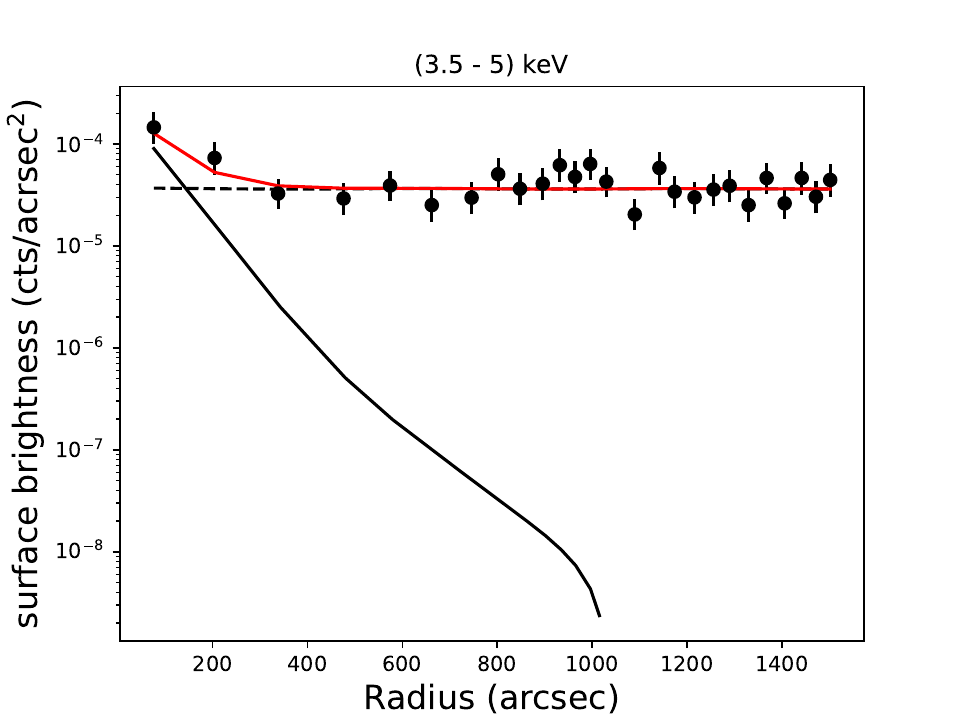}
    \caption{Surface brightness profiles of 1eRASS J004207.0-283154 in six different energy bands. The azimuthally averaged surface brightness in each band is compared with the total model (red lines), which is the summation of the constant background (black dashed lines) and the cluster model (black solid lines). Cluster models are truncated at 2 Mpc.}
    \label{fig:fitSB}
\end{figure*}

\section{Gas thermodynamics from X-ray observations}\label{sec:gasthermo}

\subsection{Generating X-ray images}

The X-ray data for the selected 22 clusters are from the first all-sky survey by eROSITA from December 2019 to June 2020. There is one cluster (1eRASS J090131.5+030055) appearing in the eFEDS cluster catalog \citep{Liu2022}, so we combine the X-ray data from both data releases to increase the exposure time. We process the data with the eROSITA Science Analysis Software System \citep[\texttt{eSASS}\footnote{Version eSASSusers\_211214\_0\_4},][]{Brunner2022}. From the calibrated photon event lists, we construct images and exposure maps in seven energy bands: [0.3–0.6], [0.6–1.0], [1.0–1.6], [1.6–2.2], [2.2–3.5], [3.5–5.0], [5.0–7.0] keV. X-ray photons are collected from all seven telescope modules (TMs) for hard bands. For the two soft bands with energy lower than 1 keV, we exclude photons from TM 5 and TM 7 due to light leak \citep{Predehl2021}. We set a fixed image size as 6 Mpc $\times$ 6 Mpc, which is sufficiently large to robustly estimate background levels. 

Figure \ref{fig:image} shows an example image for 1eRASS J004207.0-283154 in the energy range [0.3, 7] keV. It shows the exposure map-corrected counts. We mask the point sources that are identified in \citet{Merloni2024} and galaxy clusters in \citet{Bulbul2024}. The radii of the masked point sources are determined by the point spread function (PSF) scaled by their count rates \citep[ML\_RATE\_1 in][]{Merloni2024}, so that brighter sources are masked with a larger area; while the radii of masked background clusters are chosen as their half $R_{500}$. We also mask the four corners, so only the colored region is used for the following analysis.

\begin{figure*}
    \centering
    \includegraphics[scale=0.52]{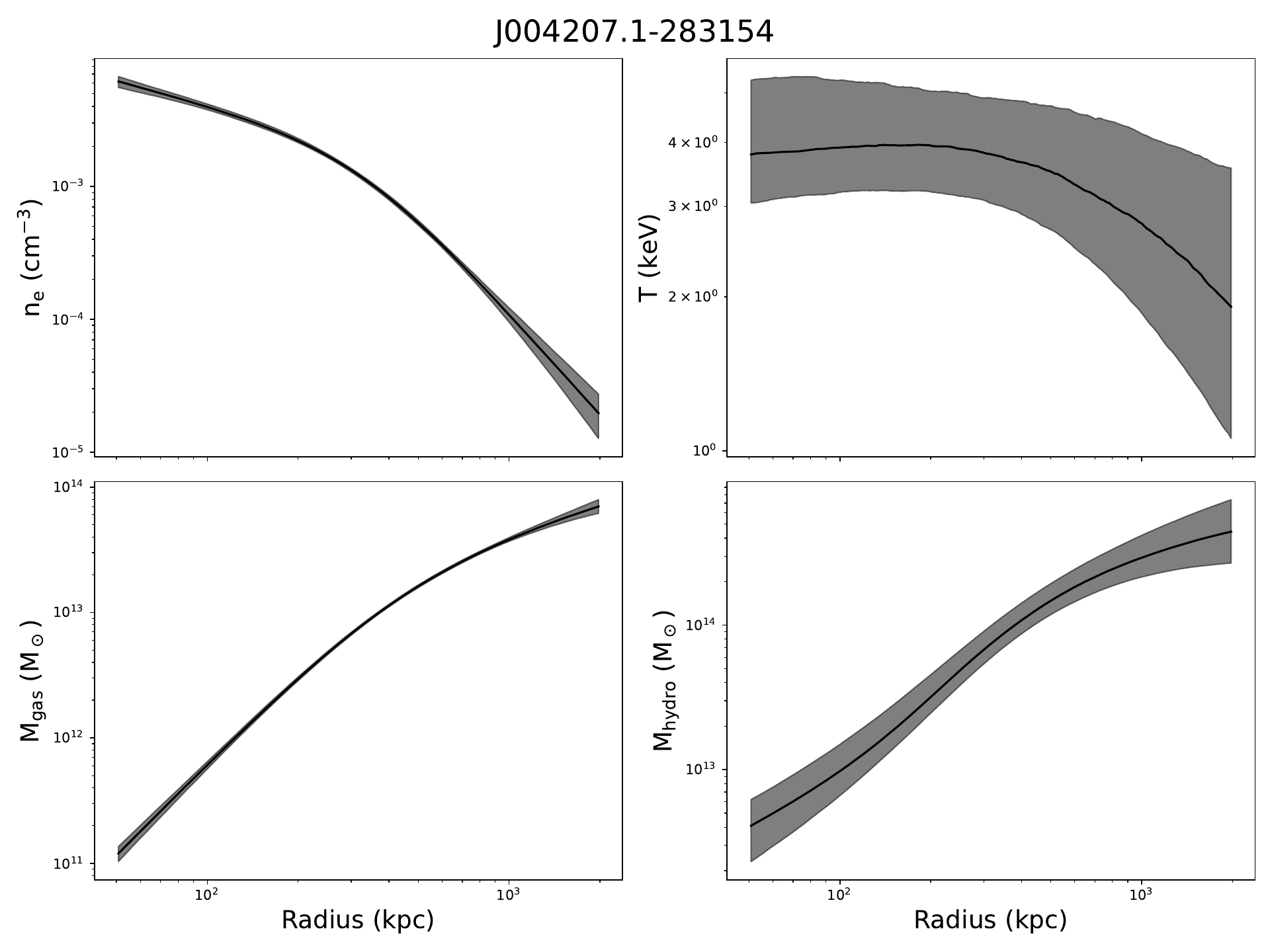}
    \caption{The electron number density (top-left), temperature (top-right), cumulative gas mass (bottom-left), and hydrostatic mass (bottom-right) profiles of 1eRASS J004207.0-283154. Solid lines and shadow regions are the median values and the corresponding 68\% confidence limits, respectively. Both the mean values and the uncertainties are directly extracted from their Markov Chains. Notice that the profiles can also be derived using the median values in the posterior distributions of the fitting parameters in Table \ref{tab:fits}. This approach could lead to slightly different profiles from that extracted from Markov Chains, but consistent within 1$\sigma$. The latter generally results in non-smooth curves. The complete figure set for all clusters is available in the Appendix (Figure \ref{fig:figset}). The data for these profiles are available at Github\protect\footnotemark.}
    \label{fig:Xrayfits}
\end{figure*}

\subsection{Fitting X-ray images}

We fit the X-ray images with the public Multi-Band Projector 2D code \citep[\texttt{MBProj2D},][]{Sanders2018, Bulbul2024}. \texttt{MBProj2D} fits multiple band images and their associated backgrounds simultaneously using a Markov Chain Monte Carlo approach. We assume the same PSF for all the seven TMs. Both the ancillary response file (ARF) and the redistribution matrix file (RMF) are implemented in the code. The Galactic hydrogen column density is also properly taken into account.

By fitting multi-band images, we measure temperature profiles, electron number density profiles and thereby gas mass profiles. Usually, the temperature profiles of galaxy clusters are measured from X-ray spectra. However, this requires long exposure time, which is hardly achievable for many eROSITA clusters. From multi-band images, we can derive roughly the same temperature profiles as that from spectra \citep{Liu2023}. We parameterize the temperature profile using the McDonald model \citep{McDonald2014},
\begin{equation}
    T(r) = T_0\frac{(r/r_{\rm cool})^{a_{\rm cool}}+T_{\rm min}/T_0}{(r/r_{\rm cool})^{a_{\rm cool}}+1}\cdot\frac{(r/r_t)^{-a}}{[1+(r/r_t)^b]^{c/b}},
    \label{eq:temperature}
\end{equation}
where cool cores within $r_{\rm cool}$ are accommodated with the temperature of $T_{\rm min}$. We fix the parameters $a=0$ and $b=2$. Therefore, the temperature profile has six free parameters. We impose the following flat priors: $0<a_{\rm cool}<2$, $0<c<2$, $-2.3<\ln(T_0/{\rm keV})<3$, $0.1<T_{\rm min}/T_0<1.5$, $0.1 < r_{\rm cool}<2000\ {\rm kpc}$, $10 < r_t<5000\ {\rm kpc}$. Part of the priors are imposed following \citet{Ghirardini2019}. We slightly narrow down some ranges considering the actual fits.

We model the electron number density profile $n_e$ using the single-$\beta$ component Vikhlinin model \citep{Vikhlinin2006},
\begin{equation}
    n_pn_e = n^2_0\frac{(x/r_c)^\alpha}{(1+x^2/r_c^2)^{3\beta-\alpha/2}}\frac{1}{(1+x^\gamma/r^\gamma_s)^{\epsilon/\gamma}},
    \label{eq:electronnum}
\end{equation}
where $n_p$ is the proton number density and assumed to be $n_p = n_e/1.21$. We fix $\gamma=3$ and impose flat priors on the six free parameters: $0<\alpha<5$, $0<\beta<5$, $0<\epsilon<5$, $-8<\ln(n_0/{\rm cm^{-3}})<2$, $100<r_s<5000\ {\rm kpc}$, $1<r_c<500\ {\rm kpc}$. Those ranges are much wider than the typical variations in the properties of galaxy clusters.

We assume a constant metallicity profile with 0.3 solar abundance using the solar value from \citet{Lodders2003}, and spatially constant background emissions in different energy bands. In order to minimize the number of free parameters, we fix the centers of galaxy clusters using the values from \citet{Bulbul2024}. As such, there are 19 fitting parameters in total: six in the density profile, six in the temperature profile, and seven for the background levels in the seven energy bands. We use the $emcee$ hammer \citep{Foreman-Mackey2013} to draw their posterior distributions. We setup 190 random walkers and run 5000 iterations with the first 1000 steps being the burnt-in phase. These are more than sufficient to achieve convergence.

We use the non-hydro modeling for clusters in \texttt{MBProj2D}, which incorporates the aforementioned density, temperature, and metallicity profiles, but does not assume any models for dark matter halos. One major improvement of \texttt{MBProj2D} over its previous version is that it fits 2D images rather than azimuthally averaged surface brightness profiles. Therefore, we can check the fit quality by examining the corresponding surface brightness profile. As an example, Figure \ref{fig:fitSB} shows the surface brightness profiles of 1eRASS J004207.0-283154 in six energy bands. We do not show the profile in the hardest band (5-7 keV) as it is dominated by background emissions and thereby remains flat. The binned surface brightness profiles in all the energy bands are well described by the total model, which is the summation of the background level and the cluster model. This verifies the image fitting.

We directly derive electron number density profiles and temperature profiles from their Markov Chains by calculating their median values, rather than using the functional forms and their best-fit parameters. As such, the median profiles may not be smooth curves (for example see Figure \ref{fig:Xrayfits}). The two approaches usually lead to slightly different results, but generally consistent within their errors. We list the best-fit parameters in Table \ref{tab:fits} so that one can easily reproduce these profiles. The top-right panel of Figure \ref{fig:Xrayfits} shows that the errors on the temperature profile are quite large. This is not because we only fit images or due to an insufficient number of photons, given \citet{Liu2023} found similarly large errors using the X-ray spectra for a luminous cluster. In fact, the temperature of ICM is generally poorly constrained.

\begin{table*}[ht!]
        \centering
        \caption{The best-fit parameters for the temperature profile (first six parameters) and the electron number density profile (last six parameters).}
        \label{tab:fits}
        \begin{tabular}{c|cccccc|cccccc}
                \hline
                1eRASS Cluster & \multicolumn{6}{c|}{Parameters for temperature profile in eq. \ref{eq:temperature}} & \multicolumn{6}{c}{Parameters for electron number density profile in eq. \ref{eq:electronnum}}\\
                 & a$_{\rm cool}$ & c & T$_0$ & T$_{\rm min}$ & r$_{\rm cool}$ & r$_t$ & $\alpha$ & $\beta$ & $\epsilon$ & n$_0$ & r$_s$ & r$_c$ \\
                     & &  & (keV) & (keV) & (kpc) & (kpc)& &&&10$^{-3}$cm$^{-3}$&(kpc)&(kpc)\\
                \hline
J004049.8-440743 & 0.80 & 0.99 & 9.40 & 1.89 & 33.90 & 498.5 & 1.69 & 0.23 & 2.41 & 6.51 & 152.3 & 135.8\\
J004207.1-283154 & 1.15 & 0.65 & 5.27 & 2.34 & 55.59 & 641.6 & 1.28 & 0.24 & 3.63 & 3.45 & 330.2 & 148.4\\
J024339.2-483339 & 0.95 & 0.80 & 12.06 & 1.58 & 2.53 & 362.7 & 2.91 & 0.32 & 1.70 & 8.38 & 222.3 & 66.5\\
J034656.2-543854 & 1.00 & 1.00 & 8.61 & 1.16 & 1.69 & 486.1 & 1.86 & 0.39 & 1.50 & 6.12 & 339.9 & 85.6\\
J043817.8-541917 & 1.32 & 0.62 & 14.24 & 1.15 & 21.58 & 713.8 & 0.60 & 0.11 & 3.14 & 13.05 & 143.0 & 92.5\\
J052806.2-525951 & 1.01 & 0.87 & 7.86 & 1.52 & 1.45 & 449.5 & 2.88 & 0.29 & 2.30 & 5.68 & 487.5 & 41.2\\
J055942.9-524851 & 1.04 & 0.64 & 10.13 & 1.28 & 2.86 & 483.6 & 0.54 & 0.47 & 1.72 & 2.05 & 745.7 & 341.7\\
J073220.0+313748 & 1.43 & 1.28 & 9.93 & 1.84 & 3.47 & 224.2 & 1.26 & 0.35 & 1.78 & 9.65 & 184.2 & 86.4\\
J073721.1+351739 & 1.08 & 0.86 & 7.96 & 1.11 & 57.51 & 715.1 & 2.50 & 0.27 & 2.78 & 7.08 & 258.9 & 56.7\\
J080056.9+360324 & 1.27 & 0.57 & 11.89 & 1.11 & 16.63 & 586.8 & 1.75 & 0.29 & 2.51 & 14.2 & 194.7 & 52.1\\
J082317.9+155700 & 1.04 & 0.66 & 6.50 & 1.65 & 23.44 & 537.8 & 1.33 & 0.34 & 1.11 & 1.24 & 333.9 & 183.6\\
J084257.5+362208 & 1.18 & 0.59 & 14.13 & 1.18 & 1.01 & 1004.4 & 0.51 & 0.23 & 2.55 & 5.97 & 237.0 & 245.4\\
J090131.5+030055 & 1.16 & 0.95 & 8.14 & 2.22 & 52.78 & 638.9 & 1.16 & 0.47 & 0.95 & 2.97 & 212.8 & 154.9\\
J101703.2+390250 & 1.37 & 1.22 & 6.56 & 1.18 & 12.48 & 363.9 & 1.51 & 0.17 & 2.48 & 6.49 & 166.1 & 108.4\\
J115518.0+232422 & 0.67 & 1.13 & 10.78 & 1.31 & 7.41 & 727.4 & 1.57 & 0.26 & 2.51 & 15.78 & 235.2 & 38.4\\
J121741.6+033931 & 1.03 & 1.06 & 7.11 & 1.60 & 97.02 & 808.7 & 1.04 & 0.20 & 3.18 & 2.34 & 357.9 & 200.5\\
J125922.4-041138 & 0.53 & 0.53 & 12.99 & 2.27 & 264.68 & 436.2 & 0.89 & 0.15 & 3.03 & 6.19 & 175.2 & 116.0\\
J130252.8-023059 & 1.40 & 0.81 & 1.76 & 2.97 & 785.38 & 943.3 & 1.88 & 0.42 & 1.40 & 1.01 & 294.3 & 242.5\\
J213056.8-645842 & 1.07 & 0.75 & 10.90 & 1.37 & 3.59 & 473.2 & 2.79 & 0.37 & 1.67 & 64.89 & 941.3 & 5.8\\
J213536.8-572622 & 1.08 & 0.56 & 6.18 & 2.09 & 36.96 & 470.7 & 1.60 & 0.34 & 2.33 & 9.70 & 330.5 & 40.3\\
J213800.9-600758 & 1.18 & 0.74 & 8.55 & 1.19 & 0.87 & 327.0 & 0.59 & 0.49 & 2.90 & 4.93 & 285.8 & 301.9\\
J235137.0-545253 & 1.18 & 0.80 & 8.28 & 1.11 & 53.51 & 426.9 & 1.62 & 0.38 & 2.35 & 3.30 & 573.5 & 142.3\\
                \hline
        \end{tabular}
        \tablefoot{The parameters are fitted using \texttt{MBProj2D} \citep{Sanders2018}. The values correspond to the medians of the Markov Chains, which can be used to reproduce our temperature and density profiles within 1$\sigma$}
\end{table*}

\footnotetext{https://pengfeili0606.github.io/data}

\subsection{Gas mass and hydrostatic mass}

The cumulative gas mass profiles can be calculated from the electron number density by
\begin{equation}
    M_{\rm gas}(<r) = 4\pi\frac{A}{Z}m_p\int_0^rn_e(r)r^2{\rm d}r,
\end{equation}
where $m_p$ is the mass of protons, $A\simeq1.4$ is the mean nuclear mass number and $Z\simeq1.2$ is the charge number of the intracluster medium (ICM) with 0.3 solar abundance \citep{Anders1989}. To derive the total mass profile using the ICM as the dynamical tracer, one assumes that the ICM is in hydrostatic equilibrium, i.e. the pressure gradient counteracts the radial gravity. One relates the pressure profile of the ICM to its temperature and electron number density using the ideal gas law, $p(r)=n_e(r)k_BT(r)$, where $k_B$ is the Boltzmann constant. As such, the hydrostatic mass is given by
\begin{equation}
    M_{\rm hydro}(<r) = -\frac{k_BTr}{G\mu m_p}\Big(\frac{{\rm d}\ln\rho_g}{{\rm d}\ln r}+\frac{{\rm d}\ln T}{{\rm d}\ln r}\Big),
    \label{eq:hydromass}
\end{equation}
where the mean molecular weight $\mu\simeq0.6$ for the assumed 0.3 solar abundance.

The bottom panels of Figure \ref{fig:Xrayfits} show example profiles of cumulative gas mass and hydrostatic mass profiles for 1eRASS J004207.0-283154. Both are calculated from the derived Markov Chains of the fitting parameters without assuming dark matter halo models. So our results are independent of halo models. The hydrostatic mass profile presents larger uncertainties than the gas mass profile. This is mainly driven by the large uncertainties on the temperature profiles. We derive the mass profiles up to 2 Mpc for all clusters, as hydrostatic equilibrium is generally broken at larger radii.

\section{Cluster dynamics from galaxy kinematics}\label{sec:galkin}

\subsection{Cluster galaxies}

In this paper, we also employ cluster galaxies as dynamical tracers to derive the total mass profiles for the selected eROSITA clusters. We use the member galaxies compiled by \citet{Kluge2024}. \citet{Kluge2024} identified cluster galaxies using both the optical and near-infrared photometry from the 9th and 10th data release of the DESI Legacy Imaging Surveys \citep{Dey2019}. Using the red-sequence Matched-filter Probabilistic Percolation cluster finder \citep[redMaPPer,][]{Rykoff2014,Rykoff2016}, the membership probabilities for individual galaxies were determined, which are dependent on color distance from red sequence models \citep[e.g.][]{Oguri2014}, spatial distributions, galaxy luminosity, and background galaxy number density. Photometric redshift for each member galaxy was also measured. Considering its large uncertainty, we only use galaxies with spectroscopic redshift to calculate velocity dispersion. \citet{Kluge2024} collected more than 4.8 million spectroscopic galaxy redshift from the literature. The major contributors are \citet{Zou2019}, \citet{Colless2001}, \citet{Coil2011}, \citet{Cool2013}, \citet{Alam2015}, \citet{Ahumada2020}. Those spectra were obtained with multiple surveys including 2dFGR \citep{Colless2001}, 6dFGR \citep{Jones2004, Jones2009}, 2SLAQ \citep{Cannon2006}, DEEP2 \citep{Davis2003,Newman2013}, SDSS DR14 \citep{Abolfathi2018}, zCOSMOS \citep{Lilly2007} etc. 

The cluster galaxies from \citet{Kluge2024} are red-sequence galaxies, which are known to be more concentrated toward the cluster center \citep{Nishizawa2018}, but also have larger velocity dispersion than blue galaxies \citep[e.g. see][]{Binggeli1987}. Therefore, using red-sequence galaxies alone does not necessarily introduce bias, as the enclosed total mass is correlated with $\sigma^2r$ \citep{Binney2008}. Instead, red-sequence galaxies are older than blue galaxies, so that they are presumably more likely relaxed. As such, employing only red-sequence galaxies as dynamical tracers is technically safer, though ideally the two populations should lead to roughly the same results if both are relaxed.

\begin{figure*}
    \centering
    \includegraphics[scale=0.45]{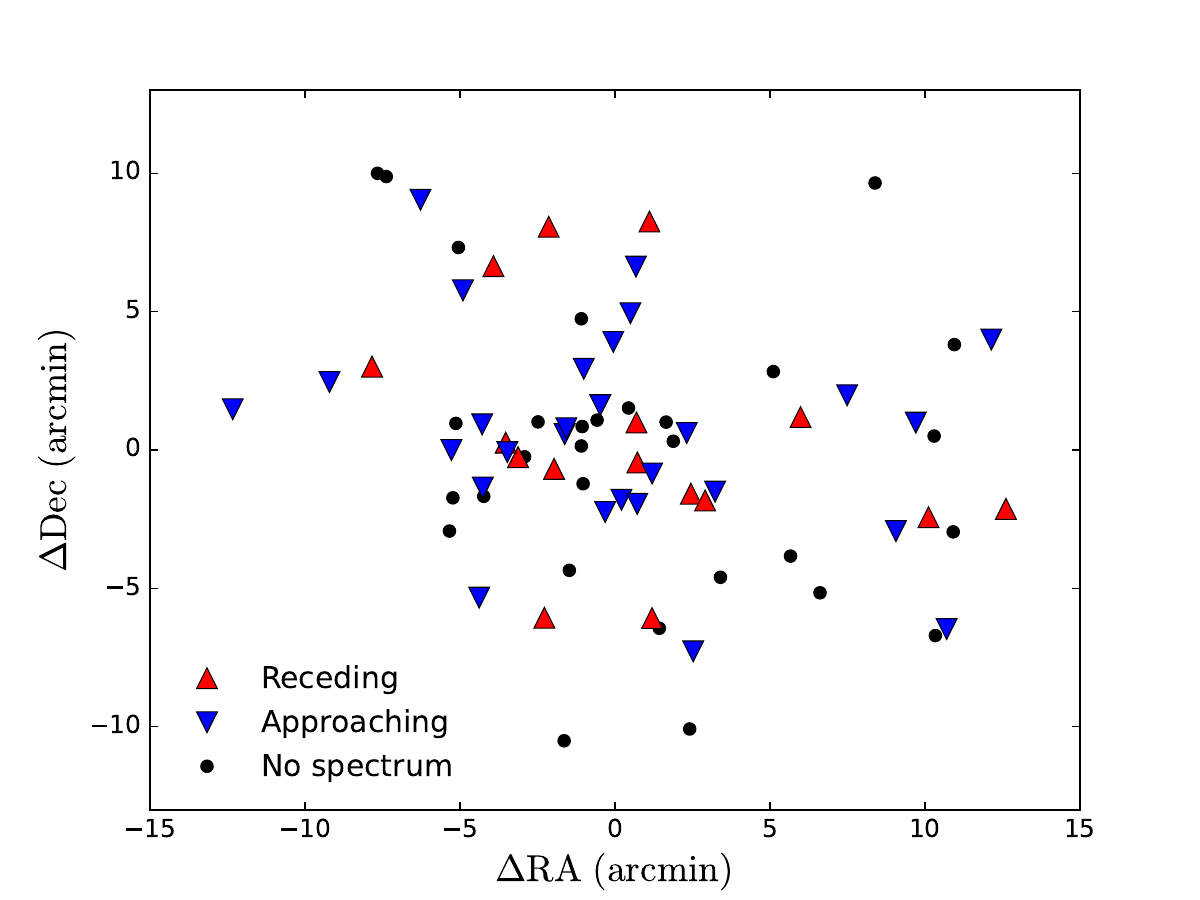}\includegraphics[scale=0.45]{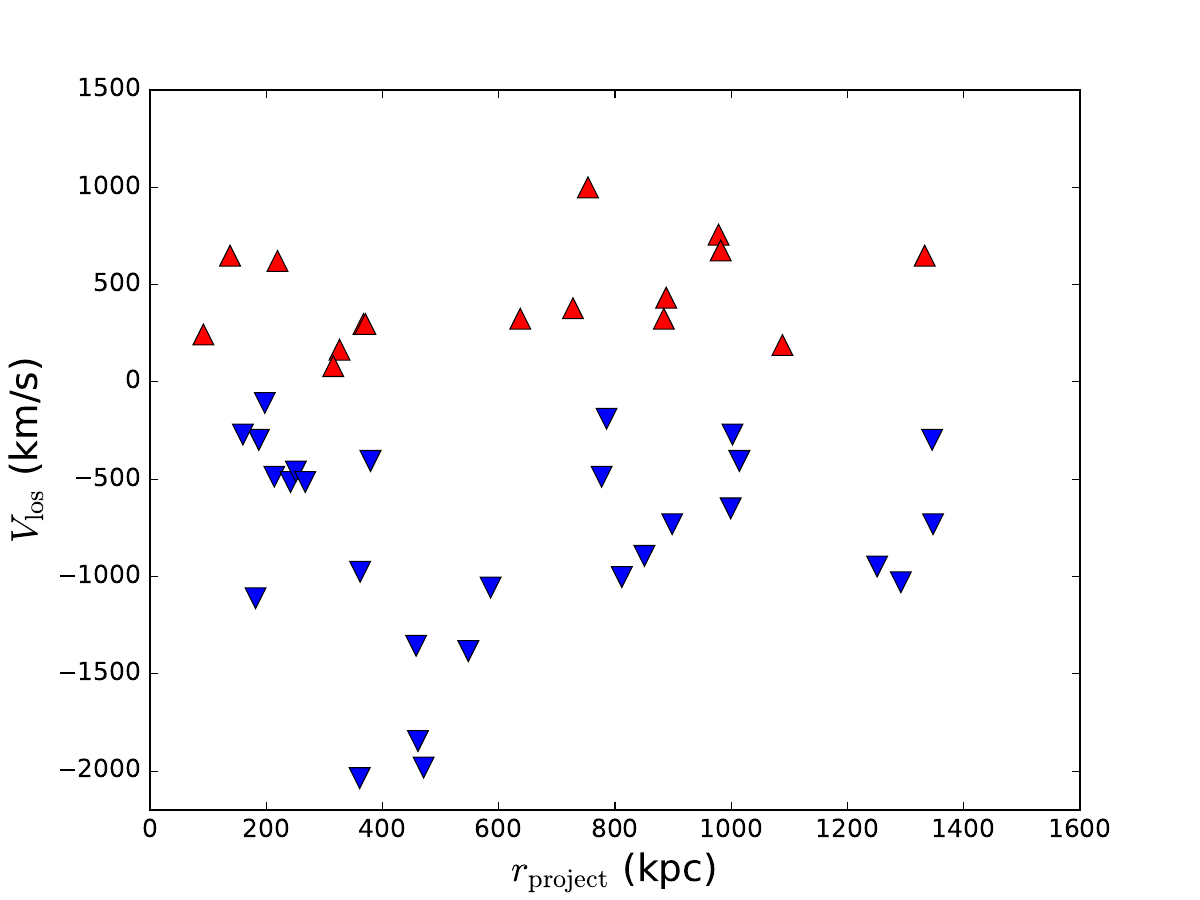}
    \caption{Left panel: the spatial distribution of the member galaxies of 1eRASS J004207.0-283154 on the projected sky. Red-upward triangles and blue-downward triangles represent receding and approaching galaxies, respectively. Black points are galaxies without spectroscopic redshift. Right panel: line-of-sight velocities of the member galaxies with respect to the BCG against the projected radii $r_{\rm proj}$. Galaxies with only photometric redshift are excluded in this plot. Cluster 004207.0-283154 presents a slightly asymmetric velocity distribution of its member galaxies. This does not affect the binned velocity dispersion, as galaxies within each bin are fitted using the generalized Gaussian function which allows a variable mean velocity (see Section \ref{sec:Jeans}).}
    \label{fig:phasspace}
\end{figure*}

\begin{figure*}
    \centering
    \includegraphics[scale=0.45]{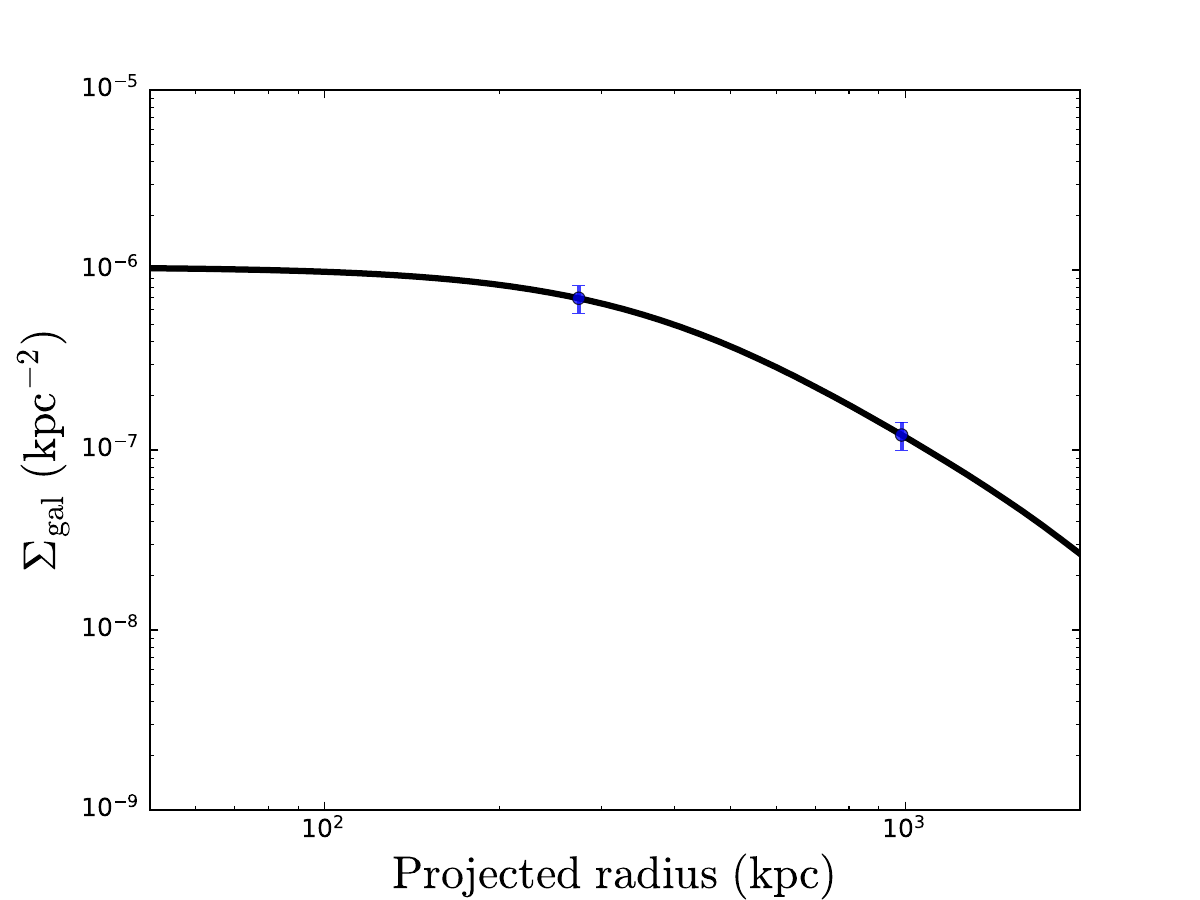}\includegraphics[scale=0.45]{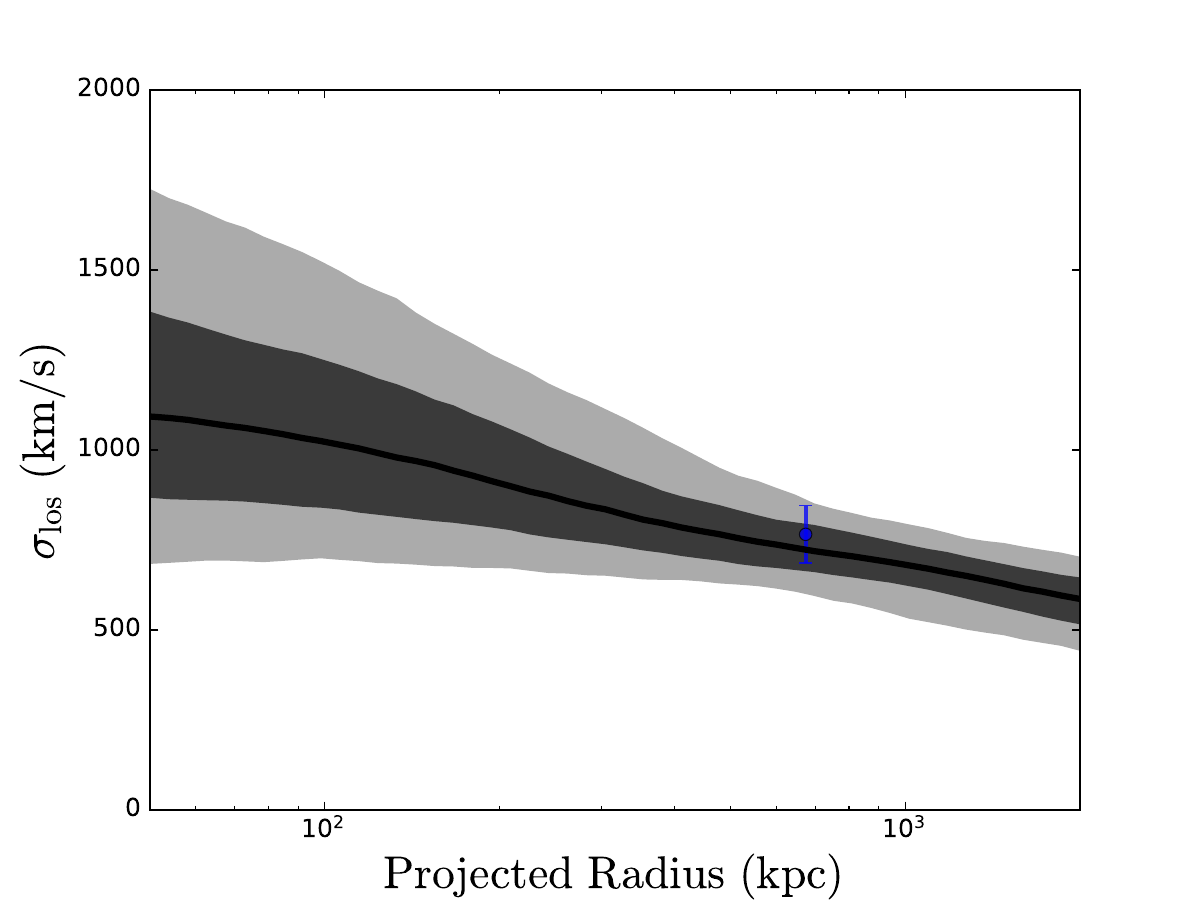}
    \caption{Left: Galaxy surface number density profile. Points with errors are for the binned  photometrically selected galaxies. The solid line is the fitted profile with three projected Plummer spheres \citep{Plummer1911}. Right: Line-of-sight velocity dispersion profile. The single point with errors is the binned velocity dispersion using spectroscopically selected galaxies, while the solid line is the expected profile based on the fitted mass profile parameterized by the coreNFWtides model \citep{Read2016}. The dark and light shadow regions present 1$\sigma$ and 2$\sigma$ limits, respectively. In both profiles, membership probabilities of galaxies are considered when binning. To be consistent with previous figures, the shown example is from 1eRASS J004207.0-283154.}
    \label{fig:Sigmasigma}
\end{figure*}

As an example, Figure \ref{fig:phasspace} plots the phase-space distribution of the member galaxies for 1eRASS J004207.0-283154. It is a general phenomenon that a significant fraction of member galaxies do not have measured spectroscopic redshift, and we only use spectroscopically confirmed galaxies to calculate the line-of-sight velocity dispersion profile. As such, there are often more bins in the galaxy surface number density profile than that for velocity dispersion (see Figure \ref{fig:Sigmasigma}). We shall stress that when calculating velocity dispersion, the cosmological effect has to be taken into account \citep{Harrison1974, Danese1980}. This is because the observed redshift $z_{\rm obs}$ is the production of two contributions,
\begin{equation}
    1+z_{\rm obs}= (1+z_{\rm cosmos})(1+z_{\rm peculiar}),
\end{equation}
where $z_{\rm cosmos}$ is the cosmological redshift, or approximately the redshift of the host cluster, and $z_{\rm peculiar}$ is the redshift due to peculiar motions of cluster galaxies. As such, the peculiar redshift difference between two member galaxies at roughly the same radii is given by \citep{Harrison1974, Kluge2024},
\begin{equation}
    \Delta z_{\rm peculiar} = \Delta z_{\rm obs}/(1+z_{\rm cosmos}).
\end{equation}
To be precise, we use the Doppler equation to calculate the relative velocity
\begin{equation}
    \Delta v_{\rm los} = c[(\Delta z_{\rm peculiar}+1)^2 - 1 ] / [ (\Delta z_{\rm peculiar}+1)^2 + 1 ],
\end{equation}
instead of $\Delta v_{\rm los}=c\Delta z_{\rm peculiar}$, though they are quite close given $\Delta z_{\rm peculiar}\ll1$ in general.

We notice that the velocity distribution of member galaxies with respect to the BCG could be slightly asymmetric. This does not affect the binned velocity dispersion, as we fit the galaxy velocity distribution within each bin using Binulator developed by \citet{Collins2021}. The Binulator fits a generalized Gaussian function allowing a varying mean value. With Binulator, velocity dispersion can be robustly determined even if the sample size is as low as 25 effective galaxies in each bin. As shown in Figure \ref{fig:Sigmasigma}, some clusters have only one binned line-of-sight velocity dispersion, so that their fitted velocity profiles and thereby dynamical mass profiles are unreliable. To avoid over-extrapolations, we do not use the whole mass profiles, but only the discrete values at the radii where the actual binned velocity dispersion is available. 

\subsection{Jeans modeling}\label{sec:Jeans}

Employing cluster galaxies as dynamical tracers makes use of Jeans modeling. This approach has been developed and thoroughly described in \citet{Li2023}. Readers interested in more details are recommended to read this paper. We only mention some main points here. The equation that relates velocity dispersion to dynamical mass is the Boltzmann equation \citep{Binney2008}. When assuming spherical symmetry, it becomes the Jeans equation as given by 
\begin{equation}
    \frac{1}{\nu}\frac{\partial}{\partial r}(\nu\sigma^2_r) + \frac{2\beta(r)\sigma^2_r}{r} = -\frac{GM(<r)}{r^2},
    \label{eq:SphereJeans}
\end{equation}
where the number density of member galaxies $\nu(r)$ and the radial velocity dispersion $\sigma_r$ can be derived by fitting their projected profiles to observations; the total mass profile $M(r)$ is the target quantity. The velocity anisotropy parameter $\beta=1-\sigma^2_t/\sigma^2_r$ relates the radial velocity dispersion $\sigma_r$ and tangential velocity dispersion $\sigma_t$. To determine its profile, one solves the projected, integrated Jeans equation \citep{vanderMarel1994, Binney1982},
\begin{equation}
    \sigma^2_{\rm los}(R) = \frac{2}{\Sigma_{\rm gal}(R)}\int_R^\infty\Big(1-\beta\frac{R^2}{r^2}\Big)\frac{\nu(r)\sigma^2_rr}{\sqrt{r^2-R^2}}{\rm d}r,
    \label{eq:sigmalos}
\end{equation}
where the projected distribution of galaxies ($\Sigma_{\rm gal}$) and line-of-sight velocity dispersion ($\sigma_{\rm los}$) are directly observable (see Figure \ref{fig:Sigmasigma}).

To solve Equation \ref{eq:SphereJeans} and \ref{eq:sigmalos}, we used the GravSphere code \citep{Read2017}, which has been tested with simulations for many different systems \citep{Read2017, Read2018, Read2021, Collins2021, Genina2020, DeLeo2024}. GravSphere parameterizes galaxy number density $\nu(r)$, velocity anisotropy $\beta(r)$, and dynamical mass profile $M(r)$ with successive functions. Galaxy number density is modeled with three Plummer spheres \citep{Plummer1911}. It can be determined by fitting its projected function, 
\begin{equation}
    \Sigma_{\rm gal} = \sum^3_{i=1}\frac{N_ia^2_i}{\pi(a^2_i+R^2)^2}.
    \label{eq:Plummer}
\end{equation}
to binned galaxy surface number density (see Figure \ref{fig:Sigmasigma}). In Equation \ref{eq:Plummer}, $N_i$ are dimensionless and chosen to satisfy $\int\nu(r){\rm d^3}\vec{r}=1$, and $a_i$ are scale radii that quantify how concentrated galaxies' distributions are. They are generally in the range of $10^{-4}<N_i<100$ and 50 kpc $<a_i<$ 2000 kpc.

The velocity anisotropy profile $\beta(r)$ is parameterized by a function with four parameters,
\begin{equation}
    \beta(r) = \beta_0 + (\beta_\infty-\beta_0)\frac{1}{1+\big(\frac{r_0}{r}\big)^\eta},
    \label{eq:beta}
\end{equation}
where $\beta_0$ and $\beta_\infty$ quantify the anisotropies at $r=0$ and $r=\infty$, respectively; $r_0$ and $\eta$ characterize the radial shape. The cumulative dynamical mass profile $M(<r)$ is described by six parameters using the cored-NFW-tides model \citep[cNFWt,][]{Read2018}, i.e. $M_{\rm cNFWt}(<r)=$
\begin{equation}
    \begin{cases}
        M_{\rm cNFW}(<r) = M_{\rm NFW}(<r)f^n & \text{if}\ r\le r_t,\\
        M_{\rm cNFW}(<r_t) + 4\pi\rho_{\rm cNFW}(r_t)\frac{r_t^3}{3-\delta}\Big[\Big(\frac{r}{r_t}\Big)^{3-\delta}-1\Big] & \text{if}\ r>r_t,\\
    \end{cases}
    \label{eq:cNFWt}
\end{equation}
where $M_{\rm NFW}$ and $M_{\rm cNFW}$ are the cumulative mass profiles of the Navarro-Frenk-White \citep[NFW,][]{Navarro1996} and cored NFW \citep[cNFW,][]{Read2016} models, respectively; $f=\tanh{\Big(\frac{r}{r_c}\Big)}$ and $n$ generate cored density profiles with $r_c$ being the core scale radius; $\rho_{\rm cNFW}(<r)=\frac{1}{4\pi r^2}\frac{{\rm d}M_{\rm cNFW}}{{\rm d}r}$; $r_t$ is the transition radius and $\delta$ is the outer density slope. As such, the cNFWt profile is a flexible function accommodating variable inner and outer slopes. 

A major advantage of the approach developed in \citet{Li2023} is the usage of two virial shape parameters \citep{Merrifield1990}, 
\begin{eqnarray}
    v_{s1} &=&\frac{2}{5}\int_0^\infty GM\nu(5-2\beta)\sigma^2_rr{\rm d}r\nonumber\\
    &=& \int_0^\infty\Sigma_{\rm gal}\langle v^4_{\rm los}\rangle R{\rm d}R,
\end{eqnarray}
\begin{eqnarray}
    v_{s2} &=&\frac{4}{35}\int_0^\infty GM\nu(7-6\beta)\sigma^2_rr^3{\rm d}r\nonumber\\
    &=& \int_0^\infty\Sigma_{\rm gal}\langle v^4_{\rm los}\rangle R^3{\rm d}R,
\end{eqnarray}
where $\langle\rangle$ denotes the mean. The two Virial Shape parameters introduce additional constraints on velocity anisotropy and dynamical mass, so that their degeneracy (see Equation \ref{eq:SphereJeans}) is ameliorated. 

The parameterized functions for $M(r)$, $\beta(r)$ and $\nu(r)$ are quite flexible. This is intended to maximized the allowed solution space, so that the the results are mainly driven by the data rather than by the chosen models. GravSphere makes use of the Markov Chain Monte Carlo method with the $emcee$ hammer by \citet{Foreman-Mackey2013} to map the posterior distributions of the parameter space in Equation \ref{eq:beta} and \ref{eq:cNFWt}. Following \citet{Li2023}, we set the below hard boundaries for the parameters in the cNFW model: $13<\log(M_{200}/M_\odot)<20$, $0.1<c_{200}<100$, 10 kpc $<r_c<$ 3000 kpc, $-0.5<n<2.0$, 10 kpc $<r_t<$ 5000 kpc, and $0<\delta<3$. For $\beta(r)$, we require $1<n<3$ and $0.5R_{\rm half}<r_0<2R_{\rm half}$, where $R_{\rm half}$ is the half mass radius. The anisotropies in the center and infinity are constrained in a symmetrized form to avoid infinite values: $-0.2<\frac{\beta_0}{2-\beta_0}<0.2$ and $-0.2<\frac{\beta_\infty}{2-\beta_\infty}<1.0$. There are in total 10 fitting parameters. We imposed flat priors within the aforementioned boundaries. We used 150 walkers and ran 10$^4$ steps to guarantee convergence.

\begin{figure}
    \centering
    \includegraphics[scale=0.4]{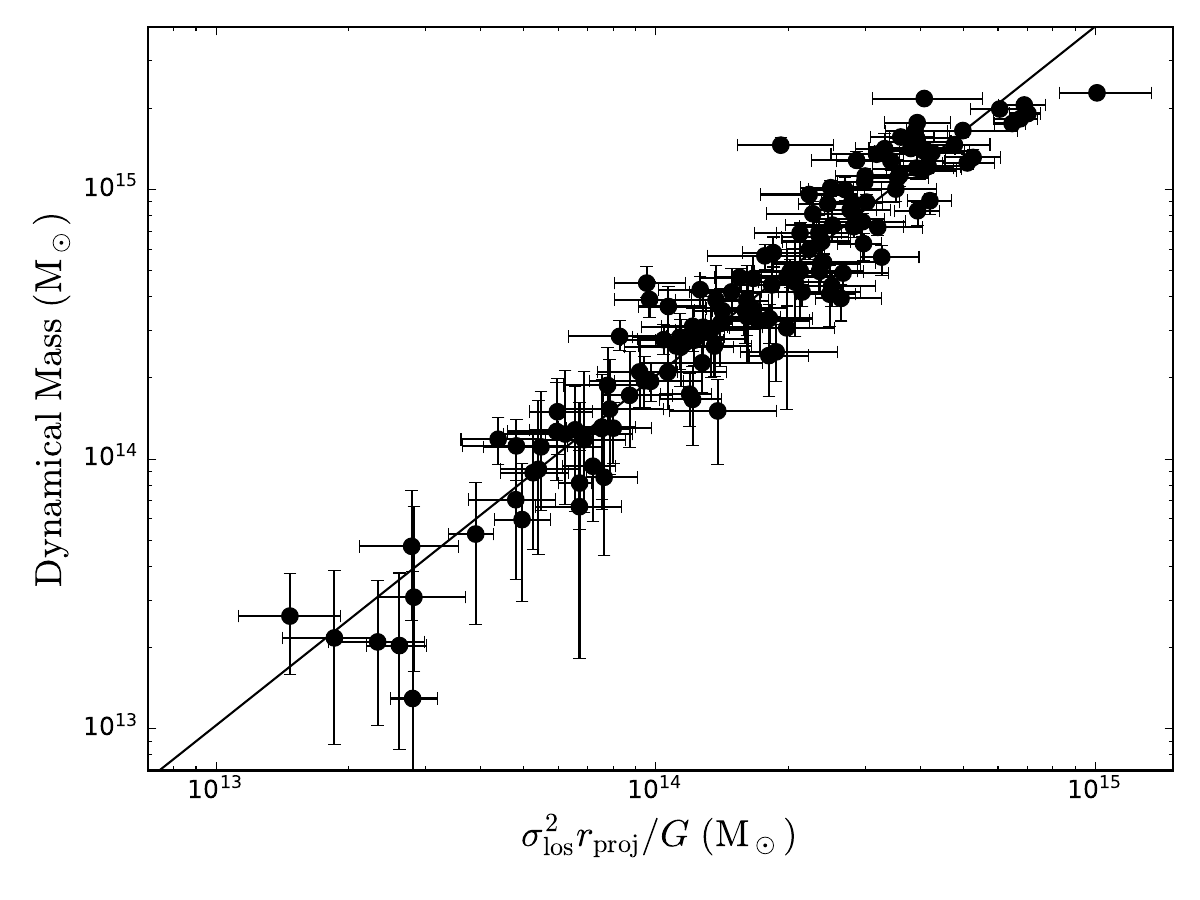}
    \caption{The enclosed dynamical mass against the product $\sigma_{\rm los}^2r_{\rm proj}/G$ for 22 eROSITA clusters and 10 HIFLUGCS clusters from \citet{Li2023}. Note that velocity dispersion is measured at rather than within a specific radius, i.e. $\sigma_{\rm los} = \sigma_{\rm los}(R=r_{\rm proj}) \neq \sigma_{\rm los}(R<r_{\rm proj})$, while the dynamical mass is cumulative, $M_{\rm dyn} = M_{\rm dyn}(r<r_{\rm proj})$, where $R$ and $r$ are 2D and 3D radii, respectively; $r_{\rm proj}$ is the projected radius. The correlation can be described by a power law (the solid line): $\log M_{\rm dyn} = (1.296 \pm 0.001)\log(\sigma_{\rm los}^2r_{\rm proj}/G) - (3.87\pm0.23)$. The total scatter is 0.14 dex ($\sim32\%$). To clarify, points with errors are the binned results at different radii, so a cluster could have more than one point.}
    \label{fig:Mvd}
\end{figure}

\subsection{The effect of interlopers}

A major concern using member galaxies as dynamical tracers is interlopers, which could be mistakenly counted as members. They could affect the galaxy number density and velocity dispersion measurements, and thereby the final mass profiles. We minimized this effect in two ways. First, we use membership probabilities to weigh the effective number of each candidate. The membership probability of each member candidate is estimated with eROMapper \citep{Kluge2024}, which is built upon the red-sequence Matched-filter Probabilistic Percolation cluster-finding algorithm \citep{Rykoff2014, Rykoff2016}. Under this scheme, a candidate with a membership probability smaller than 100\% will be assigned an effective number smaller than one. Unless the membership probability of an interloper is seriously overestimated, its contributions to galaxy number density and velocity dispersion would be negligible. Second, the line-of-sight velocity dispersion within each radial bin is not directly calculated based on the velocities of all members, but derived by fitting a generalized Gaussian function to the effective number histogram of line-of-sight velocities using Binulator \citep{Collins2021}. As such, a single miscounted interloper plays little role in the measurement of velocity dispersion. With these two measures combined, even a group of miscounted galaxies do not have significant impact on our results, as they should have small membership probabilities. As such, we do not worry about interlopers in our dynamical approach. 

\subsection{Spatially resolved mass-velocity dispersion relation}

\begin{figure*}
    \centering
    \includegraphics[scale=0.45]{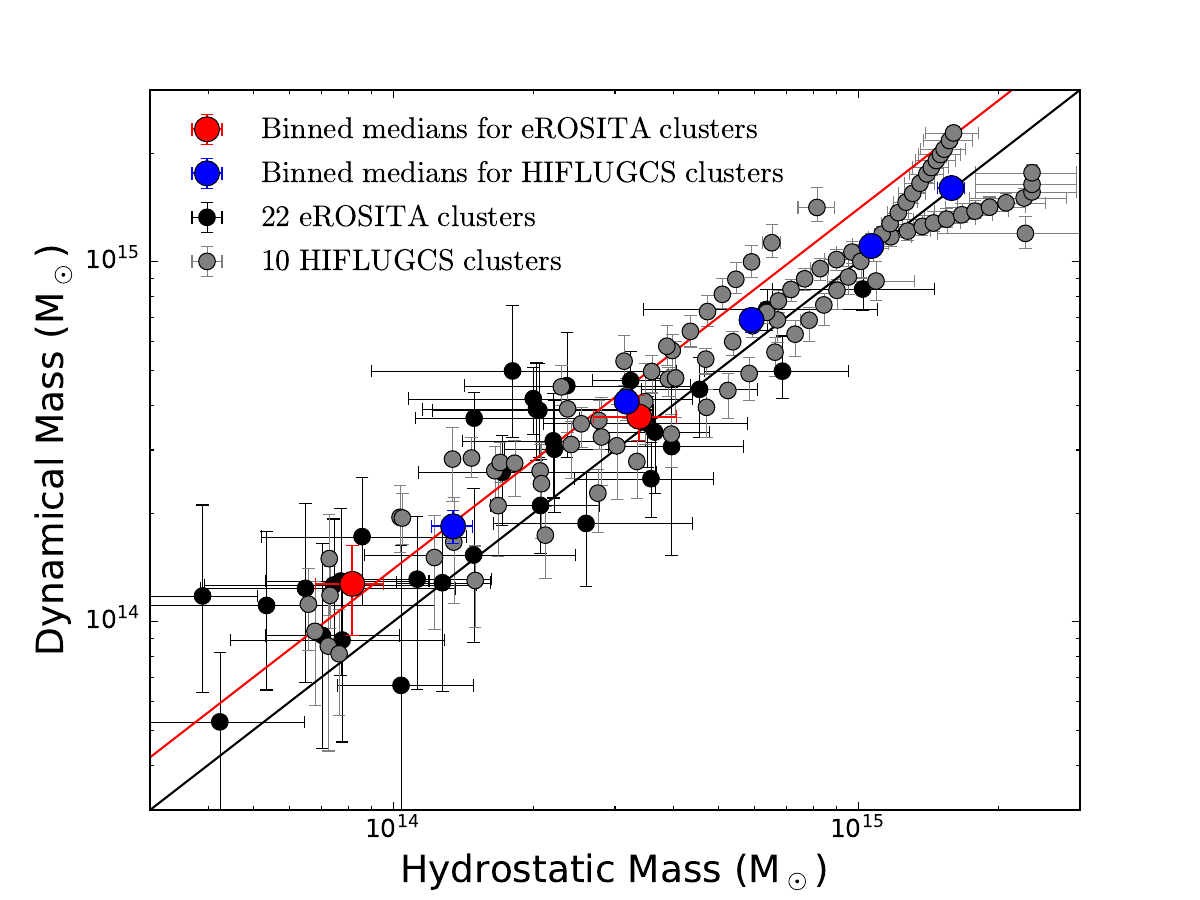}\includegraphics[scale=0.45]{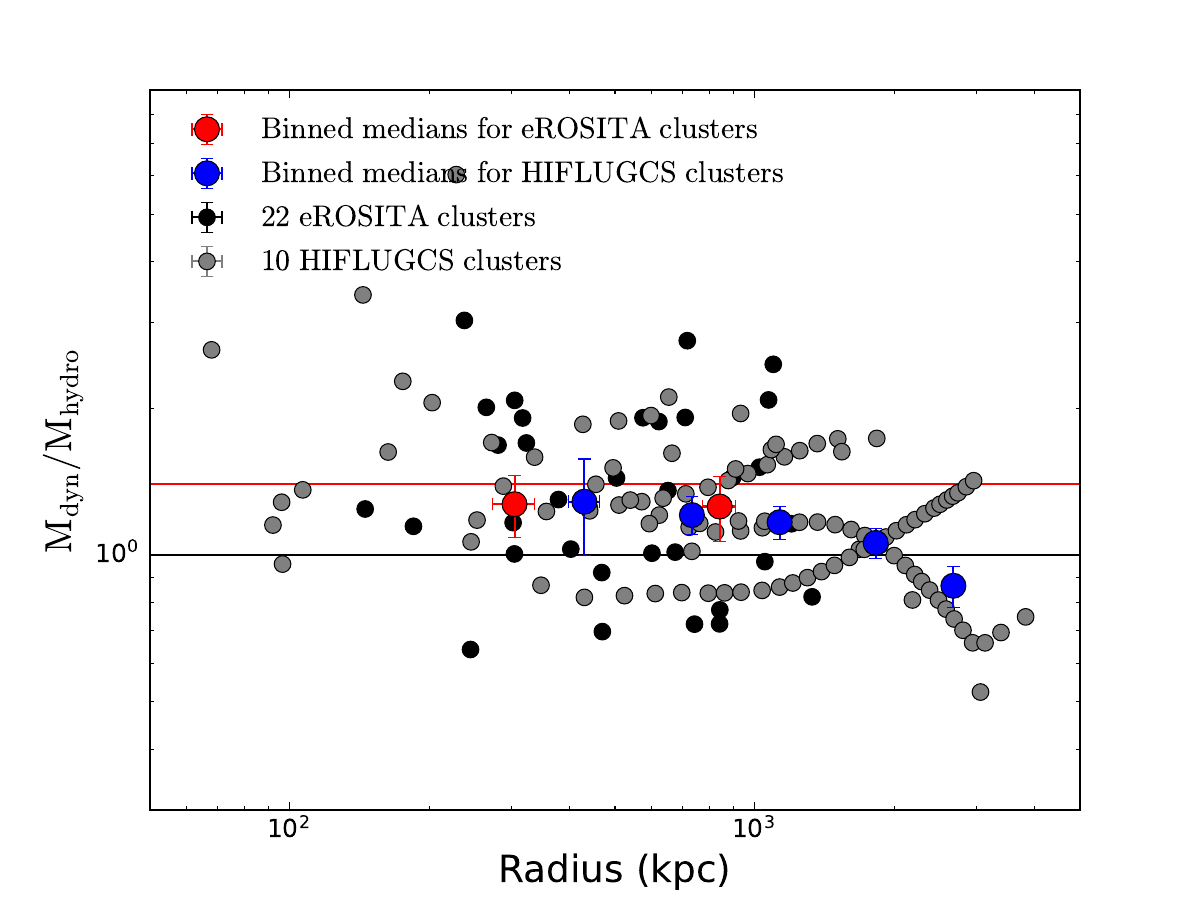}
    \caption{Comparison between dynamical mass measured from galaxy kinematics with Gravsphere \citep{Read2017} and hydrostatic mass from X-ray gas thermodynamics using \texttt{MBProj2D} \citep{Sanders2018}. Black and grey points are the binned data at different radii for the selected 22 eROSITA clusters and 10 HIFLUGCS clusters from \citet{Li2023}, respectively. To demonstrate the trend, the data points are binned as medians according to their mass (left) or radii (right): red points for eROSITA clusters and blue points for HIFLUGCS clusters. Black lines indicate the one-to-one ratio, while red lines show the 40\% higher dynamical mass than the hydrostatic mass. Note that the hydrostatic mass for the HIFLUGCS clusters is either from the X-COP database \citep{Eckert2019, Ettori2019, Gadotti2009} when available, or derived from the best-fit $\beta$ model \citep{Chen2007}.}
    \label{fig:Mass_com}
\end{figure*}

By running Gravsphere, we derived the dynamical mass profiles for the 22 eROSITA clusters. Dynamical mass is expected to be correlated with velocity dispersion \citep{Binney2008}. But the specific coefficients are unknown. By numerically solving Jeans equation with Gravsphere, we are able to determine the relation in an empirical way. To increase the statistics, we also include the 10 clusters from \citet{Li2023}, given their dynamical masses are derived in the same way. We do not distinguish clusters at different redshifts, as we derive the scaling relation directly from the Jeans equation, which does not evolve with time.

Figure \ref{fig:Mvd} plots the dynamical mass against the product of $\sigma_{\rm los}^2r_{\rm proj}/G$ for the 22 eROSITA clusters and 10 HIFLUGCS clusters. Each point represents a single value at a specific radius rather than a cluster. The 32 clusters present a tight linear correlation in logarithmic space, which can be nicely fitted by
\begin{equation}
    \log M_{\rm dyn} = (1.296 \pm 0.001)\log(\sigma_{\rm los}^2r_{\rm proj}/G) - (3.87\pm0.23),
    \label{eq:Mvd}
\end{equation}
where $\sigma_{\rm los}(R)$ is the line-of-sight velocity dispersion at $R=r_{\rm proj}$ rather than $R< r_{\rm proj}$, while $M_{\rm dyn}(<R)$ is the cumulative mass within the 3D radius $r=r_{\rm proj}$. The rms scatter around the fitted relation is 0.14 dex. This is partially contributed by the deviation from the best-fit mass profile (e.g see the right panel of Figure \ref{fig:Sigmasigma}. The best-fit line does not exactly cross the point), especially when there is more than one binned velocity dispersion. So the actual scatter of the correlation could be even smaller. Besides, clusters could have very different velocity anisotropy profiles, which are also radially dependent. This diversity is another source for the total scatter. 

Note that the slope of the fitted line is not one, as one might expect from the commonly used equation $M_{\rm dyn}=c\sigma_{\rm los}^2r_{\rm proj}/G$ \citep{Binney2008, Kluge2024}, where $c$ is the proportional coefficient. So the coefficient is not a constant, but depends on mass or radius. In other words, the correlation is approximately a power law, rather than linear. This is evident from Equation \ref{eq:SphereJeans}, which considers not only velocity dispersion, but also its radial derivative. The derivative term contributes additional dependency on velocity dispersion that steepens the slope in log space. In addition, velocity anisotropy varies radially as well, so it could also affect the slope.

Equation \ref{eq:Mvd} can be thought as a simplified version of the complicated Jeans equation. It is a spatially resolved relation, differing from the $\sigma_{\rm los}-M_{200}$ relation. The latter is frequently studied in the literature \citep[e.g.][]{Munari2013}, and evolves with redshift, since the definition of $M_{200}$ depends on the critical density of the universe. The data points that are used to determine equation \ref{eq:Mvd} range from 68 kpc to 3820 kpc in radius, and from 533 km/s to 1336 km/s in velocity dispersion. It can be used to quickly determine the cumulative dynamical mass of a galaxy cluster at different radii with an accuracy of $\sim 32\%$ (0.14 dex). By astronomical standards, this is sufficiently accurate in many situations. Despite the above relation does not rely on the assumption of the dynamical states of considered systems, it is indeed assumed that cluster galaxies are relaxed and stay in dynamical equilibrium with the whole cluster when used for mass estimations. 

\section{Comparison of different mass proxies}
\label{sec:masscomparison}

Comparing different mass estimators is useful for consistency check, which is of interest given every proxy relies on some assumption, such as the hydrostatic equilibrium for gas thermodynamics and dynamical relaxation for galaxy kinematics. Of particular interest is to investigate the so-called hydrostatic bias, which is on one hand motivated by the existence of non-thermal flow in the intracluster medium \citep{Lau2009, Vazza2009, Battaglia2012, Nelson2014, Shi2015, Biffi2016, Liu2016, Liu2018}, and on the other hand indicated by the $\sigma_8$ tension, as cluster mass from Sunyaev-Zeldovich effect used to be calibrated with hydrostatic mass \citep{Planck2016XXIV}. Multiple approaches have been employed to quantify hydrostatic bias, such as assuming a constant cosmic baryonic fraction at different cluster radii \citep{Eckert2019}, comparison with weak lensing mass \citep{Donahue2014, vonderLinden2014, Sereno2015, Hoekstra2015, Penna-Lima2017, Sereno2017}. No agreement has been reached in terms of the degree of the bias, and none of these studies found the level required for resolving the $\sigma_8$ tension. \citet{Foex2017} compared the dynamical masses measured with the Jeans, caustic and virial-theorem approaches to hydrostatic masses, and found they are generally 20\%-50\% higher. \citet{Li2023} developed a more sophisticated Jeans approach, which showed 50\% higher dynamical mass on average. Both studies have only 10 clusters, so the results remain inconclusive.

In Figure \ref{fig:Mass_com}, the left panel plots the dynamical mass against hydrostatic mass for the 22 eROSITA clusters. We also collect the 10 HIFLUGCS clusters from \citet{Li2023}. Since the hydrostatic mass for these two samples are measured with different approaches, we distinguish them with different colors. Notice that we are not comparing the total mass, but the enclosed mass at different radii. The former is often compared because of cosmological interests. The practical reason is that we do not have a sufficient number of member galaxies to ensure the robust measurement of the whole mass profiles for most of the clusters in our sample. Since we can derive successive hydrostatic mass profiles for all the clusters, we simply interpolate the masses at the radii where the binned kinematic information of member galaxies is available. Comparing enclosed mass at different radii is also beneficial, because we can investigate their radial trend.

The eROSITA clusters present a large scatter, and are distributed around the line of unity. To quantify their mean behavior, we equally divide the data points into two bins and calculate their medians. The low-mass bin shows that dynamical mass is around 40\% higher than hydrostatic mass, while the high-mass bin closely aligns with the one-to-one line. For the HIFLUGCS clusters, though we have a smaller sample, but given the large richness of each cluster, we end up with more kinematic data points. We also bin the data and find the similar behavior: the dynamical mass could be as much as 40\% higher than their hydrostatic mass at the low mass end, while they are quite close toward the high mass end. So the two samples present consistent results. Since high mass corresponds to large radii, the same trend presented in both samples could arise from their radial variations. To examine this interpretation, we plot their dynamical-to-hydrostatic mass ratios against radius in the right panel of Figure \ref{fig:Mass_com}. The trend becomes more obvious. Both samples show a clearly decreasing dynamical-to-hydrostatic mass ratios toward large radii. Since hydrostatic mass is known to be biased low, the lower dynamical mass could imply that the member galaxies of massive clusters are not relaxed. As such, this trend could be driven by the radial variation of the dynamical state of cluster galaxies. If confirmed with larger samples, it could be the evidence that cluster galaxies at large radii are still accreting towards the center. 

\begin{figure*}
    \centering
    \includegraphics[scale=0.45]{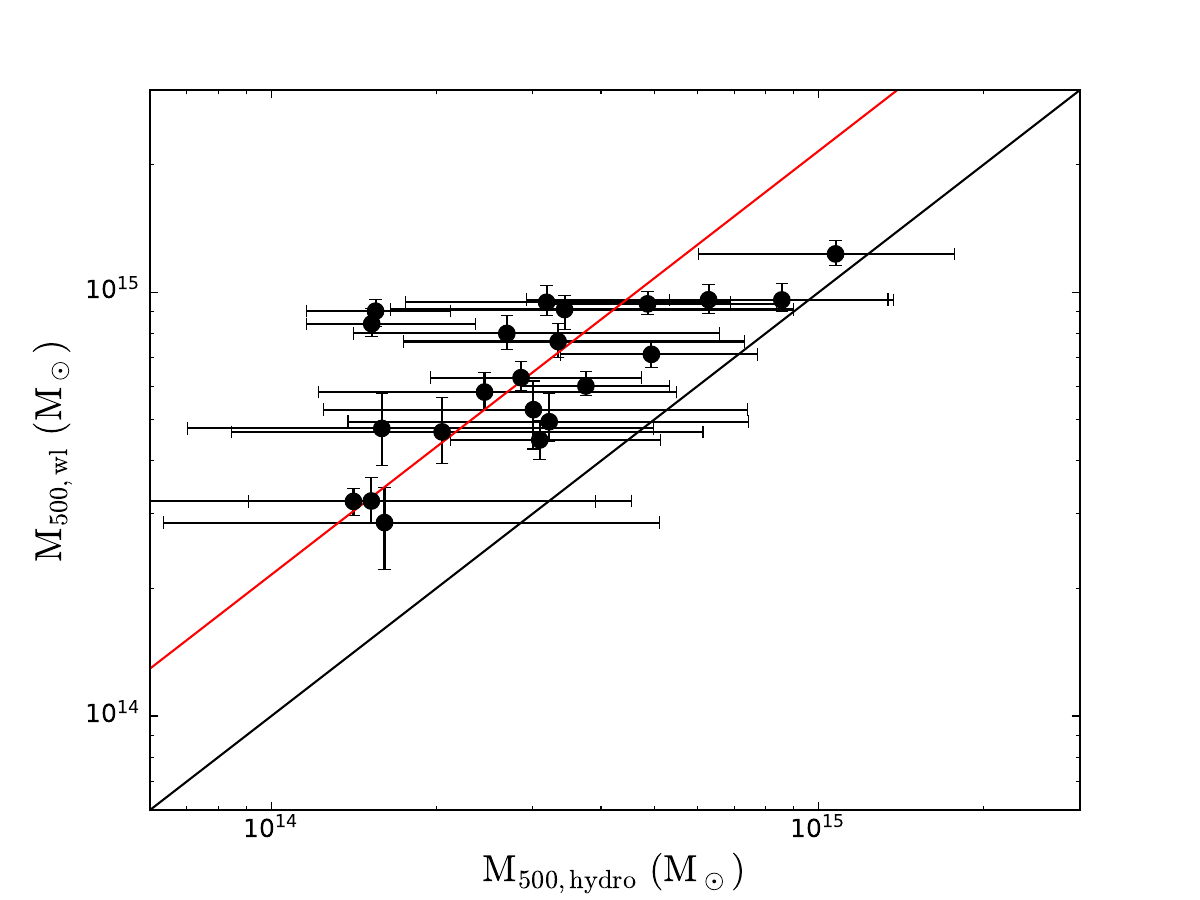}\includegraphics[scale=0.45]{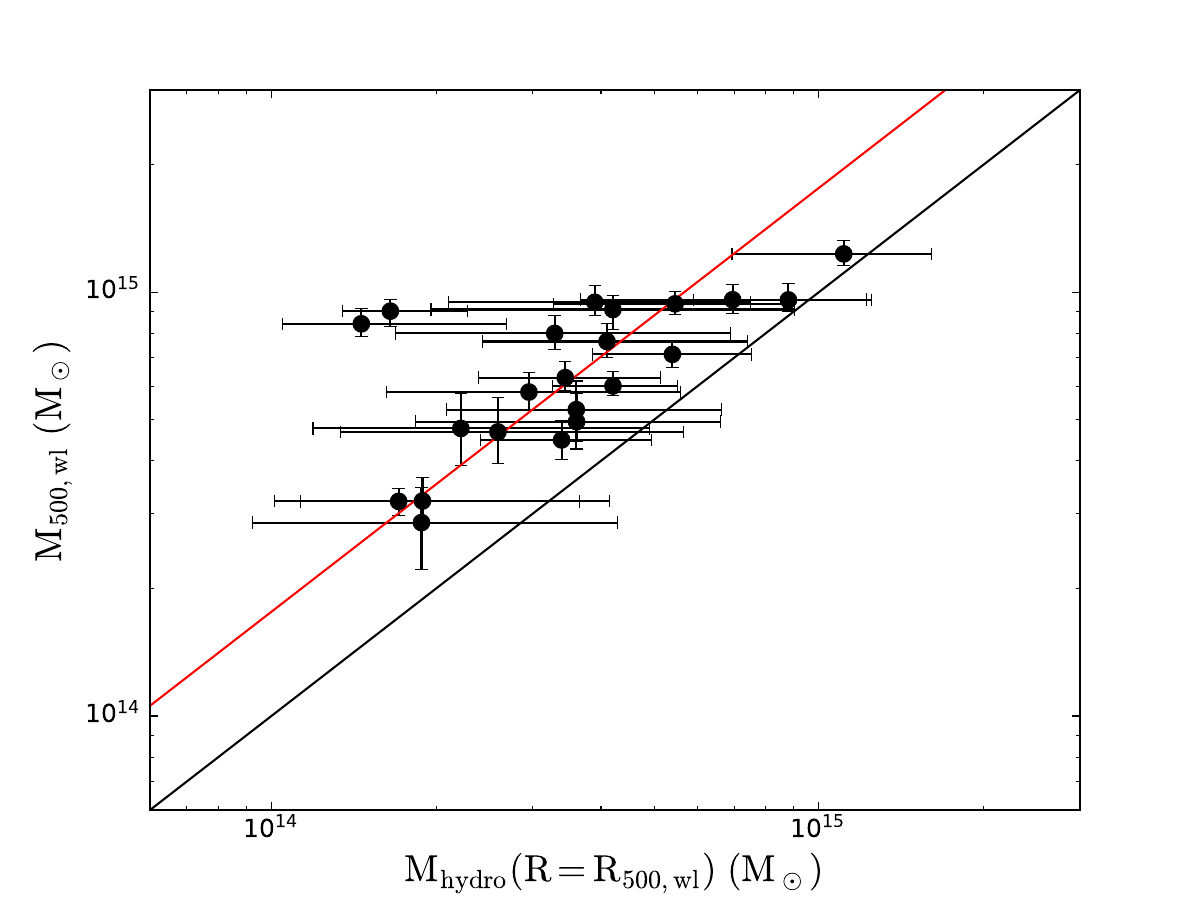}
    \caption{Comparison of masses measured with weak lensing and the hydrostatic approach. Weak lensing inferred masses and radii are from \citet{Bulbul2024}. The left panel compares the total masses M$_{500}$ independently inferred from these two mass proxies within their own R$_{\rm 500}$ (R$_{\rm 500, wl}$ and R$_{\rm 500, hydro}$); while the right panel compares the enclosed masses at the same radii, R$_{\rm 500, wl}$. Red lines indicate the medians: M$_{\rm 500, wl}$ is $\sim$110\% higher than M$_{\rm 500, hydro}$, and $\sim$71\% higher than M$_{\rm hydro}$(R=R$_{\rm 500, wl}$). Black lines are the line of unity.}
    \label{fig:M500hydro_M500wl}
\end{figure*}

Since \citet{Bulbul2024} provided the weak lensing inferred mass for each cluster, we can also compare our mass measurements with that from weak lensing. However, only the total mass enclosed within $R_{500}$ is available, since the weak lensing mass in \citet{Bulbul2024} is estimated using the X-ray count rate-cluster mass relation presented in \citet{Ghirardini2024}. The cluster masses used to build the scaling relation are from multiple lensing surveys, including DES \citep{Grandis2024}, KiDs \citep{Kuijken2019, Wright2020,Hildebrandt2021,Giblin2021}, and HSC \citep{Aihara2018, LiXC2022}. Weak lensing masses are not bias free. \citet{Ghirardini2024} calibrated the mass using Monte Carlo simulations \citep{Grandis2021} based on the surface mass density maps of massive halos presented in the cosmological TNG300 simulations \citep{Pillepich2018, Springel2018, Nelson2018, Nelson2019}. Therefore, the calibration makes the weak lensing masses slightly dependent on cosmology.

Figure \ref{fig:M500hydro_M500wl} plots the weak lensing masses against our hydrostatic masses. We do not include the comparison with dynamical masses, because for many clusters, there are not sufficient member galaxies to constrain their dynamical mass profiles, making it impossible to robustly calculate the total dynamical mass or equivalent. The left panel compares the total masses measured independently with weak lensing and hydrostatic assumption. The hydrostatic masses have significantly larger uncertainties, but they consistently present lower values. We calculate the median mass of the 22 clusters, and find weak lensing inferred mass is higher than hydrostatic mass by $\sim$116\%. Notice that the two total masses correspond to different cluster radii, so the lower hydrostatic masses are partially caused by the smaller cluster radii $R_{\rm 500, hydro}$. Interestingly, the significantly higher weak lensing masses are in line with the value of $\sigma_8$ measured by \citet{Ghirardini2024}, which is higher than almost all previous estimates in the literature. Therefore, the larger value of $\sigma_8$ might be due to the higher weak lensing masses they used.

The right panel compares weak lensing masses to the hydrostatic masses at the weak lensing inferred radius R$_{\rm 500, wl}$. Unlike the previous comparison which is interesting in cosmology, this comparison is important for testing scaling relations and studying various dark matter models \citep{Li2020}. We find the hydrostatic masses at $R_{\rm 500, wl}$ are also smaller than weak lensing masses, but the differences are much smaller, by $\sim$76\% in terms of their medians. This is expected, given $R_{\rm 500, wl}$ is systematically larger than $R_{\rm 500, hydro}$. The uncertainties on the hydrostatic masses are also smaller, because there are no contributions from the uncertainties on radii. The lower hydrostatic mass could be partially due to the lower temperature in eROSITA clusters with respect to Chandra and XMM-Newton observations \citep{Migkas2024}. But the influence may not be significant. \citet{Migkas2024} showed the difference in temperature is more profound in hard bands (1.5-7 keV), and less in soft bands (0.5-4 keV). We include even more soft bands (0.3-0.5 keV) in the image fitting. So we do not expect the possible bias could make up the big difference in the mass measurements.

Combining the two comparisons in Figure \ref{fig:Mass_com} and \ref{fig:M500hydro_M500wl}, both our dynamical and hydrostatic masses are smaller than weak lensing masses. If we trust the weak lensing mass inferred from scaling relations \citep{Ghirardini2024}, both dynamical equilibrium and hydrostatic equilibrium break down. This is possible, given that hydrostatic mass is affected by non-thermal flows and galaxies could be accreting. The question is whether they are biased by the amount we present above. Full lensing shear profiles may be required for detailed and robust investigations.

\section{Radial acceleration relation from different mass estimators}\label{sec:RAR}

Rotationally supported galaxies present an interesting correlation between centripetal acceleration ${\rm g_{obs}}$ and the acceleration due to baryonic mass distributions ${\rm g_{bar}}$\citep{McGaugh2016PRL, OneLaw, Li2018},
\begin{equation}
    g_{\rm obs} = \frac{g_{\rm bar}}{1-e^{-\sqrt{g_{\rm bar}/g_\dagger}}},
\end{equation}
where ${\rm g_\dagger}=1.2\times 10^{-10}$ m s$^{-2}$ is the transition acceleration scale. Its universality has been tested and verified in early-type galaxies \citep{Lelli2017, Shelest2020}, even with different dynamical approaches, such as gravitational lensing \citep{Mistele2024}. However, it does not seem to work in galaxy clusters, as confronted by both X-ray observations \citep[e.g.][]{Eckert2022, Liu2023} and gravitational lensing measurements \citep{Tian2020}. \citet{Tian2024} even found that galaxy clusters and BCGs show a consistent but distinct RAR. 

We noticed that these studies in galaxy clusters used a different dynamical approach with respect to disk galaxies: the former generally employs either gas thermodynamics or gravitational lensing, while the latter uses kinematics (i.e. rotation curves). To ensure consistency, \citet{Li2023} tested the RAR in galaxy clusters using galaxy kinematics for the first time, but found similar deviations from the galaxy RAR. Particularly, since galaxy distributions could be more extended than X-ray emitting gas, \citet{Li2023} were able to explore the outskirts. However, the total accelerations in the outskirts as probed by cluster galaxies are below the fiducial RAR expectations, which is contrary to that at small radii. This could be a catastrophe for the universality of the RAR. Because the observed accelerations of galaxy clusters at small radii are larger than what the RAR predicts based on observed baryonic mass \citep{Tian2020, Eckert2022, Liu2023, Li2023}, one would have to introduce additional undetected baryons in order to achieve a universal RAR. The missing baryons could be in the form of cold gas \citep{Milgrom2008}. It has also been argued that they could be non-baryonic, such as active neutrinos \citep{Sanders2003} or sterile neutrinos \citep{Angus2008}. However, the lower observed accelerations in the outskirts imply that the enclosed total mass is {\it smaller} than the mass predicted by the RAR. As such, there is no room for additional baryons to compensate for the deviation at small radii, as that would require negative baryonic mass density at large radii. However, the dynamical mass measured with galaxy kinematics relies on the assumed relaxation of member galaxies, which may be broken in the outskirt as mentioned in Section \ref{sec:masscomparison}, since galaxies could be in the process of migrating towards the cluster center. Therefore, \citet{Kelleher2024} proposed to fit clusters without the data at radii beyond 1 Mpc.

\begin{figure}
    \centering
    \includegraphics[scale=0.45]{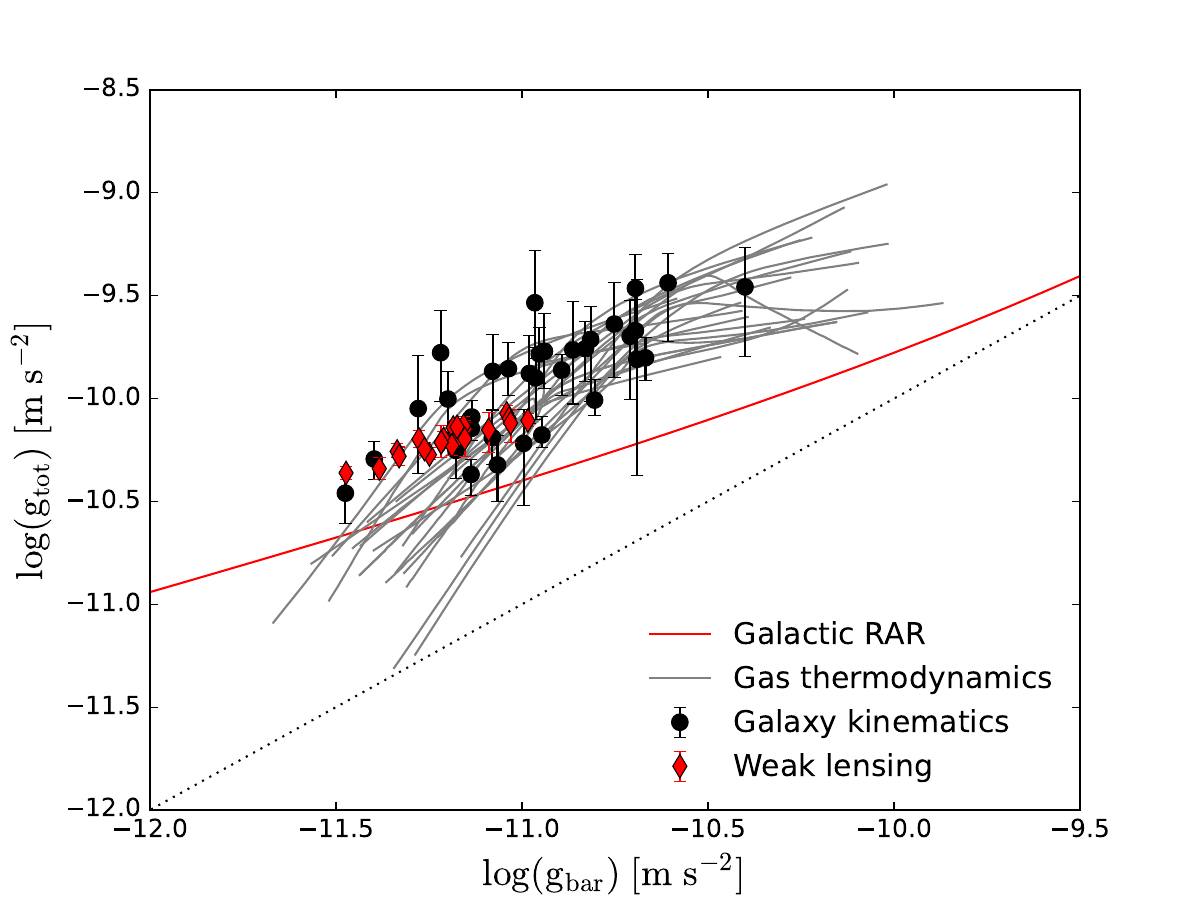}
    \caption{Radial acceleration relation of galaxy clusters constructed with X-ray gas thermodynamics (grey lines), galaxy kinematics (black points) and weak gravitational lensing (red diamonds). Grey lines are truncated within 50 kpc and 2000 kpc. The weak lensing results are only available at $R_{500}$ from \citet{Bulbul2024}, since they are estimated using the scaling relation between weak lensing mass calibrated with simulations and X-ray count rates (with optical richness considered) \citep{Ghirardini2024}. The dotted line is the line of unity, and the solid red line is the fiducial RAR defined by disk galaxies \citep{McGaugh2016PRL, OneLaw}.}
    \label{fig:RAR}
\end{figure}

Using the total mass estimated from different proxies, we can construct the RAR in galaxy clusters without being biased toward one {\it sole} assumption. With spherical symmetry, the total accelerations are simply given by  
\begin{equation}
    g_{\rm tot} = \frac{GM_{\rm tot}(<r)}{r^2}.
\end{equation}
For different mass proxies, we have different $M_{\rm tot}$ and thereby get different ${\rm g_{tot}}$. In contrast, the accelerations due to baryonic mass distributions are the same, and given by
\begin{equation}
    g_{\rm bar} = \frac{G\big(M_{\rm gas}(<r)+M_\star(<r)\big)}{r^2},
\end{equation}
since the gas mass $M_{\rm gas}$ are fixed by X-ray observations, and the stellar mass are given by multi-band observations \citep[see][]{Zou2019, Kluge2024}. For galaxies that do not have measured stellar masses, we estimated their mass using the mean stellar mass-to-light ratios of the other galaxies. The total stellar mass is then the summation of the stellar masses of all member galaxies. To get the 3D stellar mass profiles, we used the normalized best-fit three Plummer's spheres. We do not attempt to resolve the stellar mass profiles of the BCGs, given we only explore the range at radii larger than 50 kpc. 

Figure \ref{fig:RAR} shows the cluster RAR constructed with gas thermodynamics, galaxy kinematics, as well as weak gravitational lensing. The three independent mass estimators lead to roughly the same RAR, though at R$_{\rm 500,wl}$, the RAR constructed with weak lensing is slightly higher than that from the other two approaches. At small and intermediate radii, gas thermodynamics and galaxy kinematics present a RAR that is consistently higher than the fiducial line by about 0.5 dex on average. This translates into three times higher total mass than expected from the RAR. It is reminiscent of the missing mass problem in Modified Newtonian Dynamics \citep[MOND,][]{Milgrom1983}. For the first time, we confirm this problem in the most robust way that does not rely on one sole assumption. The missing mass required by MOND is about two times the observed baryonic mass. Moving to larger radii, the gap narrows down to around 0.3 dex at $R_{500}$, corresponding to two times higher total mass. This does not mean the cumulative missing mass is decreasing, as the reference mass also increases toward large radii. 
 
Since we calculated the hydrostatic mass up to 2 Mpc, the RARs it leads to are also more extended than that from other mass estimators. Eventually, they cross below the fiducial RAR line, implying less missing mass. However, at these large radii, if hydrostatic equilibrium is valid remains uncertain. Particularly as we have shown in Figure \ref{fig:M500hydro_M500wl}, the hydrostatic masses are lower than the masses inferred from weak lensing, which might suggest a breakdown in the assumed equilibrium. We also note that the cosmological parameter $\sigma_8$ \citet{Ghirardini2024} derived is higher than all other estimates in the literature (see their Figure 10). Therefore, the weak lensing mass in \citet{Bulbul2024} inferred with the scaling relation from \citet{Ghirardini2024} should be higher than the mass in other studies as well. As such, we treat the lensing results as upper limits. The current data cannot solidly confirm or rule out the possibility that adding additional baryonic mass components could make a universal RAR. We will have to look for spatially resolved mass profiles from both weak lensing measurements and galaxy kinematics. 

\section{Discussion and conclusion}\label{sec:discuss}

In this paper, we selected 22 galaxy clusters from the eRASS1 cluster catalog, requiring that there are a sufficient number of member galaxies as dynamical tracers. We measured their mass using both the thermodynamics of X-ray emitting intracluster medium and the kinematics of member galaxies. With the X-ray data from eROSITA, we generated the images in seven energy bands for each cluster, and measured their temperature profiles, gas mass profiles and hydrostatic mass profiles by fitting their images simultaneously. Using the photometric and spectroscopic redshifts of the member galaxies collected from the literature, we measured the projected galaxy number density profile and the line-of-sight velocity dispersion profile for each individual cluster, respectively. We parameterized the velocity anisotropy profile and the dynamical mass profile with flexible functions, and solved the Jeans equation using the MCMC approach. From the derived dynamical mass, we determine the correlation between cumulative total mass and velocity dispersion in Equation \ref{eq:Mvd}, which has an rms scatter of only 0.14 dex. This correlation can be used to quickly estimate the dynamical mass profile of a galaxy cluster from velocity dispersion with an accuracy of $\sim32\%$.

The comparison of the two mass proxies shows no systematic deviations at large radii. The roughly consistent mass measurements of galaxy clusters is not a good signal for the cosmological $\sigma_8$ tension, as one would expect the problem is in the calibration of cluster mass with hydrostatic bias \citep{Piffaretti2008, Meneghetti2010, Rasia2012, Rasia2014, Biffi2016, Ansarifard2020, Barnes2021}. Similar conclusions have been reached by \citet{Rines2016, Maughan2016, Foex2017, Lovisari2020} using a simpler approach based on galaxy kinematics and a smaller sample. It is well known that the $\sigma_8$ tension also occurs in weak lensing shear measurements, such as that from DES \citep{Amon2022, Secco2022}, KiDs \citep{vandenBusch2022}, HSC \citep{LiX2023}. Our results suggest that this tension also appears with galaxy kinematics, making it more general. Interestingly, the latest measurement using more than 5000 clusters of galaxies from eRASS1 \citep{Ghirardini2024} shows a value of $\sigma_8$ larger than almost all previous measurements, including that from Planck \citep{Planck2020VI, Planck2020VIII} and WMAP \citep{Hinshaw2013}. As noted earlier, this could be due to the higher weak lensing inferred cluster masses they used. 

One interesting trend we noticed is the slight decline of the dynamical-to-hydrostatic mass ratios towards large radii. It suggests that the inferred gravity from galaxy kinematics becomes smaller at larger radii with respect to the hydrostatic approach. This could be preliminary evidence that galaxies in the outskirts are accreting towards the center. We will need more cluster galaxies in the outskirts and deeper X-ray observations to confirm this behavior with higher confidence.

Another application of robust mass measurements of galaxy clusters is to test the universality of scaling relations, such as the radial acceleration \citep[RAR,][]{McGaugh2016PRL, OneLaw, Li2018}. The RAR is an empirical relation established in galaxies. It could be an emergent phenomenon during galaxy formation, similar to the abundance matching relation \citep{Moster2013, Behroozi2013, Kravtsov2018} and the mass-concentration relation \citep{Maccio2008, DuttonMaccio2014}. \citet{Navarro2017} showed that the RAR can be reproduced with the aforementioned two relations using a semi-empirical approach. \citet{Li2022} showed that the interplay between baryons and dark matter can significantly compress the dark matter halos of massive galaxies, leading to significant deviations to the RAR. So one might need to carefully consider the adiabatic contraction in order to reproduce the RAR and maintain stable dark matter halos simultaneously, given massive galaxies generally experience significant baryonic compression \citep{Li2022A&A, Li2023IAU}. The RAR could also be an indicator of modified gravity, particularly MOND \citep{Milgrom1983}, and its small scatter in disk galaxies seems to imply this interpretation \citep{Li2018, Desmond2023}. 

The key to distinguishing between these two interpretations is to test its universality, as galaxy formation is specific to galaxies, while a gravity theory should be universal in all systems. Previous studies using the hydrostatic approach \citep[e.g.][]{Eckert2022} and gravitational lensing \citep{Tian2020, Tian2024} consistently found galaxy clusters present higher total accelerations than what the RAR predicts with observed baryonic mass. These two approaches are fundamentally different from that used in rotating galaxies, where the RAR was initially established. To ensure consistency, \citet{Li2023} used a kinematic approach that employs cluster galaxies as dynamical tracers, but found similar deviations in the inner regions of clusters. In this paper, we constructed the RAR in galaxy clusters with three approaches. With a larger sample, we further confirm the missing mass problem in galaxy clusters as required by MOND using three independent approaches for the first time. The particularly interesting part in \citet{Li2023} is in the outskirt, where clusters cross below the RAR, suggesting too much mass is predicted by the RAR. If confirmed, it would rule out a universal RAR as there would be no room for adding additional baryonic mass components in galaxy clusters. Unfortunately, the cluster sample we used in this paper does not have a sufficient number of galaxies as dynamical tracers in the outskirts, so that we cannot further investigate this behavior. Future observations with SDSS-V \citep{Almeida2023} and the 4-meter Multi-Object Spectroscopic Telescope \citep{deJong2012} will enable further studies.

\begin{acknowledgements}
This work is based on data from eROSITA, the soft X-ray instrument aboard SRG, a joint Russian-German science mission supported by the Russian Space Agency (Roskosmos), in the interests of the Russian Academy of Sciences represented by its Space Research Institute (IKI), and the Deutsches Zentrum für Luft- und Raumfahrt (DLR). The SRG spacecraft was built by Lavochkin Association (NPOL) and its subcontractors, and is operated by NPOL with support from the Max Planck Institute for Extraterrestrial Physics (MPE).
\\
The development and construction of the eROSITA X-ray instrument was led by MPE, with contributions from the Dr. Karl Remeis Observatory Bamberg \& ECAP (FAU Erlangen-Nuernberg), the University of Hamburg Observatory, the Leibniz Institute for Astrophysics Potsdam (AIP), and the Institute for Astronomy and Astrophysics of the University of Tübingen, with the support of DLR and the Max Planck Society. The Argelander Institute for Astronomy of the University of Bonn and the Ludwig Maximilians Universität Munich also participated in the science preparation for eROSITA. The eROSITA data shown here were processed using the eSASS/NRTA software system developed by the German eROSITA consortium.
\\
P.L. acknowledges the support from the Alexander von Humboldt Foundation when staying at AIP where part of the work was done. A.L., M.K., E.B. V.G., C.G., and X.Z. acknowledge financial support from the European Research Council (ERC) Consolidator Grant under the European Union’s Horizon 2020 research and innovation program (grant agreement CoG DarkQuest No 101002585). Y.T. is supported by Taiwan National Science and Technology Council NSTC 110-2112-M-008-015-MY3. M.S.P. acknowledges funding of a Leibniz-Junior Research Group (project number J94/2020).
\end{acknowledgements}

\bibliographystyle{aa}
\bibliography{PLi}

\onecolumn

\begin{appendix}
\FloatBarrier
\section{Additional information for the cluster sample}

\begin{table}[ht!]
        \centering
        \caption{Alternative names and photon counts in the band of (0.2, 2.3) keV for the selected 22 clusters. }
        \label{tab:clusternames}
        \begin{tabular}{ccc}
                \hline
                1eRASS Cluster & Other names & Photon counts\\
                \hline
J004049.8-440743 & ACT-CL J0040.8-4407, PSZ2-G309.43-72.86, & 123 \\
& SPT-CL J0040-4407 & \\
J004207.1-283154 & Abell2811, MCXC-J0042.1-2832, & 944\\
& ACT-CL J0042.1-2832, PSZ2-G358.21-87.49, &\\
& SPT-CL J0042-2831, XCLASS-23788 & \\
J024339.2-483339 & ACT-CL J0243.5-4833, PSZ2-G265.10-59.50, & 188 \\
& SPT-CLJ0243-4833 & \\
J034656.2-543854 & ACT-CL J0346.9-5438, SPT-CL J0346-5439 & 125 \\
J043817.8-541917 & ACT-CL J0438.2-5419, PSZ2-G262.73-40.92, & 988\\
& SPT-CL J0438-5419, XCLASS-2656 & \\
J052806.2-525951 & ACT-CL J0528.1-5259, SPT-CL J0528-5300 & 47 \\
J055942.9-524851 & ACT-CL J0559.7-5249, PSZ2-G260.63-28.94, & 93 \\
& SPT-CL J0559-5249, XCLASS-2450 & \\
J073220.0+313748 & Abell586, MCXC-J0732.3+3137, & 361 \\
& PSZ2-G187.53+21.92, XCLASS-2775 & \\
J073721.1+351739 & Abell590 & 205\\
J080056.9+360324 & Abell611, MCXC-J0800.9+3602, & 163 \\
& XCLASS-2775 & \\
J082317.9+155700 & Abell657 & 98\\
J084257.5+362208 & Abell697, MCXC-J0842.9+3621, & 324 \\
& PSZ2-G186.37+37.26, XCLASS-2467 & \\
J090131.5+030055 & ACT-CL J0901.5+0301, eFEDS-J090131.1+030056 & 1322\tablefootmark{a} \\
J101703.2+390250 & Abell963, MCXC-J1017.0+3902, & 406 \\
& PSZ2-G182.59+55.83, XCLASS-1637 & \\
J115518.0+232422 & Abell1413, MCXC-J1155.3+2324, & 919 \\
& PSZ2-G226.18+76.79, XCLASS-1701 & \\
J121741.6+033931 & MCXC-J1217.6+0339, XCLASS-1967 & 2725 \\
J125922.4-041138 & Abell1651, ACT-CL J1259.3-0410, & 2684 \\
& MCXC-J1259.3-0411, PSZ2-G306.77+58.61 & \\
J130252.8-023059 & Abell1663, ACT-CL J1302.8-0230, & 532 \\
& MCXC-J1302.8-0230, XCLASS-580 & \\
J213056.8-645842 & SPT-CL J2130-6458 & 79 \\
J213536.8-572622 & ACT-CL J2135.6-5726, SPT-CL J2135-5726 & 46 \\
J213800.9-600758 & PSZ2-G333.89-43.60, SPT-CL J2138-6008 & 183 \\
J235137.0-545253 & ACT-CL J2351.6-5452, SPT-CL J2351-5452 & 62 \\
                \hline
        \end{tabular}
        \tablefoot{The names are from the cluster catalogs of Abell \citep{Abell1989}, MCXC \citep{Piffaretti2011}, X-CLASS \citep{Koulouridis2021}, eFEDS \citep{Liu2022}, ACT \citep{Hilton2021}, SPT \citep{Bocquet2019, Bleem2020}, or SPZ2 \citep{Planck2016SPZ2}. The photon counts are within $R_{500}$ and collected from \citet{Bulbul2024}.\\\tablefoottext{a}{Total photon counts from eRASS1 and eFEDS.}}
\end{table}
\clearpage

\section{Electron number density, temperature, gas mass and hydrostatic mass profiles for all clusters from multi-band X-ray image fits}

\begin{figure*}[h]
\centering
\includegraphics[scale=0.5]{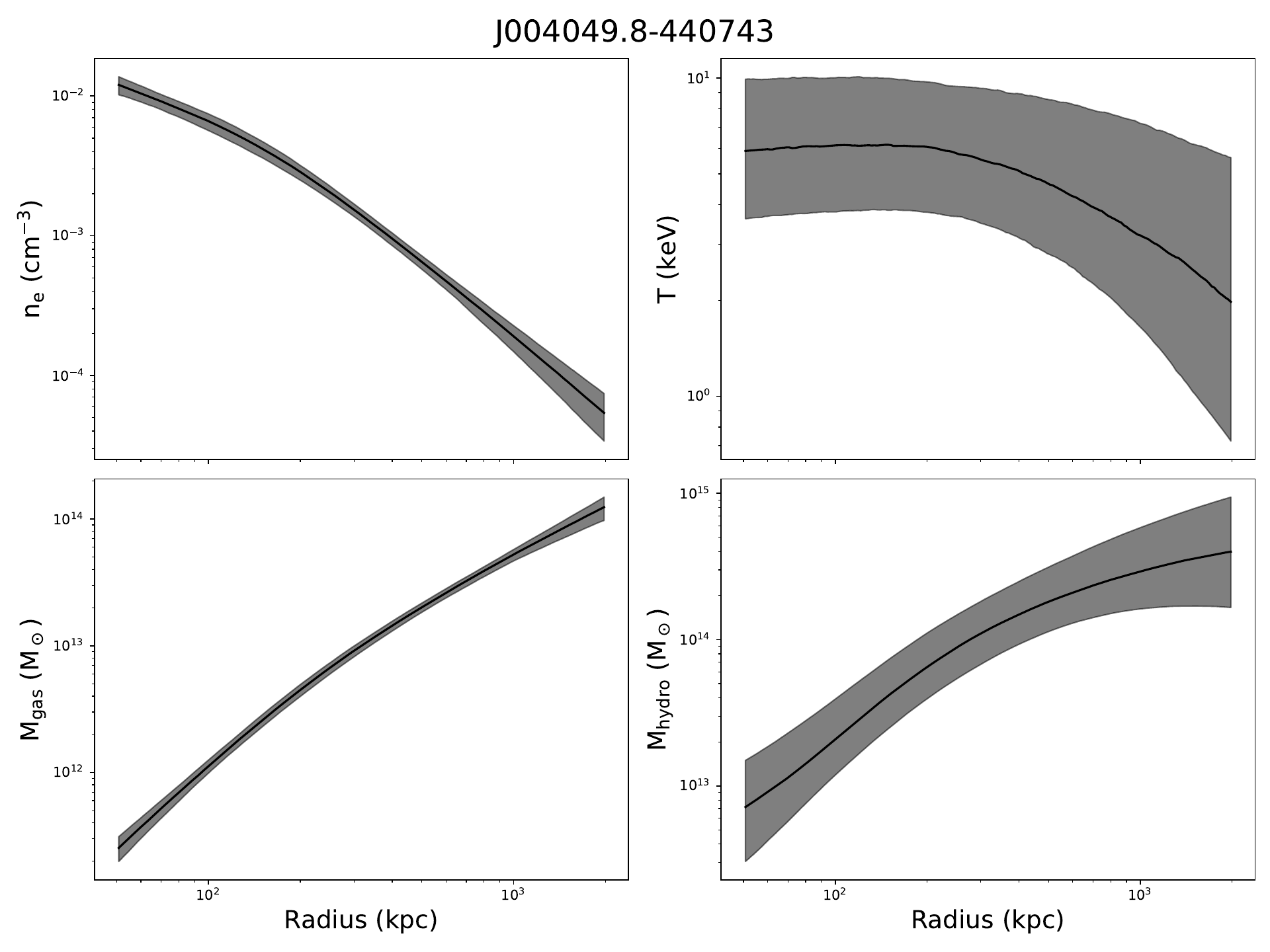}
\caption{Same as Figure \ref{fig:Xrayfits} but for the remaining 21 clusters.}
\end{figure*}

\renewcommand{\thefigure}{B.1}

\begin{figure}
\centering
\includegraphics[scale=0.5]{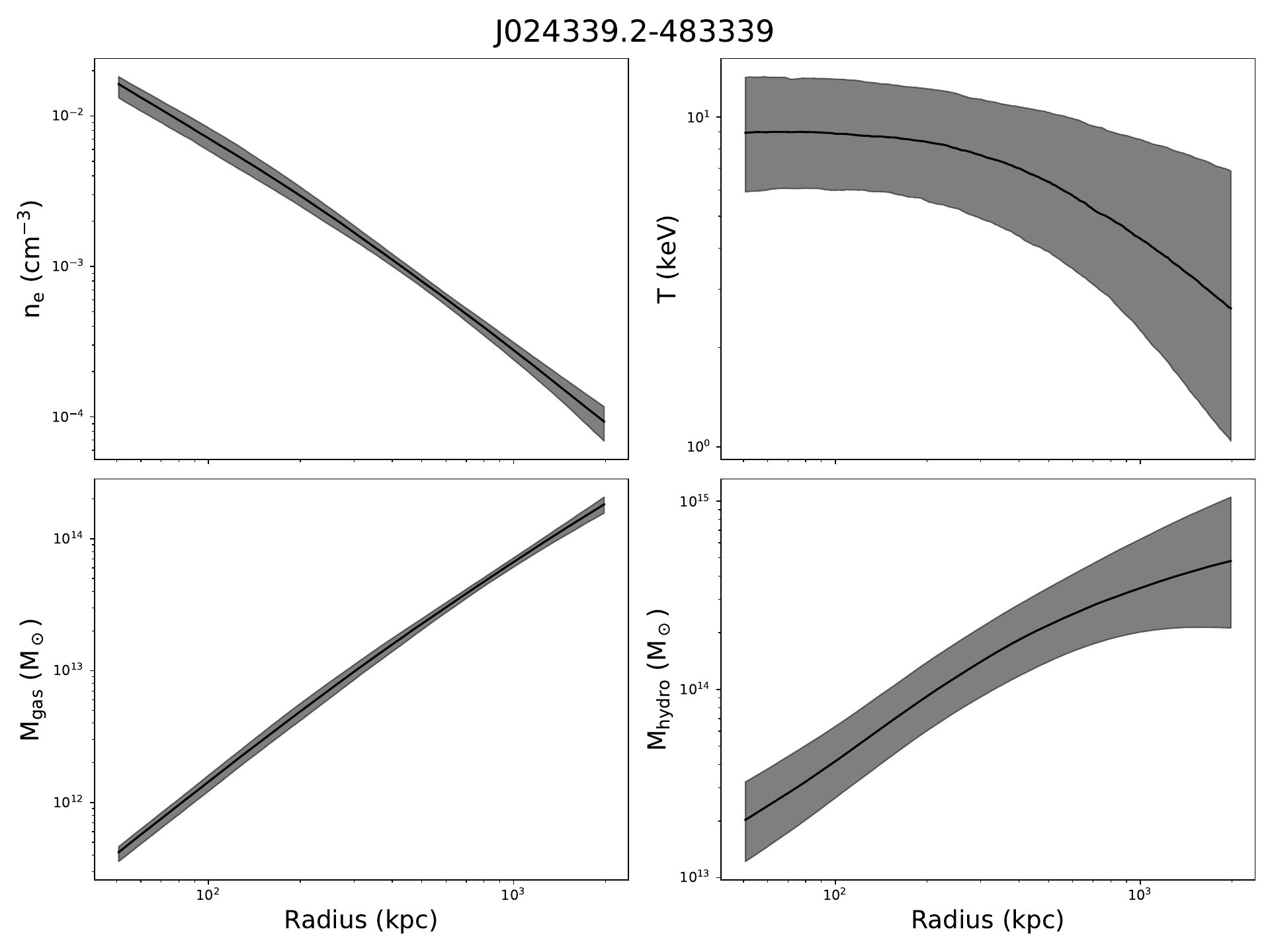}
\includegraphics[scale=0.5]{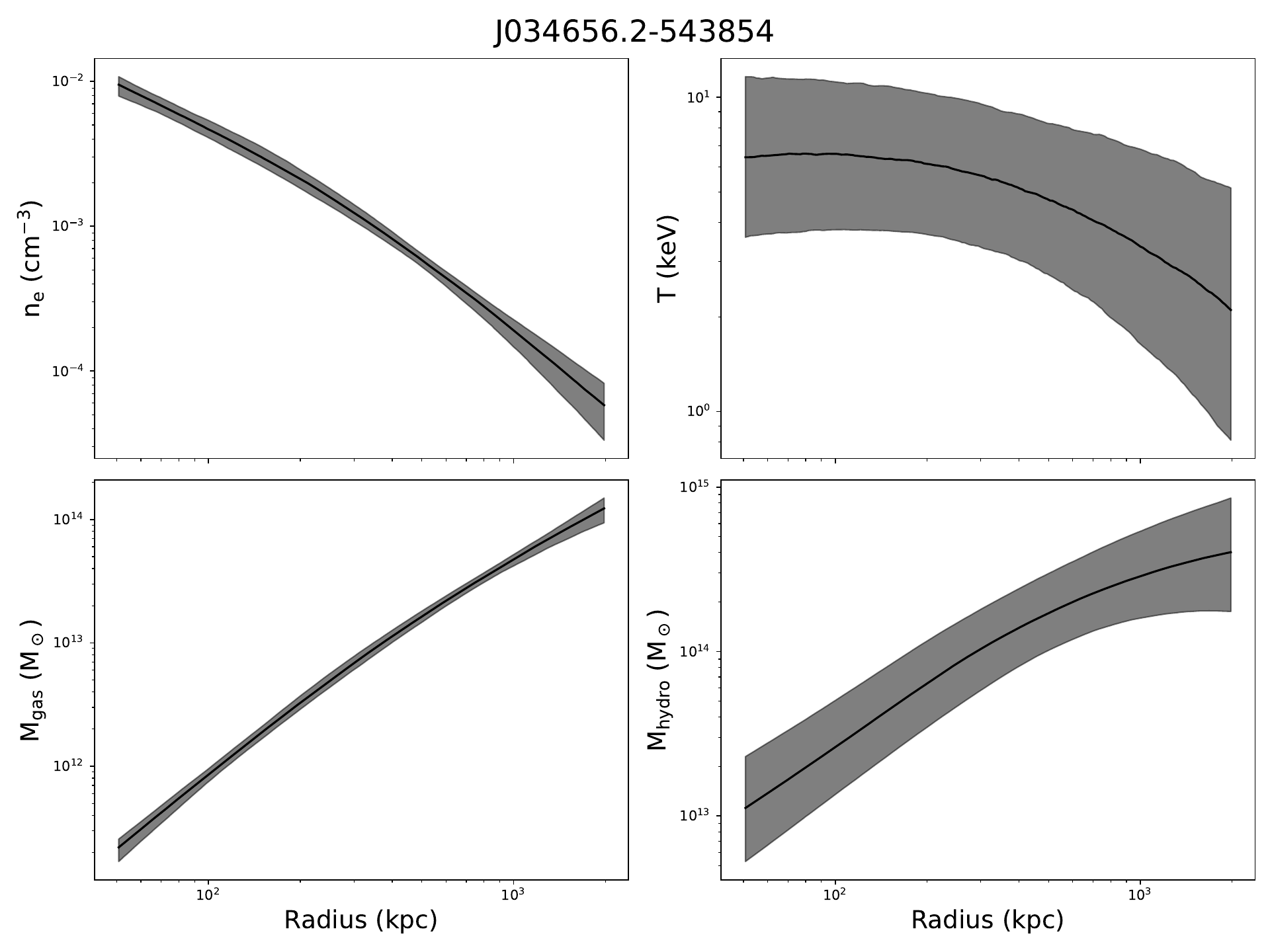}
\caption{Continued.}
\label{fig:figset}
\end{figure}

\begin{figure}
\centering
\includegraphics[scale=0.5]{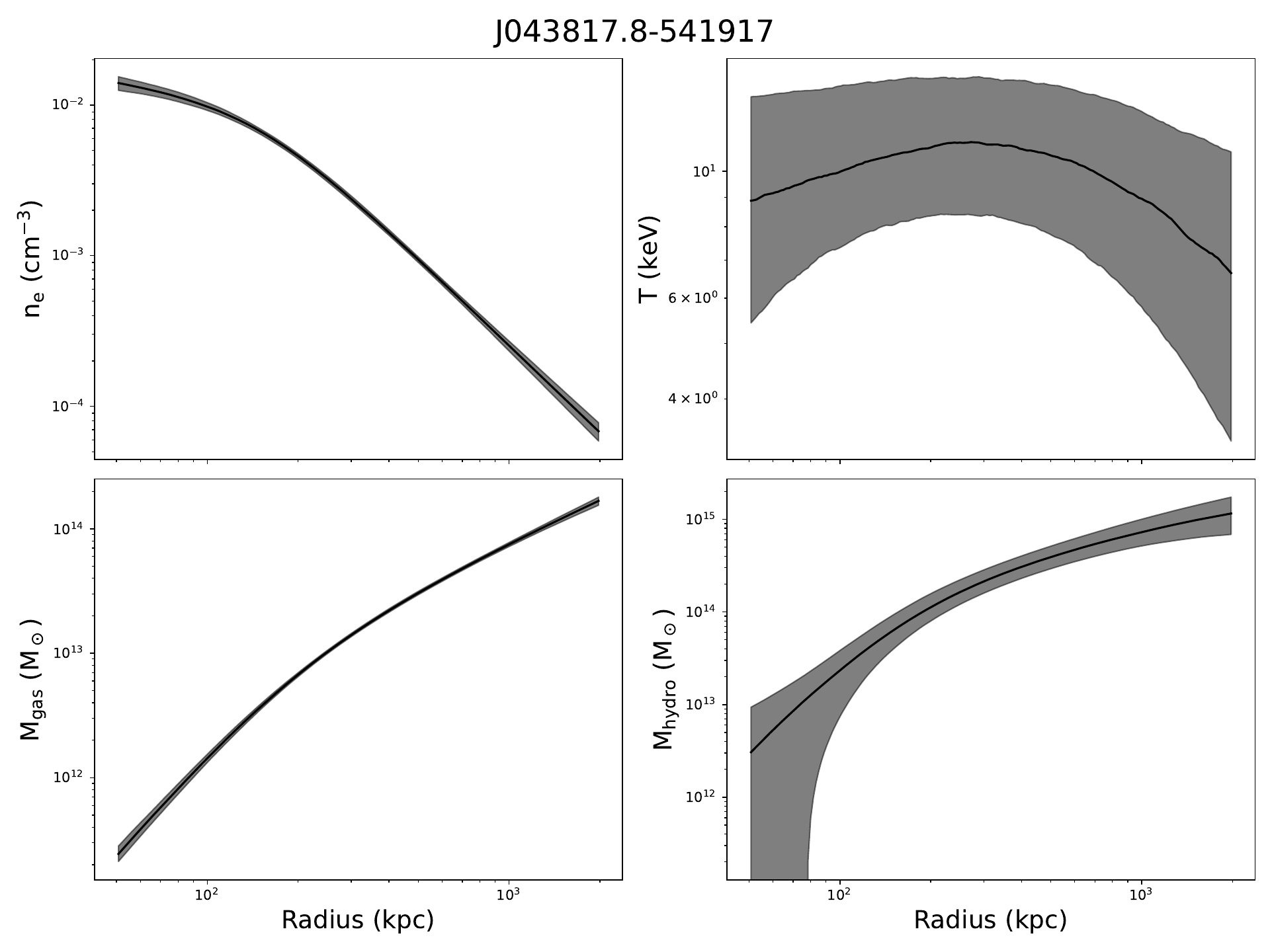}
\includegraphics[scale=0.5]{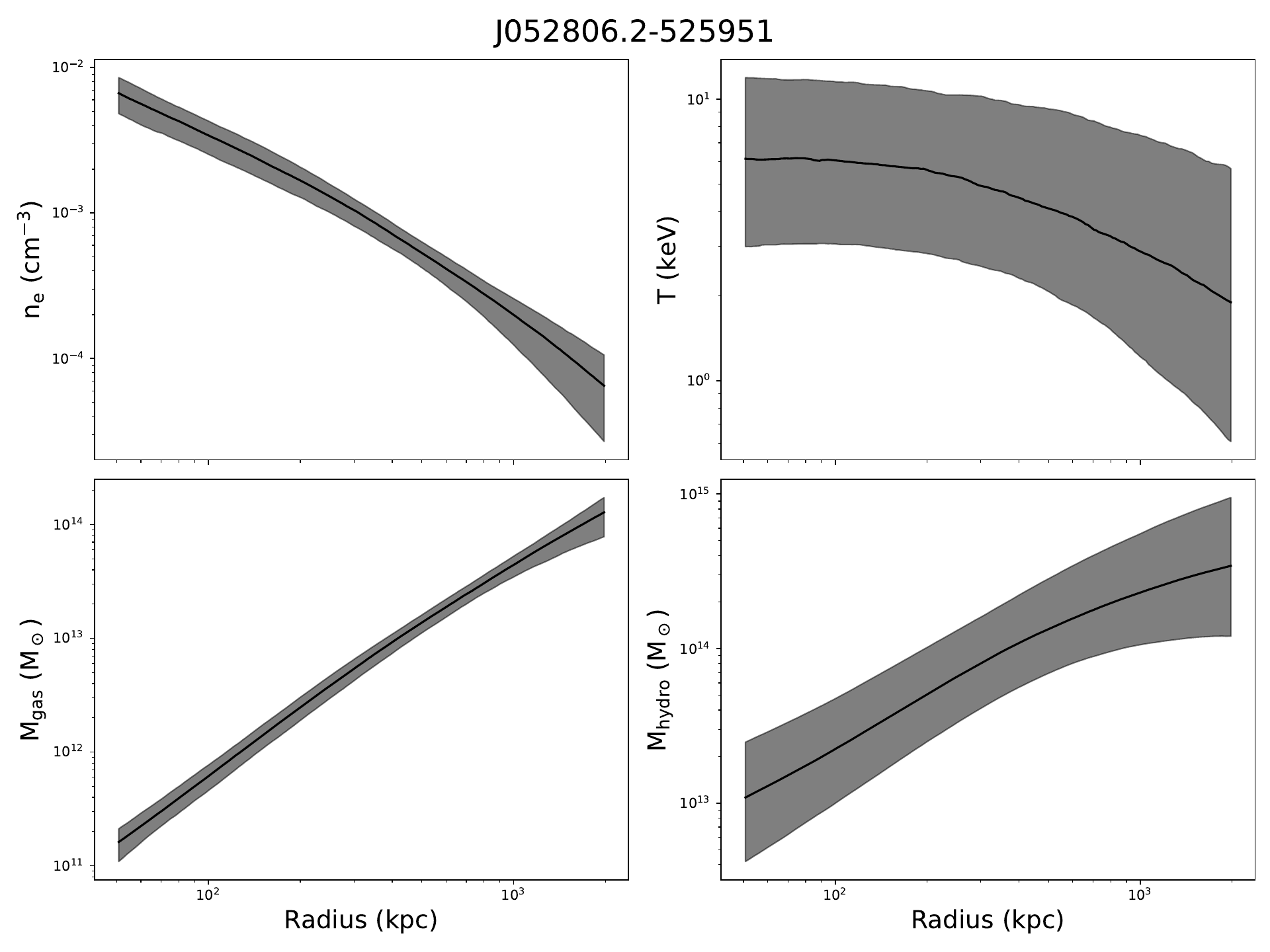}
\caption{Continued.}
\end{figure}

\begin{figure}
\centering
\includegraphics[scale=0.5]{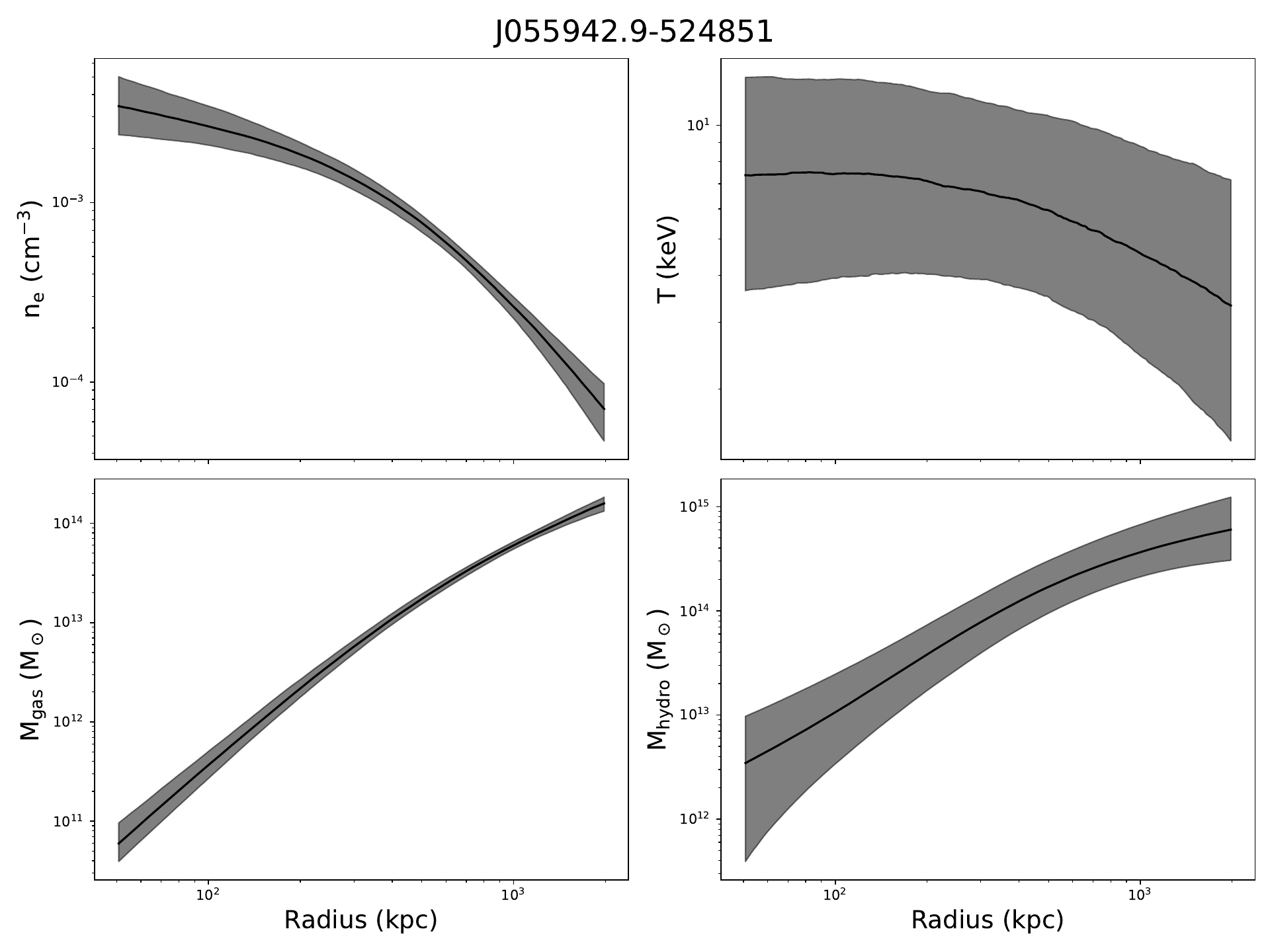}
\includegraphics[scale=0.5]{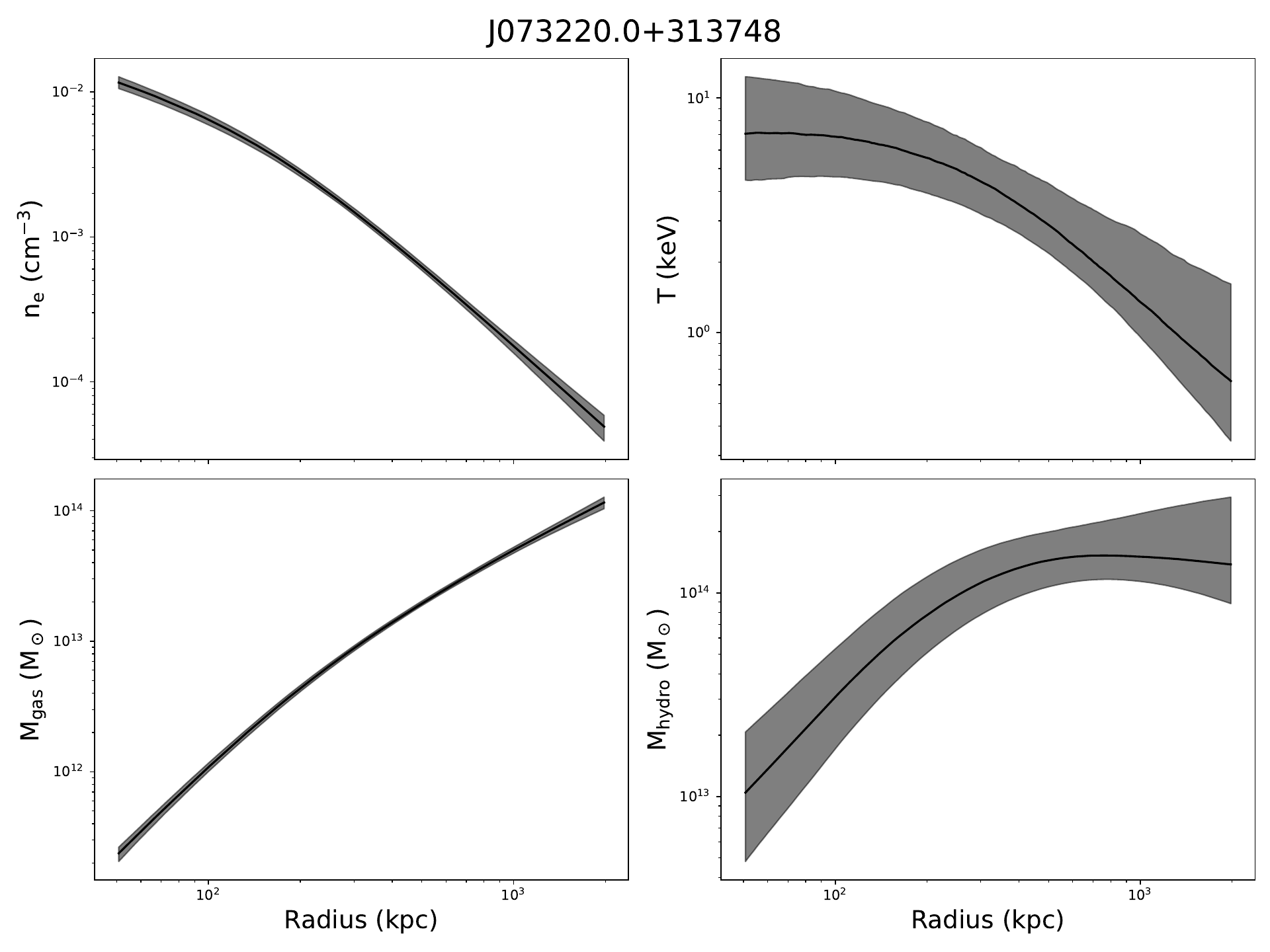}
\caption{Continued.}
\end{figure}

\begin{figure}
\centering
\includegraphics[scale=0.5]{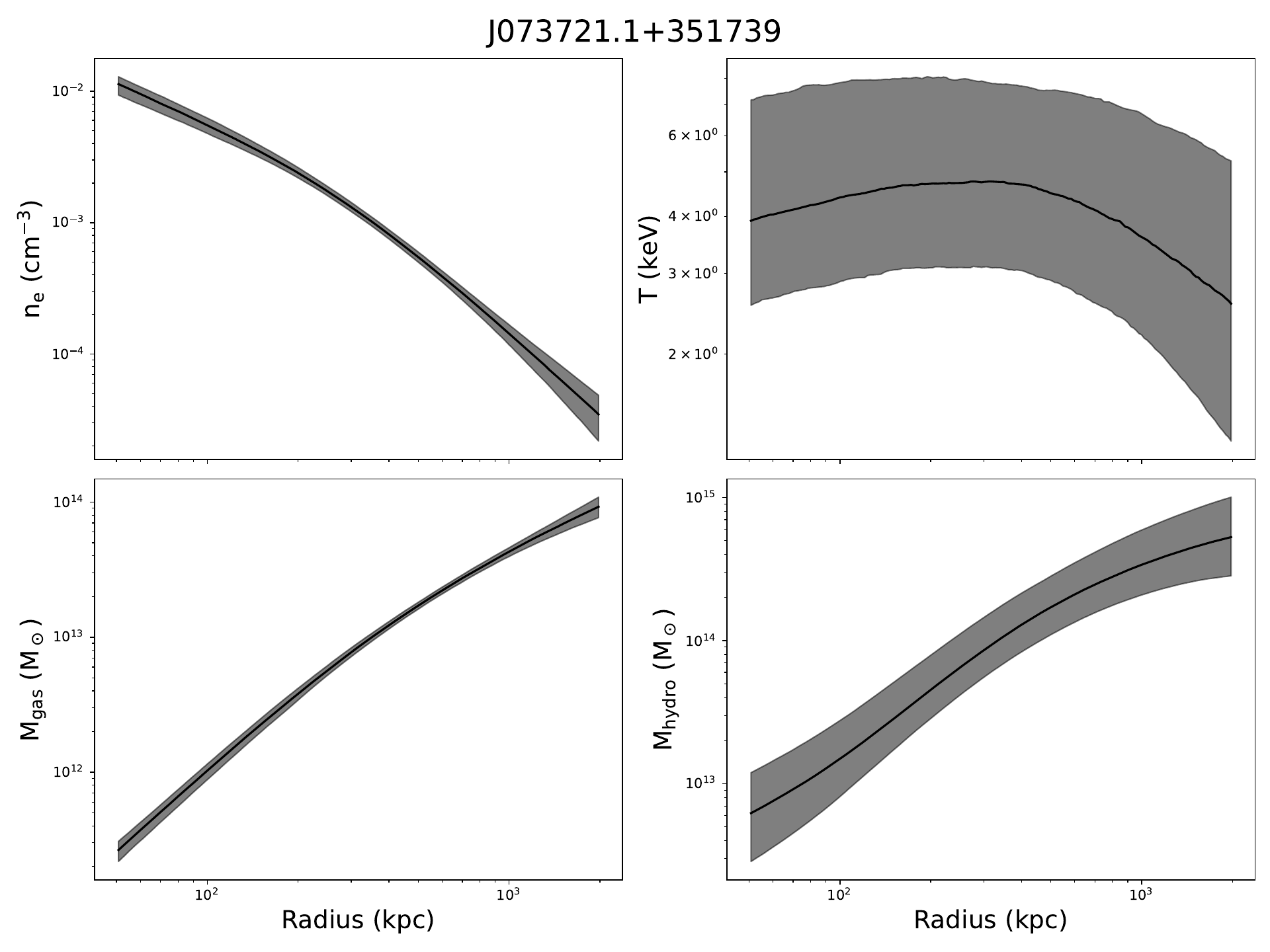}
\includegraphics[scale=0.5]{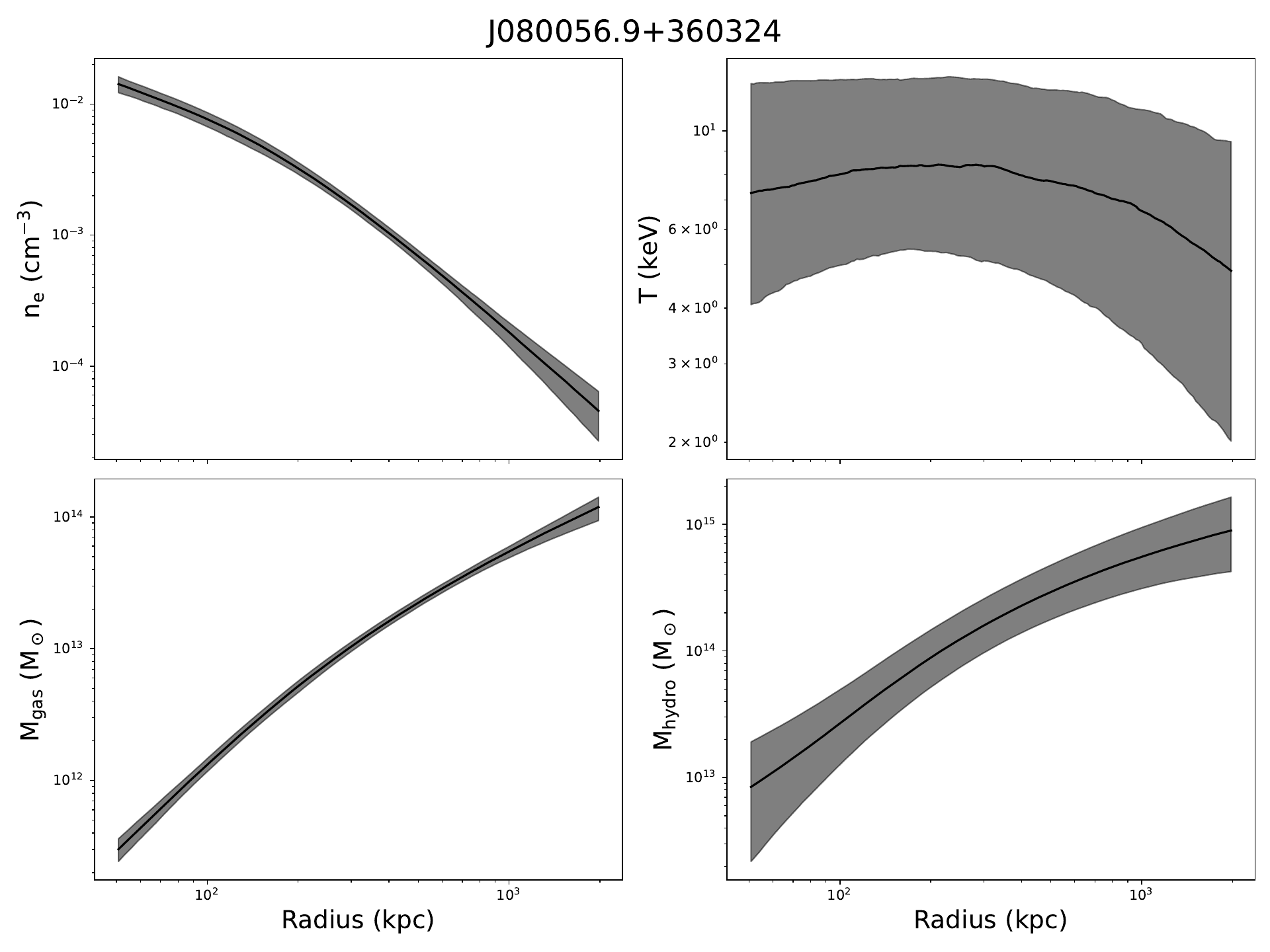}
\caption{Continued.}
\end{figure}

\begin{figure}
\centering
\includegraphics[scale=0.5]{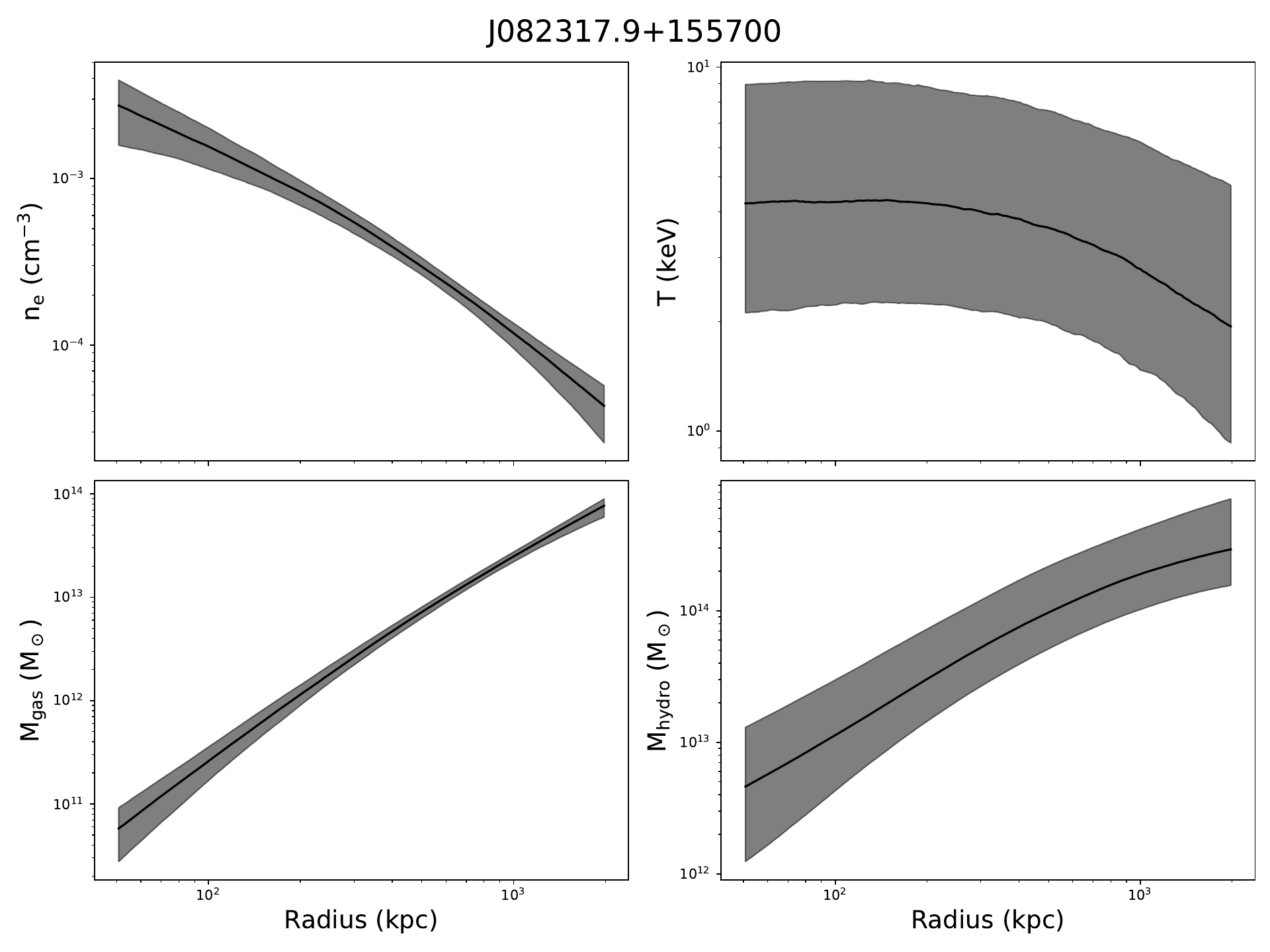}
\includegraphics[scale=0.5]{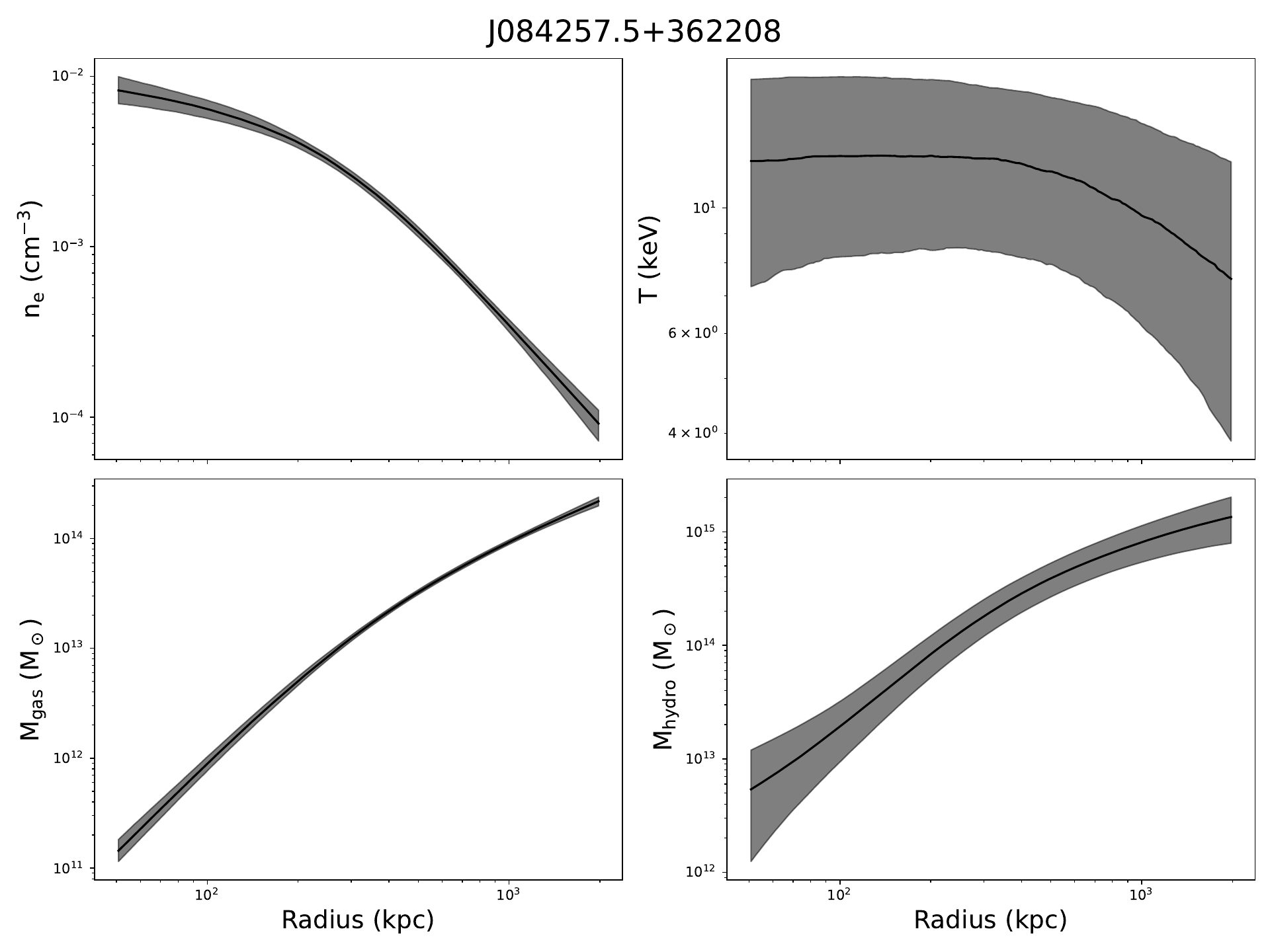}
\caption{Continued.}
\end{figure}

\begin{figure}
\centering
\includegraphics[scale=0.5]{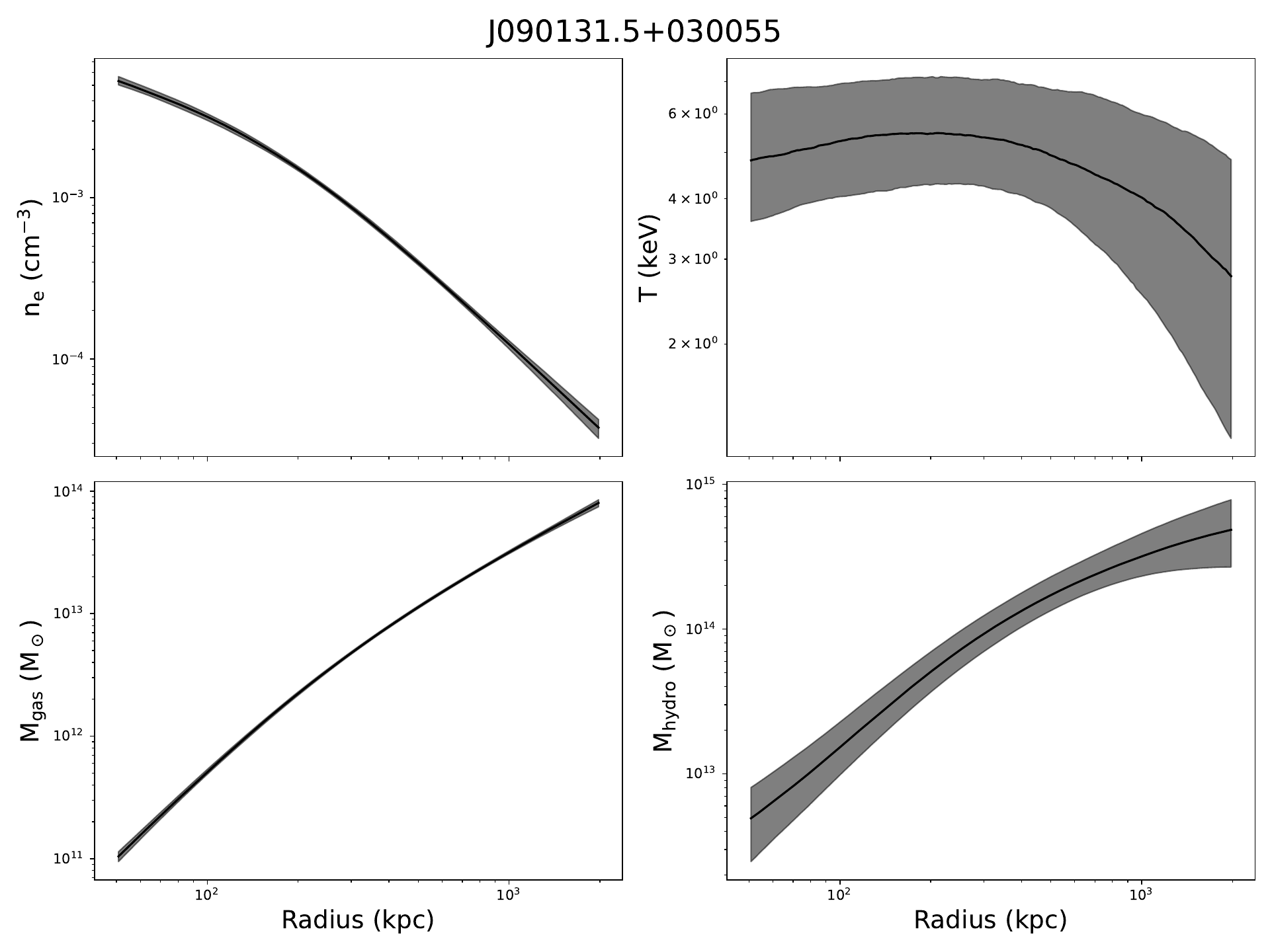}
\includegraphics[scale=0.5]{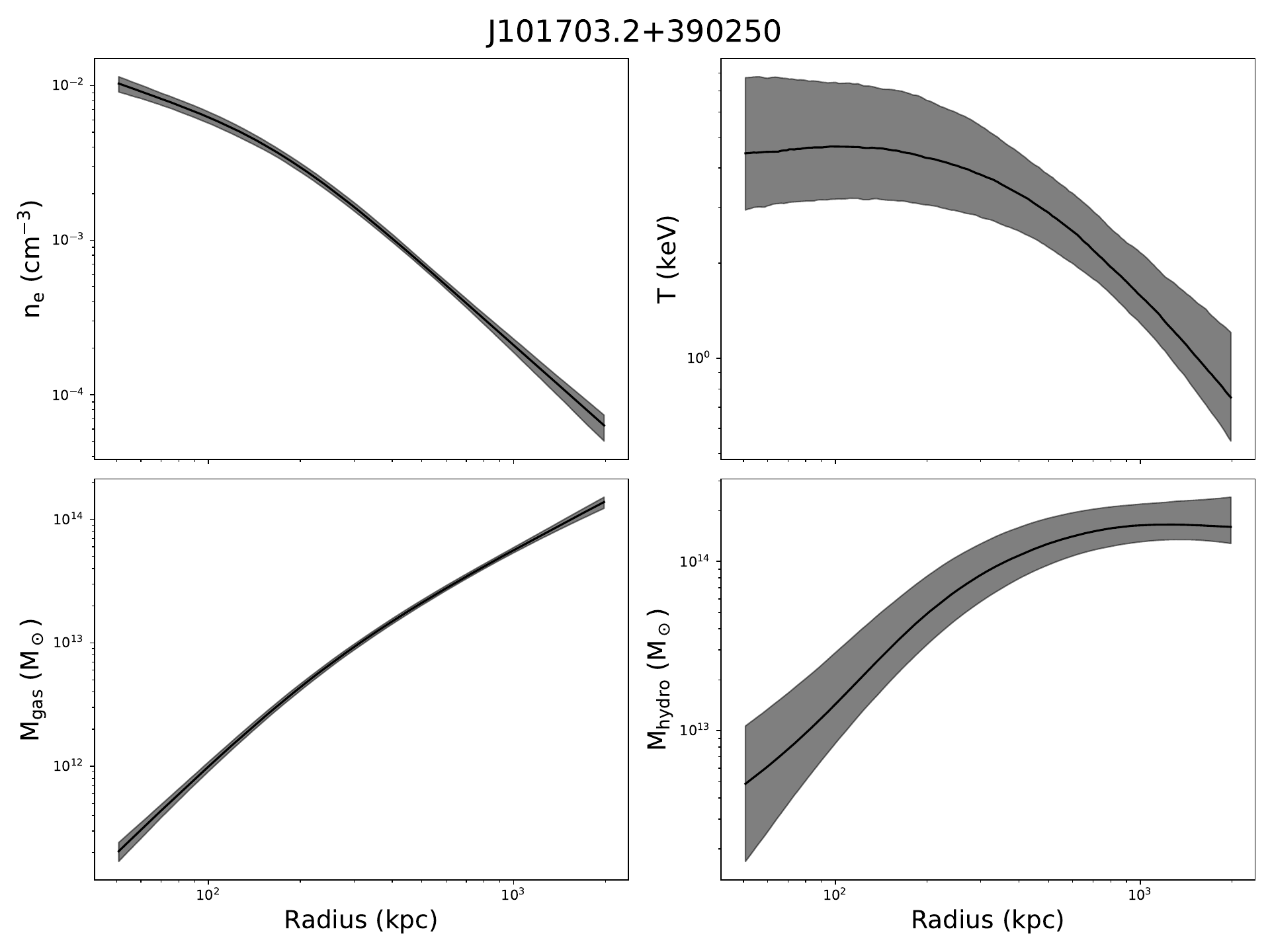}
\caption{Continued.}
\end{figure}

\begin{figure}
\centering
\includegraphics[scale=0.5]{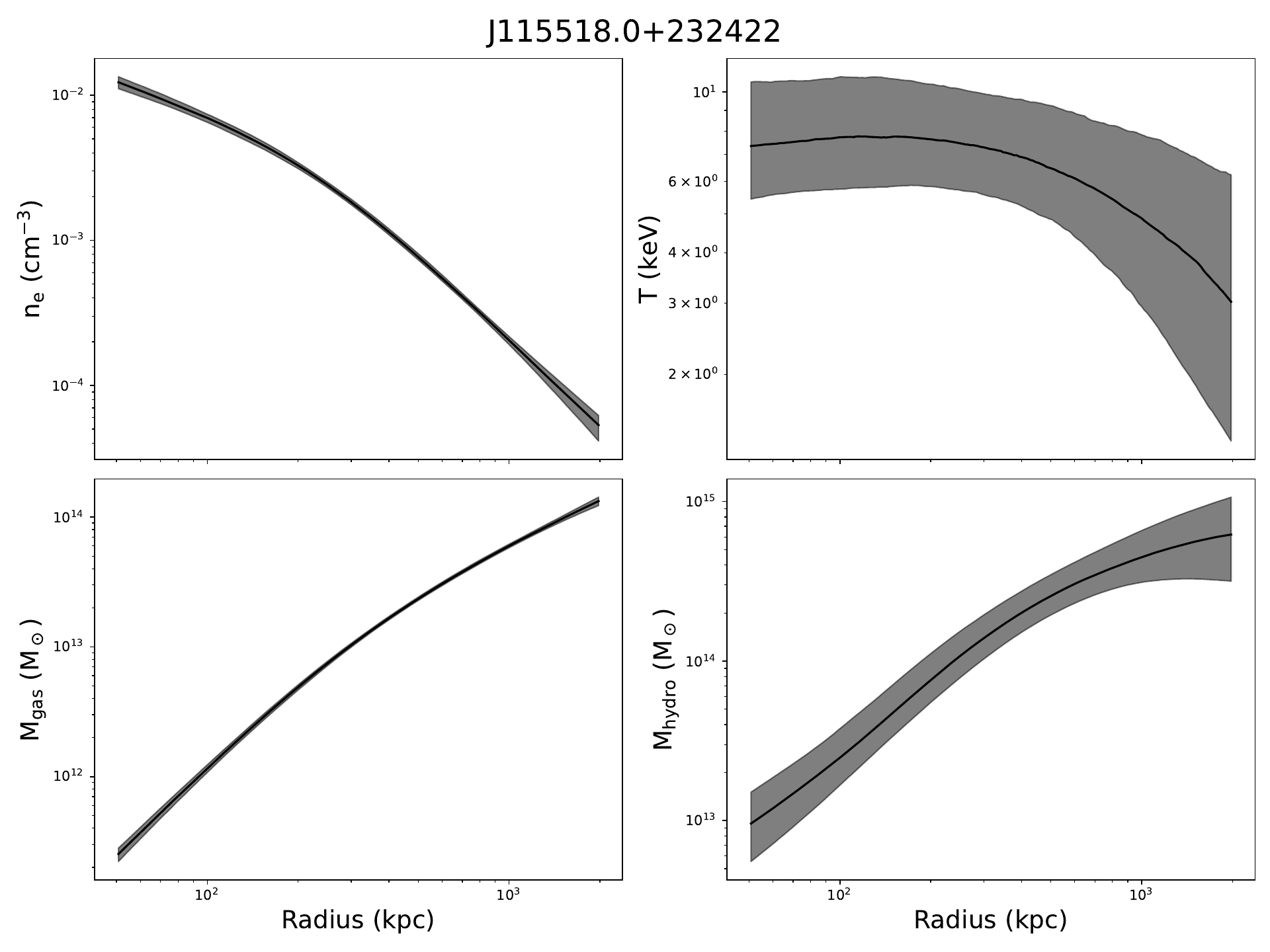}
\includegraphics[scale=0.5]{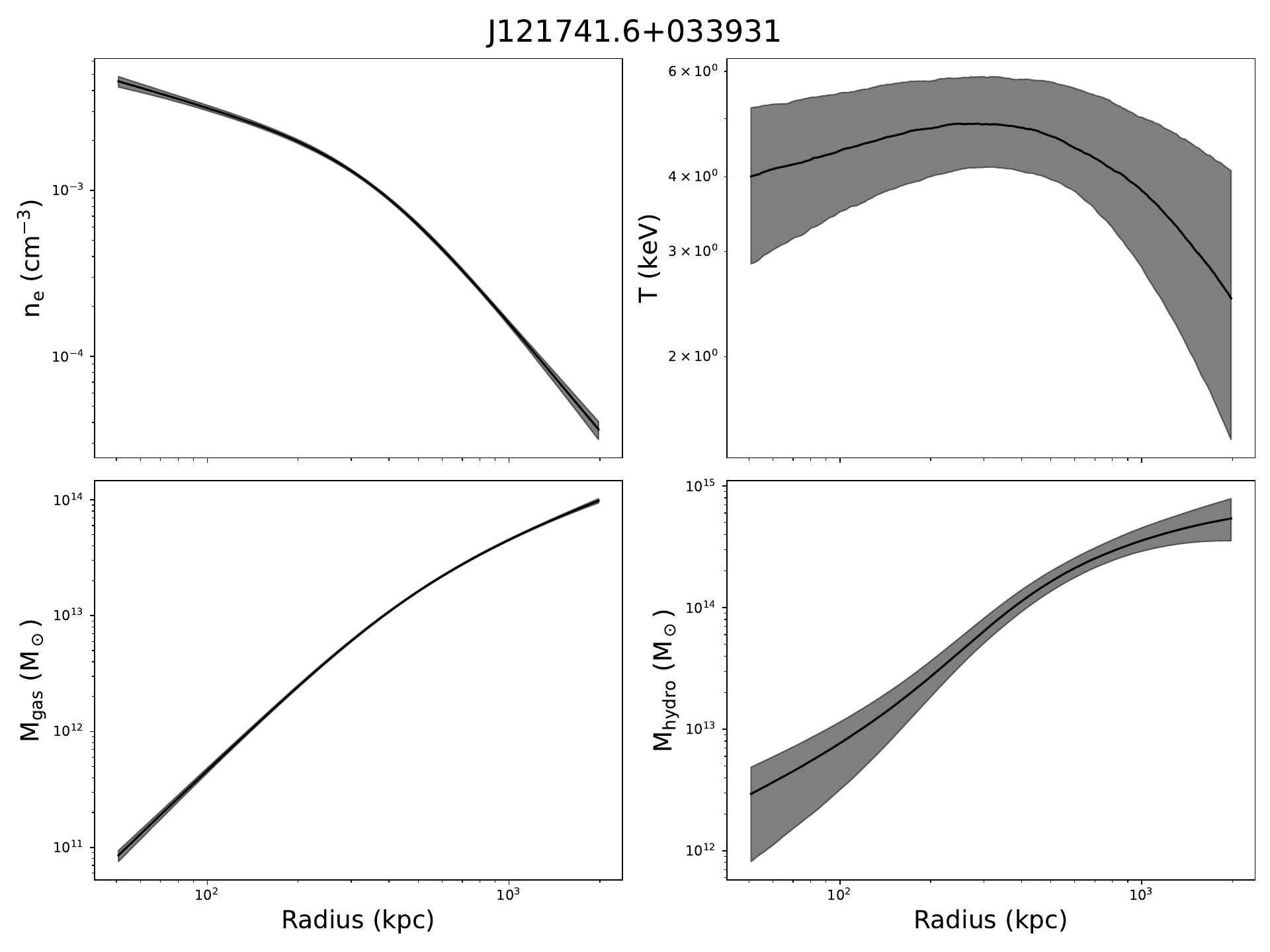}
\caption{Continued.}
\end{figure}

\begin{figure}
\centering
\includegraphics[scale=0.5]{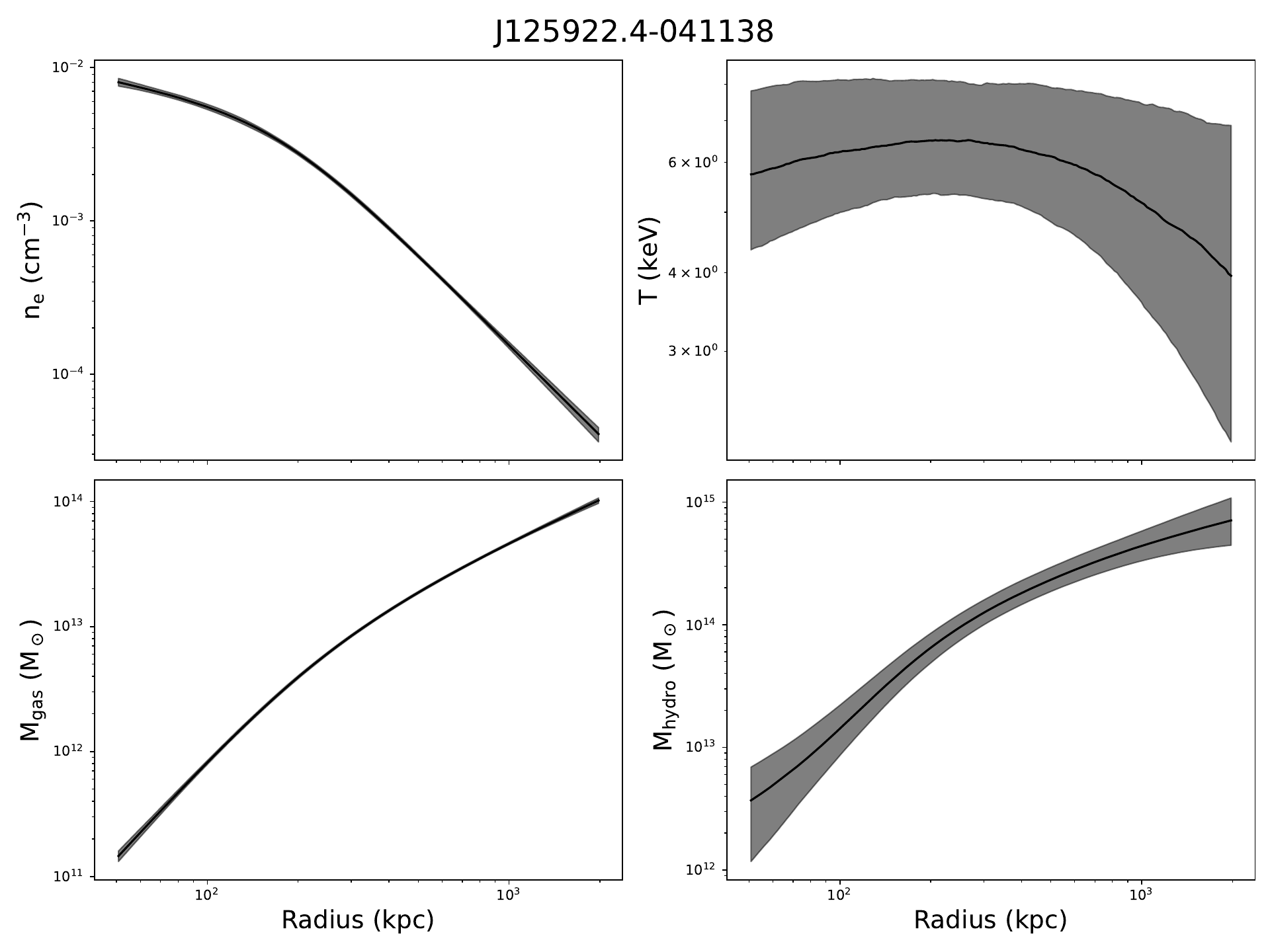}
\includegraphics[scale=0.5]{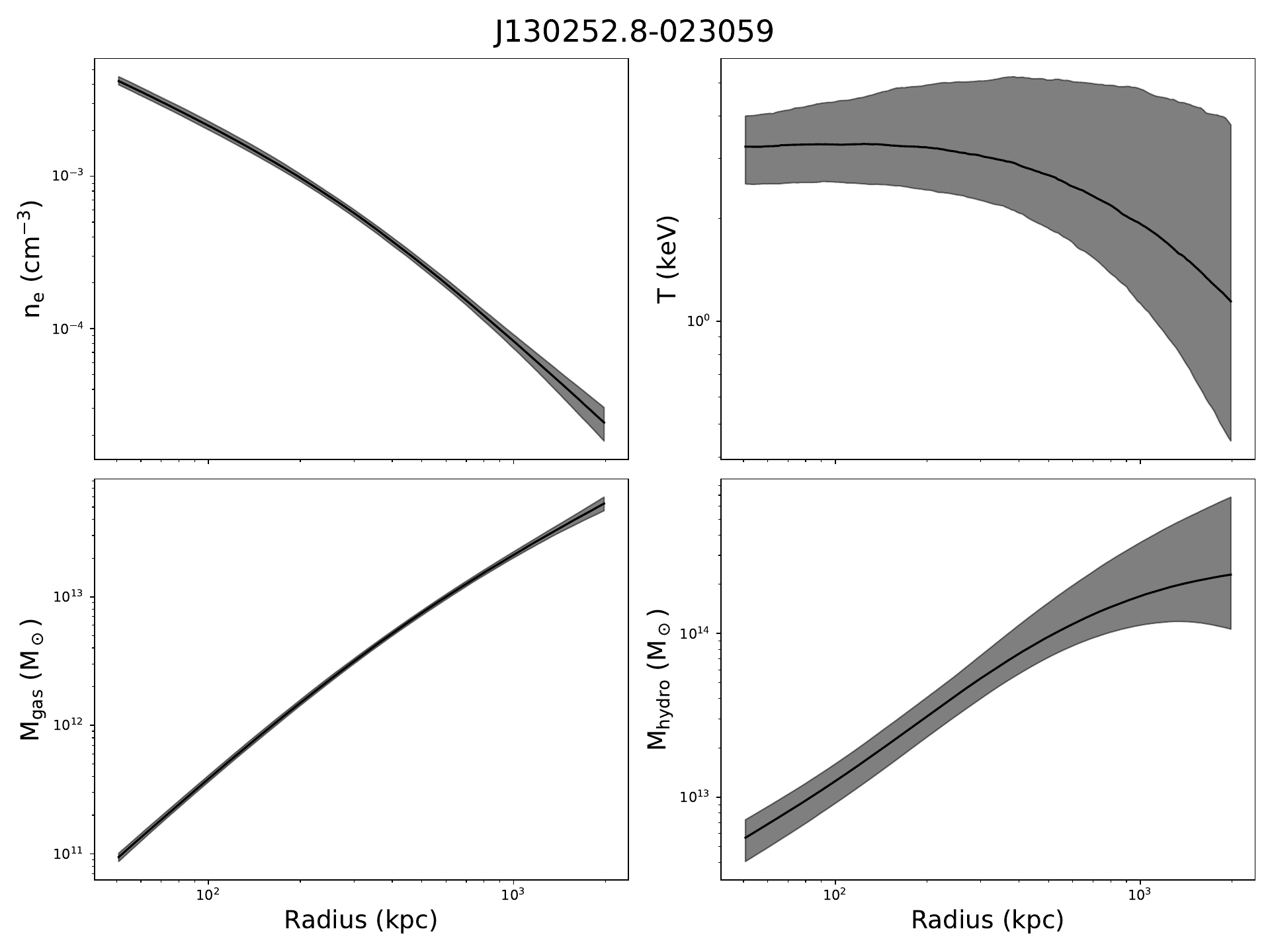}
\caption{Continued.}
\end{figure}

\begin{figure}
\centering
\includegraphics[scale=0.5]{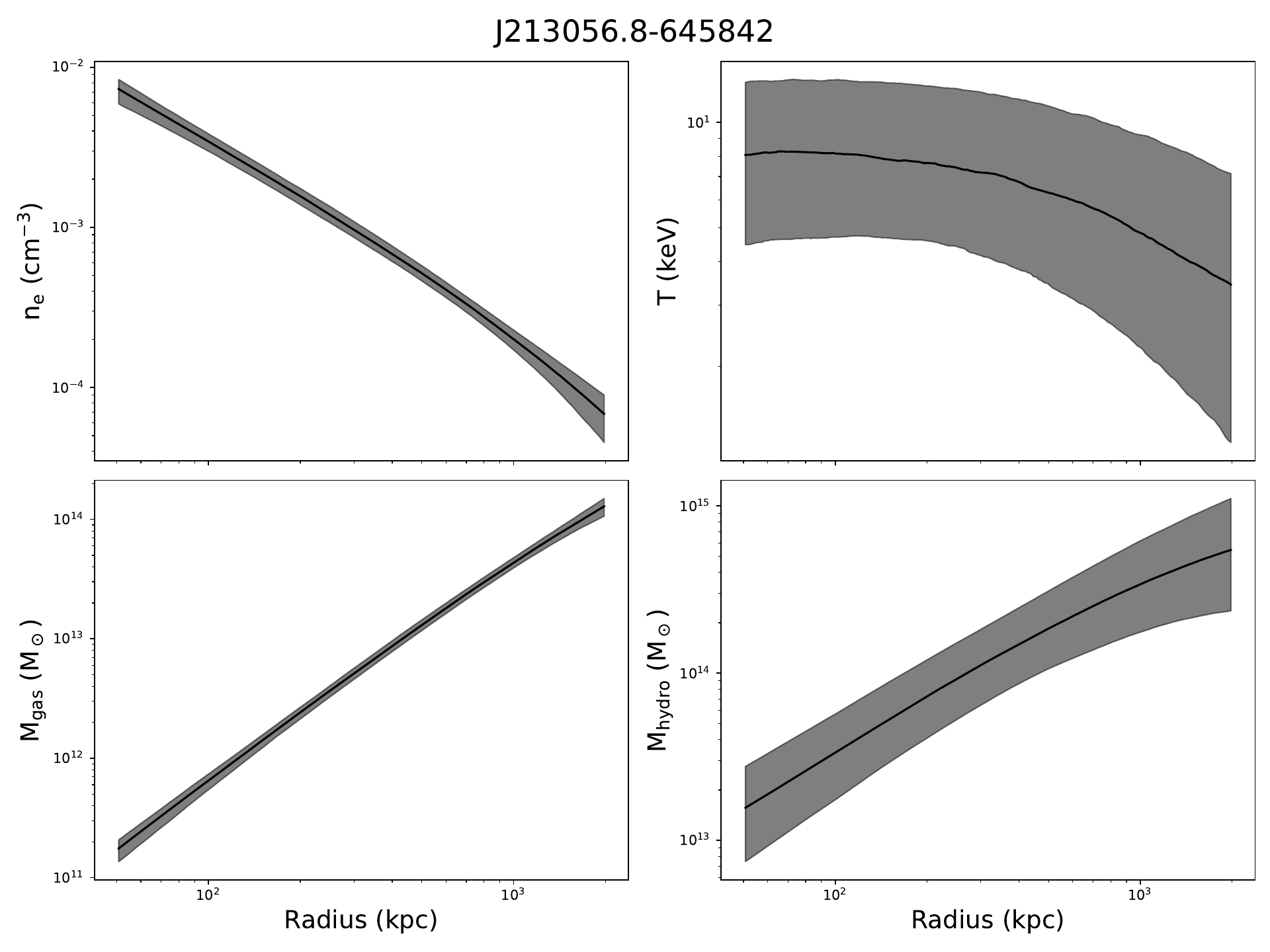}
\includegraphics[scale=0.5]{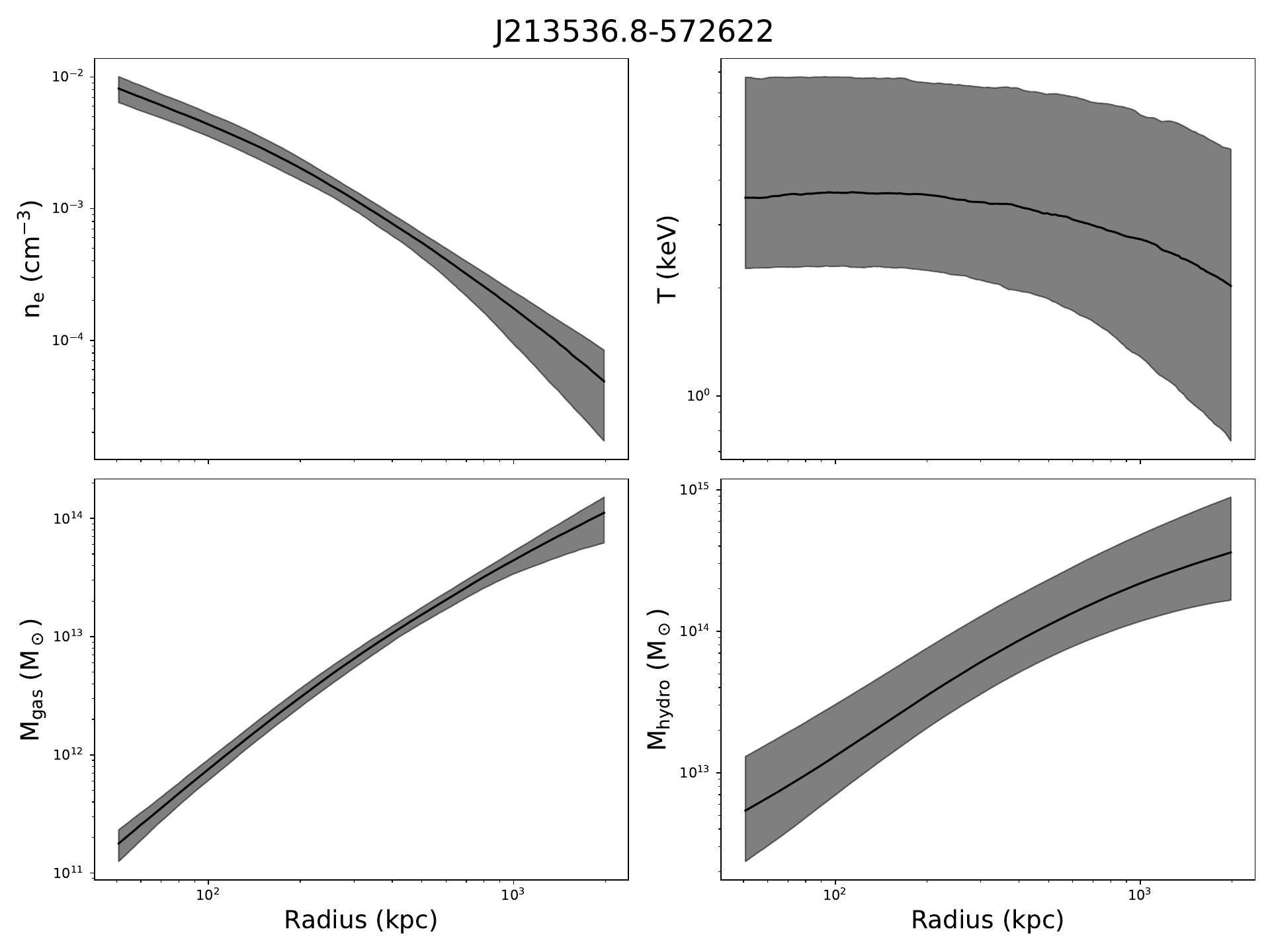}
\caption{Continued.}
\end{figure}

\begin{figure}
\centering
\includegraphics[scale=0.5]{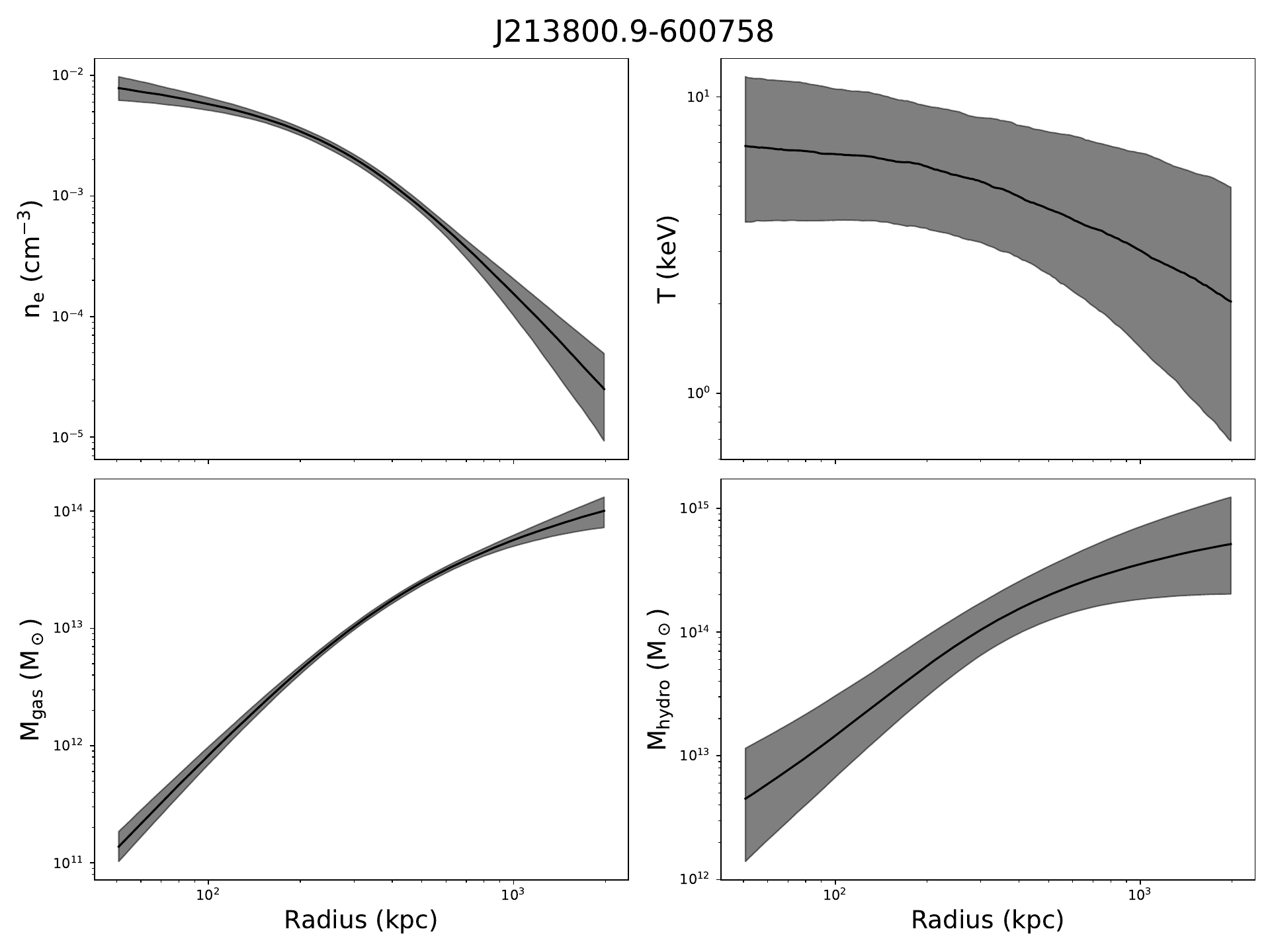}
\includegraphics[scale=0.5]{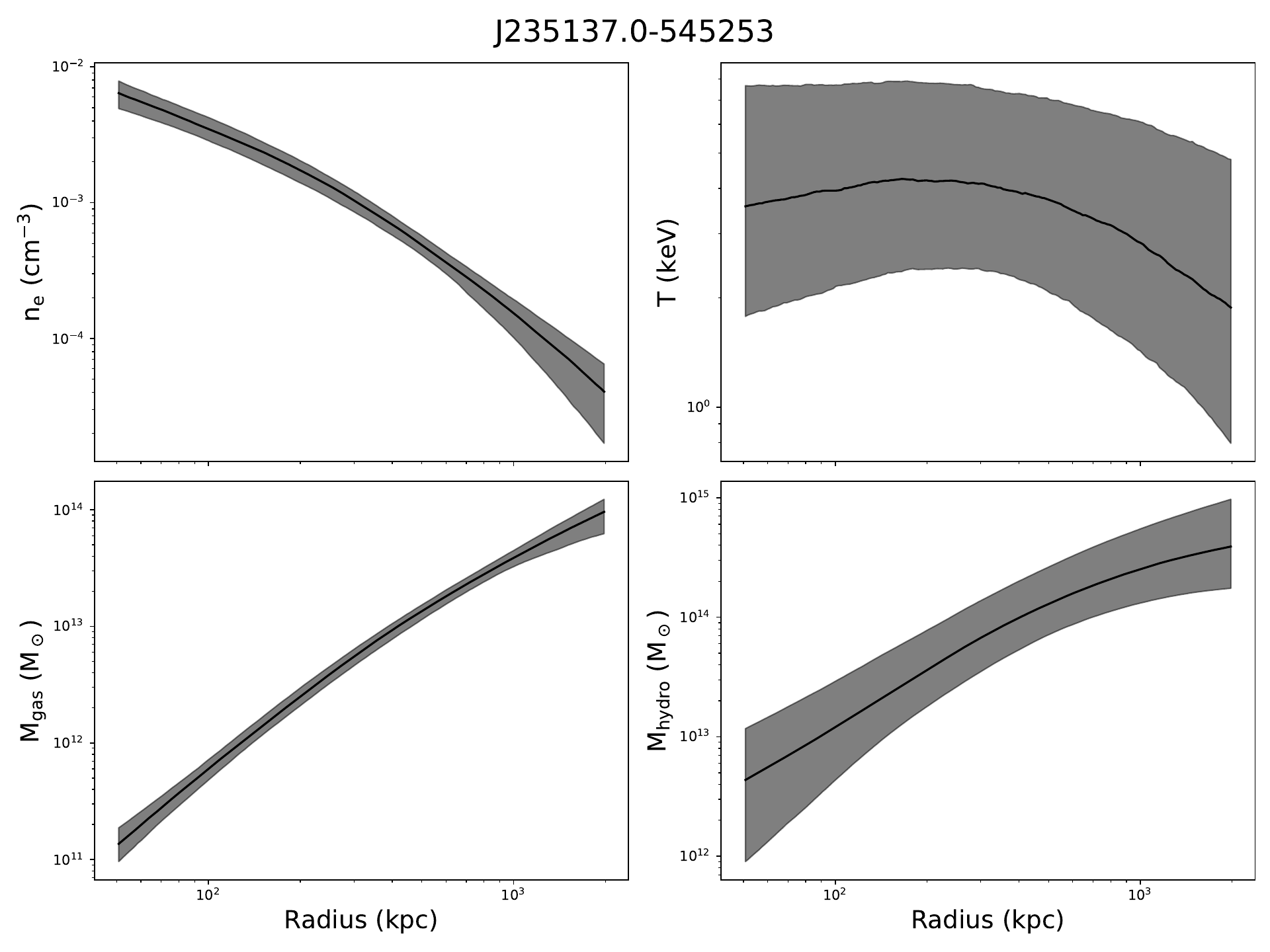}
\caption{Continued.}
\end{figure}

\end{appendix}

\end{document}